\documentclass[twocolumn,numberedappendix,appendixfloats]{emulateapj} 
\usepackage{epsfig} 
\usepackage{multirow} 
\usepackage{graphicx} 
\usepackage{epstopdf} 
\usepackage{amsmath, amsthm, amssymb} 
\usepackage{natbib} 
\usepackage{rotating} 

\newdimen\colwidth\colwidth=3.39in
\newdimen\rotwidth\rotwidth=4.5in
\newdimen\srotwidth\srotwidth=3.9in
\newdimen\ccolwidth\ccolwidth=5.0in

\def\deg{\degr}
\def\dex#1{10$^{#1}$}
\def\tdex#1{$\times$10$^{#1}$}
\def\kms{km\,s$^{-1}$}
\def\cmm#1{cm$^{-#1}$}

\def\l{$\lambda$}
\def\ll{$\lambda\lambda$}
\def\fu{erg\,cm$^{-2}$\,s$^{-1}$\,\AA$^{-1}$}
\def\Msun{M$_\odot$}

\def\Lstar{$L_*$}
\def\HST{{HST}}

\def\STIS{{STIS}}

\def\COS{{COS}}
\def\MAST{{MAST}}
\def\ch#1{\colhead{#1}}
\def\e{$\pm$}
\def\Lya{Ly$\alpha$}

\def\Lyd{Ly$\delta$}
\def\Lye{Ly$\epsilon$}
\def\Lyz{Ly$\zeta$}

\def\HI{\ion{H}{1}}
\def\CII{\ion{C}{2}}
\def\CIII{\ion{C}{3}}
\def\CIV{\ion{C}{4}}
\def\NI{\ion{N}{1}}

\def\OI{\ion{O}{1}}
\def\OVI{\ion{O}{6}}

\def\AlII{\ion{Al}{2}}
\def\SiII{\ion{Si}{2}}
\def\SiIII{\ion{Si}{3}}

\def\PII{\ion{P}{2}}
\def\SII{\ion{S}{2}}

\def\FeII{\ion{Fe}{2}}

\def\NiII{\ion{Ni}{2}}
\def\Fsizehist{1}

\def\Fboxmap{2}

\def\Fgalhist{3}

\def\Fqsomap{4}
\def\Fqsomapvel1{401}
\def\Fqsomapvel2{402}
\def\Fshifts{5}
\def\Ferror{6}
\def\Fsimulgals{7}
\def\Fgalprofile{8}

\def\Fsimullocs{9}
\def\Fsimulegb{10}
\def\Fionization{11}
\def\Fdensityhist{12}
\def\Fspec{13}
\def\Fbcomp{14}
\def\FimpEWb{15}
\def\Fcrossviewp{16}
\def\Fcrossegb{17}

\def\Fdetfrac{18}

\def\Tobslist{1}
\def\Timpact{2}
\def\Tmeasure{3}

\def\Sfilament{2}
\def\Sgalsample{2.1}
\def\Sfiladef{2.2}
\def\Sobserve{3}

\def\SfilAGN{3.2}
\def\SCOSdata{4}

\def\Ssimulations{5}

\def\Sresults{6}
\def\Sspectra{6.1}

\def\Scrosscuts{6.3}

\def\Sdetfrac{6.5}

\def\vminfilmap{2900}
\def\vmaxfilmap{4300}
\def\vminfiltab{2400}
\def\vmaxfiltab{4800}
\def\ext{pdf}

\begin{document}
\title{Nearby galaxy filaments and the \Lya\ forest: confronting simulations and the UV background with observations\footnotemark[*]}

\author{%
Bart P.\ Wakker\altaffilmark{1},
Audra K. Hernandez\altaffilmark{1},
David French\altaffilmark{1},
Tae-Sun Kim\altaffilmark{2},
Benjamin D. Oppenheimer\altaffilmark{3},
Blair D.\ Savage\altaffilmark{1}}

\altaffiltext{1}{Supported by NASA/NSF, affiliated with Department of Astronomy, University of Wisconsin, Madison, WI 53706; wakker@astro.wisc.edu, savage@astro.wisc.edu}
\altaffiltext{2}{Osservatoria Astronomico di Trieste, Via G.B. Tiepolo, 11, 34143, Trieste, Italy}
\altaffiltext{3}{CASA, Department of Astrophysical and Planetary Sciences, University of Colorado, Boulder, CO 80309, USA}

\footnotetext[*]{Based on observations taken by the NASA/ESA Hubble Space
Telesocpe, obtained at the Space Telescope Science Institute, which is operated
by the Association of Universities for Research in Astronomy, Incorporated,
under NASA contract NAS5-26555.}

\begin{abstract}
Simulations of the formation of large-scale structure predict that dark matter,
low density highly ionized gas, and galaxies form 10--40~Mpc scale filaments.
These structures are easily recognized in the distribution of galaxies. Here we
use \Lya\ absorption lines to study the gas in 30x6~Mpc filament at
$cz$$\sim$3500~\kms, defined using a new catalogue of nearby
($cz$$<$10,000~\kms) galaxies, which is complete down to a luminosity of about
0.05\,\Lstar\ for the region of space analyzed here. With \HST\ spectra of 24
AGN we sample the gas in this filament. All of our sightlines pass {\it outside}
the virial radius of any known filament galaxy. Within 500~kpc of the filament
axis the detection rate is $\sim$80\%, but no detections are seen more than
2.1~Mpc from the filament axis. The width of the \Lya\ lines correlates with
filament impact parameter and the four BLAs in our sample all occur within
400~kpc of the filament axis, indicating increased temperature and/or
turbulence. Comparing to simulations, we find that the recent Haardt \& Madau
(2012) extragalactic ionizing background predicts a factor 3--5 too few ionizing
photons. Using a more intense radiation field matches the hydrogen density
profile within 2.1~Mpc of the filament axis, but the simulations still
overpredict the detection rate between 2.1 and 5~Mpc from the axis. The baryonic
mass inside filament galaxies is 1.4\tdex{13}\,\Msun, while the mass of filament
gas outside galaxy halos is found to be 5.2\tdex{13}\,\Msun.
\end{abstract}

\parskip=0pt


\section{Introduction} 
\par The current paradigm for the formation of large-scale structure holds that
after the Big Bang dark matter is concentrated by gravity into sheets, filaments
and halos. Observational (e.g.\ Fukugita \& Peebles 2006, Shull et al.\ 2012,
Danforth et al.\ 2014) and theoretical (e.g.\ Cen \& Ostriker 1999; Dave et al.\
2001; Smith et al.\ 2011, Cen 2013) studies suggest that the baryons are carried
along, with a small fraction ($\sim$10\% at the present time) forming galaxies
inside dark matter halos. The remaining baryons stay in the circumgalactic (CGM)
and intergalactic (IGM) medium, with at $z$=0 about 30$\pm$10\% in the form of
photoionized \HI\ at \dex4~K (Penton et al.\ 2002; Lehner et al.\ 2007; Danforth
\& Shull 2008; Shull et al. 2012), and 40\% to 60\% at higher temperatures in
the Warm-Hot Intergalactic Medium (WHIM). All these different processes result
in the Cosmic Web of dark matter, galaxy and gas filaments.
\par The galaxies represent a small fraction of the baryons. As they are
luminous and energetic, our understanding is relatively advanced. However, to
fully understand the development of structure we should also understand the 10
times more numerous (but more difficult to observe) baryons {\bf outside of
galaxies} still in the IGM.
\par Hydrodynamical simulations can be used to describe the evolution of \Lya\
lines from high redshifts until the present. They predict the gradual
disappearance of the \Lya\ forest from $z$=3 to $z$=0, the increase in
temperature of the IGM over time, and the nature of the association of
intergalactic gas with galaxies (Dav\'e et al.\ 1999). Simulations also predict
that the optical depth of \HI\ absorption is closely tied to the underlying
overdensity of dark matter (e.g.\ Croft et al.\ 1998; Schaye 2001), and give a
specific expectation for the hydrogen column density profile perpendicular to
the dark matter filaments. \par Observationally, the decrease in the number
counts and column density of \Lya\ absorbers over time provides evidence for the
theoretical picture of the IGM (e.g. Weymann et al.\ 1998; Kim et al.\ 2007,
2013; Rudie et al.\ 2013; Rahmati et al.\ 2013). Other {\it statistical}
evidence comes from the spatial association between strong (log $N$(\HI)$>$14)
\Lya\ absorbers and galaxies (Morris \& Januzzi 2006; Ryan-Weber 2006; Stone et
al.\ 2010) and from the power spectrum of the \Lya\ absorbers (see e.g.
Paschos et al.\ 2009; Lee et al.\ 2015 and references therein).
\par To turn the simulated hydrogen density into a prediction for the observable
\HI\ column density requires a prescription for the intensity of the ionizing
flux in the extragalactic background (EGB). Until recently, the most widely used
prescription was provided by Haardt \& Madau (2001), based on the spectra and
number density of quasars and galaxies and the evolution of those quantities
over time. They updated their model in 2012 (Haardt \& Madau 2012), but this
later version has much lower (a factor 3.7) ionizing flux at $z$=0.
\par At $z$=0 Dav\'e et al.\ (2010) compared the simulated column density
distribution of \Lya\ absorbers to the observations of Lehner et al.\ (2007),
showing a relatively good fit. However, with the updated version for the EGB
given by Haardt \& Madau (2012), and a larger sample of low redshift \Lya\
absorbers (Danforth et al.\ 2014), Kollmeier et al.\ (2014) found a factor of
five times too few photons are produced to ionize the \Lya\ forest to the
observed levels, by comparing the expected and observed column density
distribution of \Lya\ absorbers. In this paper we will confirm this deficit
using a different method. The discrepancy has been further analyzed, with Khaire
\& Srianand (2015) arguing that the quasar contribution is higher by a factor of
two, and additionally suggesting a 4\% escape fraction of Lyman continuum
photons from star-forming galaxies to match the opacity inferred from the \Lya\
forest. Shull et al.\ (2015) use different simulations to reduce the \HI\ column
densities, probably due to extremely strong feedback and argue that the Haardt
\& Madau (2012) quasar emissivity combined with $\sim$5\% escape fraction will
produce the observed \Lya\ forest. 
\par Most of the effort to understand absorption lines from the IGM has gone
into work on the circumgalactic medium of galaxies (the gas within
$\sim$300~kpc), using ensemble studies of individual galaxy halos (see e.g.\
Morris et al.\ 1993; Lanzetta et al.\ 1995; Prochaska et al.\ 2006; Stocke et
al.\ 2006; Wakker \& Savage 2009; Rudie et al.\ 2012; Tumlinson et al.\ 2013;
Stocke et al. 2013 and many references in those papers). Yet, the most visually
striking prediction of the hydrodynamic simulations is the presence of
large-scale filamentary structure. Such filamentary structure is clearly evident
in the distribution of galaxies, but this gas has not been directly studied.
Crosscuts through simulations (see e.g.\ Fig.~2 in Luki\'c et al.\ 2015) clearly
show these filaments in the \HI\ distribution.
\par The possible presence of filaments can be suggested by looking at the
distribution of \Lya\ absorbers in a single sightline. Morris et al.\ (1993)
analyzed a deep galaxy sample within a degree of the 3C\,273 sightline, combined
with a shallower sample out to 10~Mpc. They found that the distribution of
galaxies and absorbers as function of redshift shows definite peaks (indicating
the presence of galaxy filaments), and that absorbers are not randomly
distributed with respect to the galaxies; they also found absorbers that seemed
to have no associated galaxies within 1~Mpc. In the same manner, Tripp et al.\
(1998) also showed that both absorbers and galaxies cluster around certain
redshifts for the sightlines toward H\,1821+643 and PG\,1116+215.
\par Studies of the transverse distribution of \Lya\ absorbers with respect to
the sightline are rare. Tejos et al.\ (2014) looked at the correlation between
galaxies and absorbers in six fields toward QSOs with $z$$\sim$0.7. However,
they could only study the correlation function out to a few Mpc from each
sightline. From this they concluded that 50\% of weak \HI\ absorbers reside in
what they call ``galaxy voids'', i.e.\ no substantial dark matter halo is
present within a few Mpc of the absorber. However, they did not really map the
distribution of absorbers relative to the structures seen in the galaxy
distribution.
\par At redshifts beyond about 10,000~\kms\ mapping out the galaxies over large
regions around individual sightlines is observationally expensive. The galaxies
are relatively faint but still need to be mapped over many degrees, whereas
instrumental limitations typically results in galaxy maps covering only about
one degree around a sightline and larger-scale mapping is eschewed in favor of
mapping multiple sightlines. At the lowest redshifts ($cz$$<$5,000~\kms\ or so),
however, it is possible to get a mostly complete galaxy sample down to low
luminosities (0.1\,\Lstar\ or lower) across most of the sky, since this
luminosity corresponds to a galaxy with $m$$\sim$17. Most of the sky has been
covered this deeply, especially in regions covered by the footprint of surveys
such as the Sloan Digital Sky Survey (SDSS) and the 6dF Galaxy Survey (6dFGS). 
\par The potential of this approach was first shown in a study by Narayanan et
al.\ (2010). They combined \Lya\ and \OVI\ absorption at $cz$$\sim$3000~\kms\ in
the direction of the Seyfert galaxy Mrk\,290 to show the presence of gas at a
temperature of 1.4\tdex{5}~K at an impact parameter of 475~kpc to the nearest
large galaxy, NGC\,5987. However, they also showed that this galaxy is located
in a 30\deg\ ($\sim$30~Mpc) long T-shaped filament. With two other sightlines
through the filament showing absorption at similar velocities, while three
off-filament sightlines yielded non-detections, the interpretation of the
absorber toward Mrk\,290 as originating in the outer halo of NGC\,5987 was
thrown into doubt.
\par To follow up on this realization, we used the Cosmic Origins Spectrograph
(\COS) on the Hubble Space Telescope (\HST) to obtain spectra for 17 additional
AGN sampling the filament found by Narayanan et al.\ (2010). Supplementing these
with seven archival sightlines, we describe in this paper how we used this data
to develop a new approach to constraining the properties of the \Lya\ forest
that uses the {\it spatial} distribution of the absorbers in relation to the
location of the dark matter/galaxy filaments. To do so we first need to set up
several independent pieces: 
\par\noindent (a) Sect.~\Sfilament\ presents the galaxy filament found by
Narayanan et al.\ (2010) and the details of the method that we used to
objectively define the filament axis.
\par\noindent (b) Sect.~\Sobserve\ presents the selection of targets and our
definition of a ``filament impact parameter''.
\par\noindent (c) In Sect.~\SCOSdata\ we have a diversion in which we describe
our method to correct the wavelength scale of \COS\ spectra
\par\noindent (d) Sect.~\Ssimulations\ concentrates on the description of
simulations from which \HI\ column densities can be derived, needed to interpret
our results.
\par We combine all these items in Sect.~\Sresults, where we give the
measurements on the spectra, discuss the galaxy/AGN impact parameters and the
distribution of equivalent widths, column densities and line widths as function
of separation to the filament axis, both in terms of just observations and in
terms of comparisons to the simulations. Specifically, we look at the influence
of the assumed intensity of the extragalactic background radiation and compare
the predictions for the run of detection fraction as function of filament impact
parameter to the observations.


\section{The Galaxy Filament} 

\begin{figure} \figurenum{\Fsizehist}
\begin{center}$\begin{array}{c} \includegraphics[width=\colwidth, angle=0]{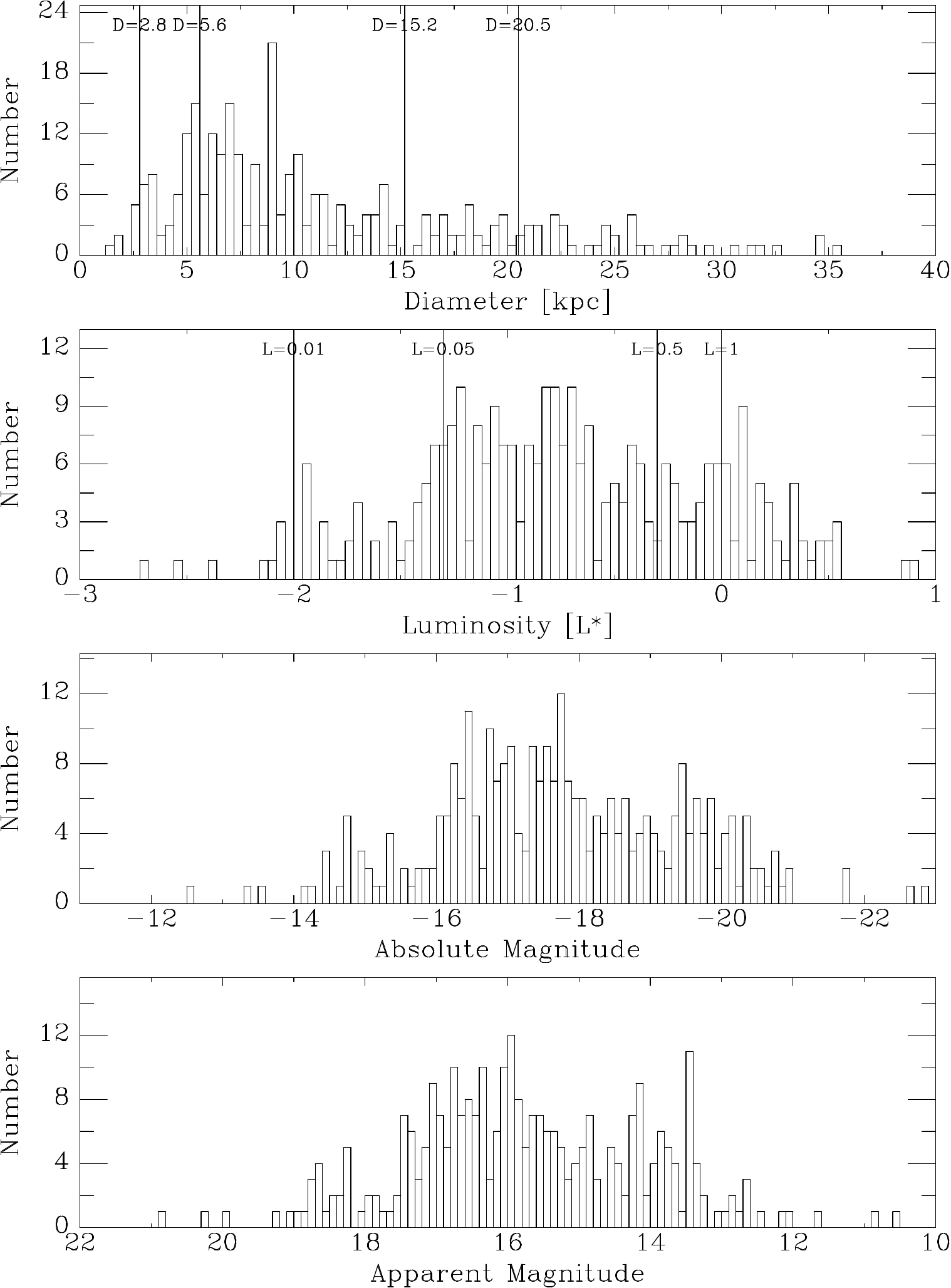} \end{array}$ \end{center} 
\caption{%
Histogram of implied galaxy diameters, observed apparent magnitudes and derived
luminosities and absolute magnitudes for the galaxies in the filament. This plot
includes all 365 galaxies in the sky area shown in Fig.~\Fboxmap. Diameters are
mostly based on the 2MASS K-band angular diameters (see text), combined with the
estimated galaxy distances. Luminosities follow from the empirical relation
between luminosity and diameter found by Wakker \& Savage (2009), which gives
the luminosity to within a factor of 2. In the upper two panels vertical lines
indicate galaxies with $L$=0,01, 0.05, 0.5 and 1~\Lstar.}
\end{figure}

\begin{figure*} \figurenum{\Fboxmap}
\begin{center}$\begin{array}{c} \includegraphics[width=\rotwidth, angle=270]{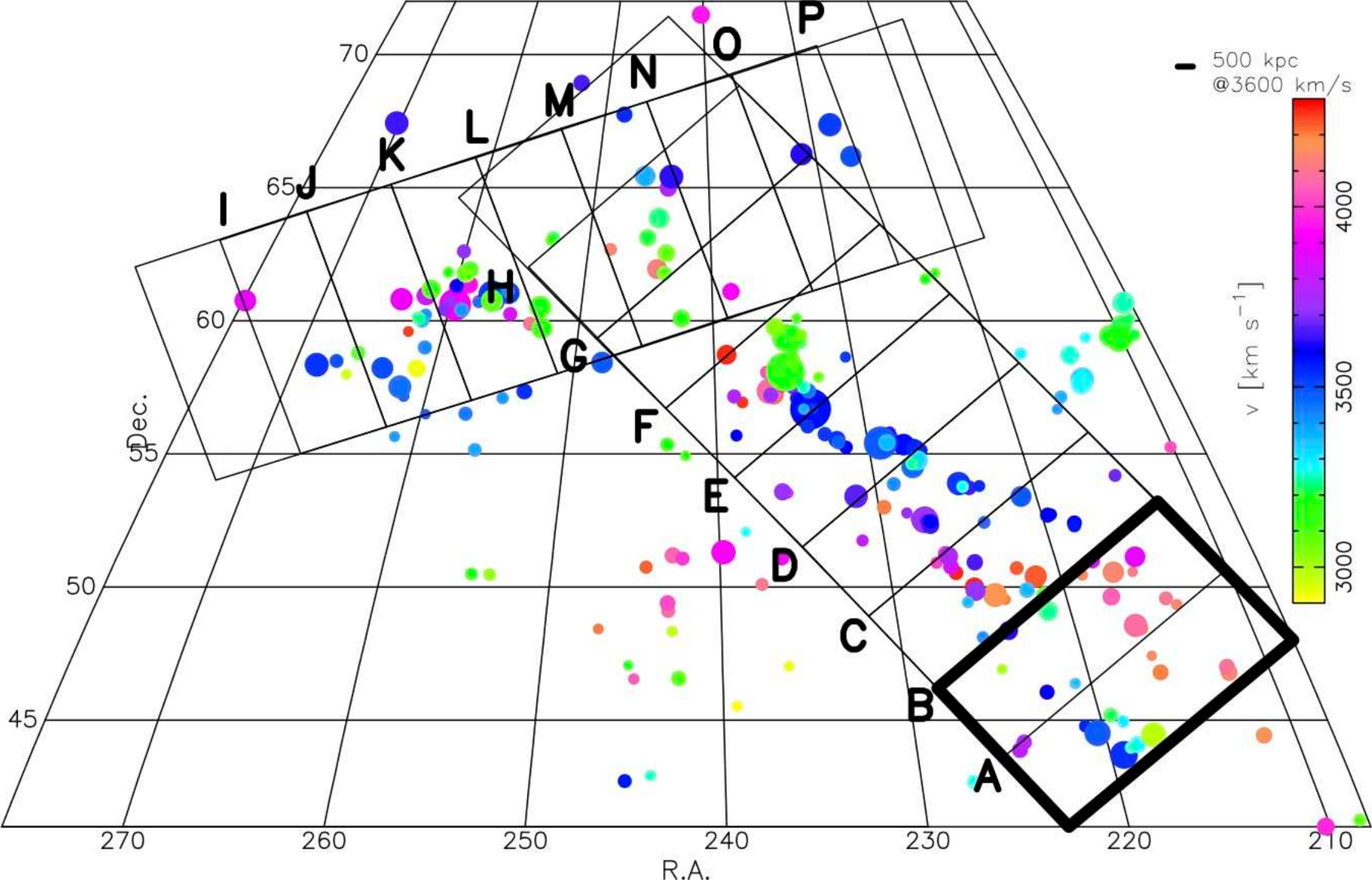} \end{array}$ \end{center} 
\caption{%
Distribution of galaxies (colored circles) in the galaxy filament. The sizes of
the galaxy circles are proportional to their area, while the color indicates
their velocities, following the scale bar on the right. The labeled rectangles
show the strip boxes used to calculate the filament axis. Box A is outlined with
a thick line as a reminder that these boxes overlap by 50\%.
}\end{figure*}

\begin{figure} \figurenum{\Fgalhist a}
\begin{center}$\begin{array}{c} \includegraphics[width=\colwidth, angle=0]{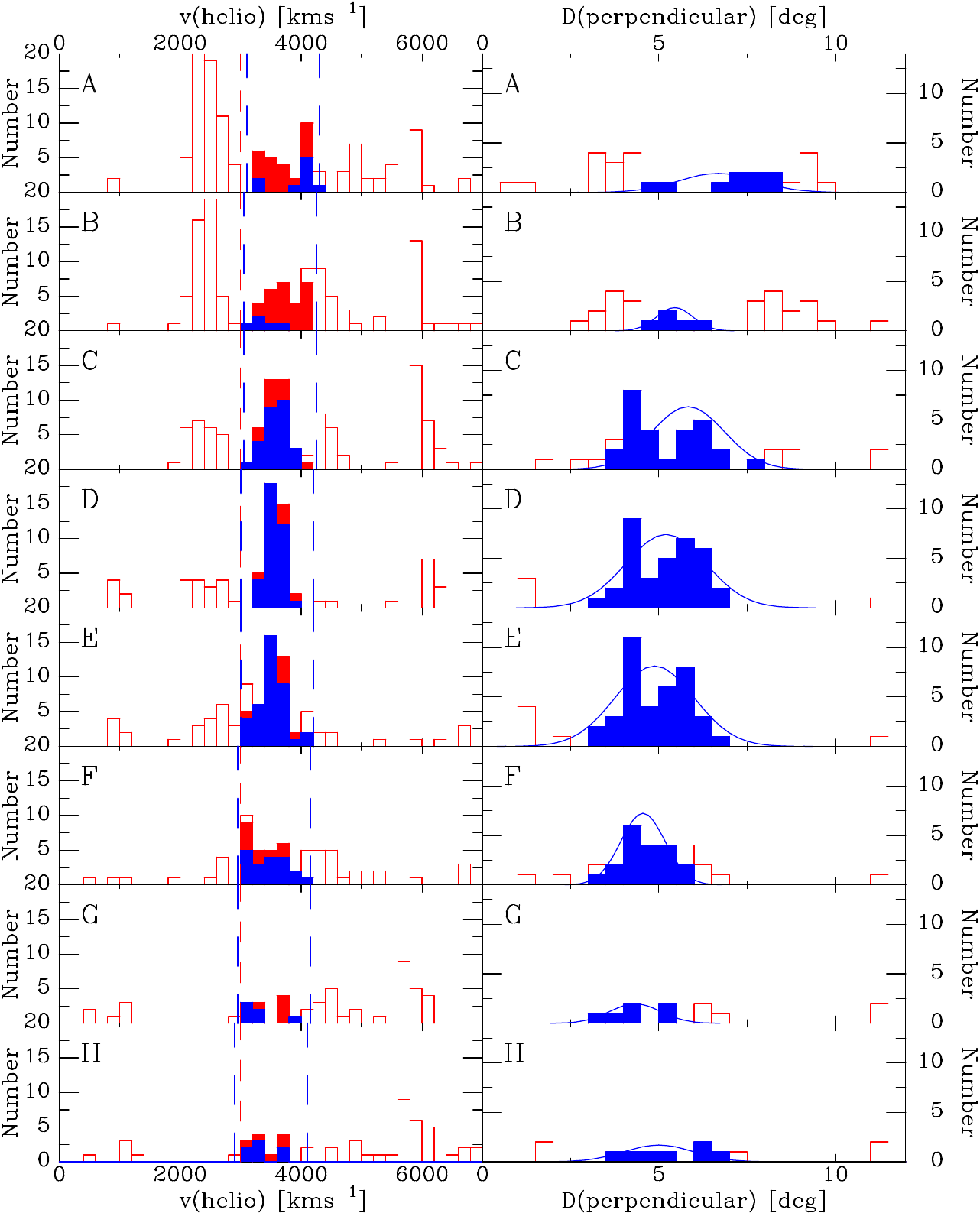} \end{array}$ \end{center} 
\caption{%
The red histograms in the left panels give the velocity distribution of the
galaxies in each strip (same label as in Fig.~\Fboxmap, given in the upper left
corner of each panel). The filled blue bins show the galaxies considered part of
the filament after applying {\it both} the velocity and edge-separation
selection, i.e. within $\pm$500~\kms\ of the central velocities and inside a
selected edge-separation range. The red dashed vertical lines in the left panel
give the velocity range included in Fig.~\Fboxmap, while the blue dashed
vertical lines show the range of velocities used to select the galaxies in the
filament. The right panels gives the distribution of distance to the edge of the
strip for galaxies in the velocity range selected for each strip. The red
histograms indicate galaxies in the proper velocity range, but not in the proper
range of edge-separations. For strips A--H the edge of the strip runs along the
bottom, from (RA,Dec)=(223,41) to (261,65), while for strips I--P this edge runs
from (270,54) to (218,64) (see Fig.~\Fboxmap). The distribution of angular
separations from this edge is determined by fitting by a gaussian if there are
12 or more galaxies, by the first and second moment if there are fewer than 12.
The resulting centers are used to define the filament axis; these fits are shown
by the solid blue line.
}\end{figure}

\begin{figure} \figurenum{\Fgalhist b}
\begin{center}$\begin{array}{c} \includegraphics[width=\colwidth, angle=0]{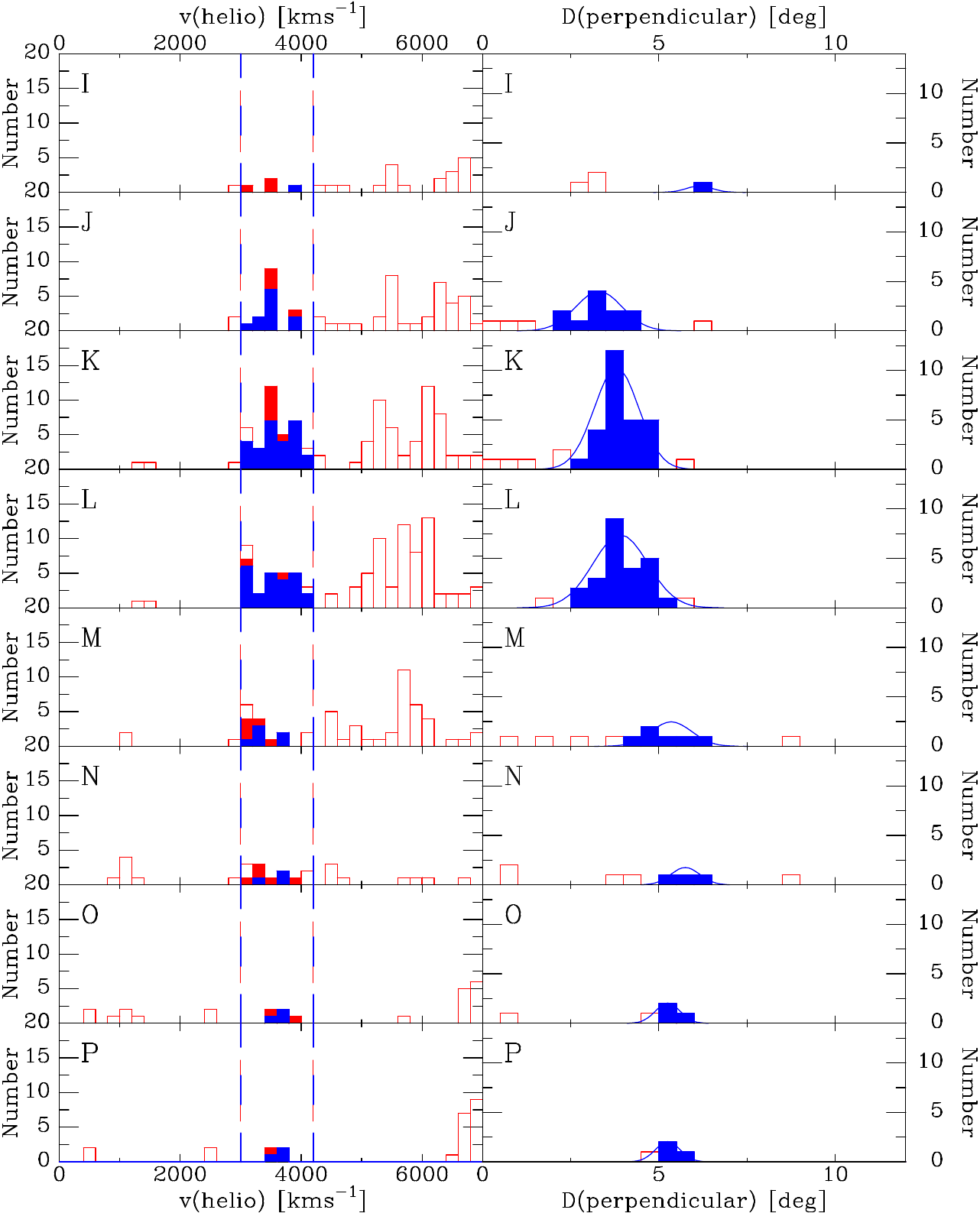} \end{array}$ \end{center} 
\caption{%
continued.
}\end{figure}


\subsection{Galaxy sample and impact parameters} 
\par In this section we describe the galaxy filament we study and discuss the
method we use to derive a ``filament impact parameter'', in addition to the
usual ``galaxy impact parameter'' that is commonly used when studying
intergalactic absorption lines.
\par We base our definition of the galaxy filament on the NASA Extragalactic
Database (NED), first retrieving all information for all galaxies with
$cz$$<$10,000~\kms, finding their location on the sky, redshift,
redshift-independent distance (if available), angular diameter, position angle,
inclination and type. This dataset will be described in more detail in a future
paper (French et al., in preparation).
\par Galaxy diameters were derived from angular diameters and individual galaxy
distances. All angular diameter measurements were retrieved from NED. Diameters
are preferentially taken from 2MASS K$_s$ ``total'' surface brightness
extrapolation diameter measurements (for about 50\% of all galaxies in our
sample). For galaxies with no 2MASS measurement, we plot different magnitudes
against K$_s$ ``total'' values, and make a simple least squared linear fit
between these magnitudes to estimate K$_s$. For $\sim$20\% of the sample SDSS
(Sloan Digital Sky Survey) diameters are available, while for $\sim$27\% no
diameter is given in NED. The remaining 3\% of diameters are based on other
surveys. To derive the 2MASS K$_s$ diameters, the ``total" aperture radius
$r_{tot}$ is defined to be the point at which the surface brightness extends
down to 5 disk scale lengths (see Jarrett et al.\ 2003 for a full description).
The $r_{tot}$ value is derived as: $r_{tot} = r' + a (ln 148)^b$, where $r'$ is
the starting point radius ($>$5'' - 10'' beyond the nucleus, essentially beyond
the PSF and nuclear or core influence) and $a$ and $b$ are the scale length
parameters from a Sersic exponential function: $f = f_0\,exp(-r/a)^{(1/b)}$. The
fit extends to the point at which the mean surface brightness in the elliptical
annulus has S/N$<$2.
\par Figure~\Fsizehist\ shows the distribution of diameters and apparent
magnitudes of all galaxies in the region we analyze (R.A.\ 208\deg\ to 276\deg,
declination 41\deg\ to 72\deg\ and recession velocity \vminfilmap\ to
\vmaxfilmap~\kms; see Sect.~\Sfiladef). It also shows estimated luminosities and
absolute magnitudes, which are based on the empirical scaling between luminosity
($L$/\Lstar) and diameter $D$ that was found by Wakker \& Savage (2009): log$L$
= 2.31 log$D$$-$3.03. This relation yields luminosities to within a factor
$\sim$2. It gives $L$=0.05\,\Lstar\ for $D$=5.6~kpc and $L$=1\,\Lstar\ for
$D$=20.5~kpc. Using the fact that $M_B$=$-$19.57 for a \Lstar\ galaxy (Marzke et
al.\ 1994), $L$=0.05\,\Lstar\ corresponds to an apparent magnitude of 17.2 at a
redshift $cz$=3500~kms. It would be preferable to use observed magnitudes and
distances to derive galaxy luminosities, but the data in NED is extremely
inhomogeneous. We are working on a proper derivation of luminosity (French et
al.\ in preparation), but preliminary numbers show that the distribution of
luminosities will remain very similar.
\par Fig.~\Fsizehist\ shows that the number of galaxies increases down to a
diameter of about 5~kpc (corresponding to a luminosity of 0.05$\sim$\Lstar,
i.e.\ about the luminosity of the SMC), with a break at about that
diameter/luminosity. We conclude that our galaxy sample is mostly complete above
$L$=0.05\,\Lstar. That this limiting luminosity is relatively low is due to the
fact that this area of the sky is in the footprint of the Sloan Digital Sky
Survey (SDSS); about half of the galaxies in the filament had their redshifts
first measured by this survey.
\par We also estimate the virial radius of each galaxy, using the
parametrization discussed by Stocke et al.\ (2013). This relates a galaxy's
luminosity to its virial radius. Fig.~1 of Stocke et al.\ (2013) shows a number
of relations between virial radius and luminosity, which they use to derive an a
representative relation. This has a steeper slope at high luminosity than at low
luminosity ($L$$<$0.1\,\Lstar). Since almost all of our galaxies are brighter
than 0.1\,\Lstar\, we use the plot in Stocke et al.\ (2013) to approximate their
average relation between virial radius and luminosity as log\,$R_{\rm vir}$ =
0.3\,log\,$L$ + 2.25. Combining this with the Wakker \& Savage (2009) empirical
relation between galaxy diameter and luminosity shows that a galaxy's virial
radius can be approximated as log\,$R_{\rm vir}$=0.69\,log\,$D$+1.24. 

\subsection{Defining filament axes} 
\par Figure~\Fboxmap\ shows all galaxies with R.A.\ between 208\deg\ and
276\deg, declination between 41\deg\ and 72\deg\ and recession velocity between
\vminfilmap\ and \vmaxfilmap~\kms. It is obvious that with this selection most
galaxies (especially the ones with $cz$ between $\sim$3000 and $\sim$3700~\kms)
are distributed along two mostly-linear structures, which we interpret as a
T-shaped galaxy filament.
\par Several papers have been written about algorithms to automatically find
filaments in galaxy datasets (e.g.\ Sousbie 2011, Cautun et al.\ 2013, Tempel et
al.\ 2014, Chen et al.\ 2015, and references in these papers). However, we do
not have these codes available. Many are optimized for finding filaments in
simulations, but not in real (messier) galaxy data, which have issues like
varying luminosity limits and large distance uncertainties when using the Hubble
constant to convert velocities to distances. Until we can assess the usefulness
of these various codes and algorithms, we proceed using the simpler, less
general, but more intuitive method we developed ourselves, which is described
below.
\par\noindent 1) First we draw two rectangular boxes on the sky map and divide
these into 6~Mpc ``strips'' along the filament axis, using a distance of 50~Mpc,
which is based on an estimated central velocity of 3500~\kms. These strips
overlap by 50\%, as shown in Fig.~\Fboxmap.
\par\noindent 2) Next, we make histograms of the velocities of the galaxies in
each strip. These are shown on the left side of Fig.~\Fgalhist. This makes clear
that there is a concentration of galaxies near $\sim$3500~\kms, which is
especially obvious in strips C through F and J through P.
\par\noindent 3) Using these histograms, we fit a gaussian to the velocity
distribution of the galaxies in the filament, separately for each strip,
determining a mean and a dispersion. The mean velocities as function of location
along the filament axis are then used to derive a linear relation between the
central filament velocity and position along the filament. The typical FWHM
around the mean is about 600~\kms, which (combined with Fig.~\Fgalhist) we used
to choose a velocity width of 1000~\kms\ for each strip. That is, we selected
all galaxies in each strip within $\pm$500~\kms\ of the central filament
velocity. The resulting velocity ranges are indicated by the blue filled bins in
Fig.~\Fgalhist.
\par\noindent 4) With the filament's velocity structure now constrained, we find
the location of the filament axis by deriving the angular separation of each
galaxy from the long edge of the guide box, i.e., the outside boundary of the
strips shown in Fig.~\Fboxmap. This was done for each strip separately, as shown
in the right-hand panels of Fig.~\Fgalhist. We then calculated the centroid and
dispersion of this perpendicular distribution (using a gaussian fit if there are
more than 12 galaxies, and a moment calculation if there are fewer). Combining
the center of the strip along the long axis with the centroid of the galaxy
distribution perpendicular to the axis defines a point on the filament axis. The
axis itself is defined by connecting these points, creating 3\,Mpc long axis
segments.
\par We note that the lower filament seen in Fig.~\Fboxmap\ shows a slight
velocity gradient, with the central filament velocity increasing from 3300~\kms\
in the lower right to 3850~\kms\ in the upper left. The upper segment has a much
smaller gradient, going from 3750~\kms\ on the left to 3550~\kms\ on the right.
The best single velocity range that includes all galaxies in the filament runs
from \vminfilmap\ to \vmaxfilmap~\kms, which is the range shown in
Fig.~\Fboxmap. We note, however, that for our analysis we include the effects of
the velocity gradient.
\par Finally, we note that the filament is mostly parallel to the plane of the
sky (which is why it can be easily recognized in a map selecting galaxies in a
limited velocity range). The 550~\kms\ gradient corresponds to a change in
distance of $\sim$8~Mpc, contrasting with a projected length of $\sim$27~Mpc.


\section{Observations} 

\subsection{Datasets} 
\par To sample the selected galaxy filament shown in Fig.~\Fboxmap, we
constructed a set of 24 targets. First, we correlated the V\'eron-V\'eron QSO
catalog with the GALEX database, to find all AGNs with UV flux larger than
1.5\tdex{-15}\,\fu. This gave a list of 75 objects in the region of interest. We
selected most of the brighter ones, but kept a few fainter targets located in
strategic directions. The final set of 17 targets were observed using the {\it
Cosmic Origins Spectrograph} (\COS) on the {\it Hubble Space Telescope} (\HST,
program 12276) which has a resolution of about 20~\kms, allowing us to resolve
the \Lya\ absorbers. That sample is supplemented by seven targets in the same
region for which archival data are available, mostly also obtained using \COS,
but in one case (3C\,351.0) with the {\it Space Telescope Imaging Spectrograph}
(\STIS). The \STIS\ data are described by Tripp et al.\ (2008), and we use the
same dataset for 3C\,351.0 as they did. The \COS\ instrument is described in
Green et al.\ (2012) and we use the x1d spectra available from the Multimission
Archive at Space Telescope (\MAST), with modifications (see Sect.~\SCOSdata).
\par The locations and program information for each target are given in
Table~\Tobslist. The observations in program 12276 were taken with central
wavelengths set to 1291\,\AA\ and 1327\AA, resulting in continuous wavelength
coverage from $\sim$1135 to $\sim$1465\,\AA. For six targets the exposure time
is about 2~ks, while for ten targets two orbits were used (giving 5~ks
exposures). One target was exposed for 3 orbits (8.4~ks). When observing for
multiple orbits, we also used two different FP-POS settings in the second orbit
(central wavelength 1327\,\AA). The other seven targets were part of eight other
programs, using a variety of exposure times and grating settings. The different
central wavelengths allow an assessment of detector glitches and fixed-pattern
noise, as the same wavelength will fall on two or three different places on the
detector. The CALCOS pipeline (v2.19.7) was used to process the raw data,
producing one-dimensional extracted spectra.


\begin{deluxetable*}{lrrrlllllc} \tablenum{1} \tablecolumns{9} \tablewidth{0pt}
\tabletypesize{\scriptsize} \tablecaption{Observations} \tablehead{%
\ch{Target} & \ch{R.A.} & \ch{Dec.} & \ch{$z$} & \ch{Program} & \ch{PI}  & \ch{Obs. ID} & \ch{Obs. date} & \ch{T$_{\rm exp} [ks]$} \\
\ch{(1)}    & \ch{(2)}  & \ch{(3)}  & \ch{(4)} & \ch{(5)}     & \ch{(6)} & \ch{(7)}     & \ch{(8)}       & \ch{(9)} }
\startdata
3C309.1          & 14 59 07.6 & +71 40 20 & 0.9050 & 12486 & Bowen      & LBP240 & 2012 03 27 &   9.4\\ 
3C309.1          &            &           & 0.9050 & 12486 & Bowen      & LBP245 & 2012 03 29 &   9.4\\ 
3C351.0          & 17 04 41.4 & +60 44 31 & 0.3719 &  8015 & Jenkins    & O57901 & 1999 06 27 &  19.8\\ 
3C351.0          &            &           & 0.3719 &  8015 & Jenkins    & O57903 & 1999 06 29 &  14.4\\ 
3C351.0          &            &           & 0.3719 &  8015 & Jenkins    & O57902 & 2000 02 10 &  18.6\\ 
3C351.0          &            &           & 0.3719 &  8015 & Jenkins    & O57904 & 2000 07 25 &  26.5\\ 
4C63.22          & 15 23 45.9 & +63 39 24 & 0.2040 & 12276 & Wakker     & LBI601 & 2011 08 14 &   1.9\\ 
FBS1526+659      & 15 27 28.5 & +65 48 10 & 0.3450 & 12276 & Wakker     & LBI603 & 2011 05 02 &   2.0\\ 
H1821+643        & 18 21 57.2 & +64 20 36 & 0.2970 & 11484 & Hartig     & LABO06 & 2009 07 26 &   0.6\\ 
H1821+643        &            &           & 0.2970 & 12038 & Green      & LBGL33 & 2012 07 06 &  11.5\\ 
Kaz447           & 17 03 28.9 & +61 41 09 & 0.0773 & 12276 & Wakker     & LBI618 & 2011 10 05 &   5.2\\ 
Mrk290           & 15 35 52.3 & +57 54 09 & 0.0296 & 11524 & Green      & LB4Q02 & 2009 10 28 &   3.9\\ 
Mrk486           & 15 36 38.3 & +54 33 33 & 0.0389 & 12276 & Wakker     & LBI607 & 2011 12 18 &   5.0\\ 
Mrk817           & 14 36 22.1 & +58 47 40 & 0.0315 & 11505 & Noll       & LACD01 & 2009 08 04 &   2.2\\ 
Mrk817           &            &           & 0.0315 & 11524 & Green      & LB4Q01 & 2009 12 28 &   1.3\\ 
Mrk876           & 16 13 57.2 & +65 43 10 & 0.1290 & 11686 & Arav       & LB4F05 & 2010 04 09 &   3.1\\ 
Mrk876           &            &           & 0.1290 & 11524 & Green      & LB4Q03 & 2010 04 08 &   9.5\\ 
PG1626+554       & 16 27 56.2 & +55 22 32 & 0.1330 & 12029 & Green      & LBGB01 & 2011 06 15 &   3.3\\ 
RBS1483          & 15 19 21.6 & +59 08 24 & 0.0781 & 12276 & Wakker     & LBI612 & 2011 06 25 &   1.9\\ 
RBS1503          & 15 29 07.5 & +56 16 06 & 0.0990 & 12276 & Wakker     & LBI614 & 2010 11 26 &   2.0\\ 
RX\,J1500.5+5517  & 15 00 30.8 & +55 17 09 & 0.4048 & 12276 & Wakker     & LBI602 & 2011 05 11 &   8.4\\ 
RX\,J1503.2+6810  & 15 03 16.5 & +68 10 06 & 0.1140 & 12276 & Wakker     & LBI609 & 2010 12 31 &   1.9\\ 
RX\,J1508.8+6814  & 15 08 52.8 & +68 14 07 & 0.0586 & 12276 & Wakker     & LBI608 & 2010 12 05 &   1.9\\ 
RX\,J1608.3+6018  & 16 08 20.5 & +60 18 28 & 0.1780 & 12276 & Wakker     & LBI610 & 2010 11 21 &   5.2\\ 
RX\,J1717.5+6559  & 17 17 38.0 & +65 59 39 & 0.2927 & 12276 & Wakker     & LBI619 & 2011 06 22 &   5.4\\ 
SBS1458+535      & 14 59 49.6 & +53 19 09 & 0.3380 & 12276 & Wakker     & LBI611 & 2011 10 21 &   5.0\\ 
SBS1503+570      & 15 04 55.6 & +56 49 20 & 0.3589 & 12276 & Wakker     & LBI617 & 2011 10 19 &   5.2\\ 
SBS1521+598      & 15 21 53.8 & +59 40 21 & 0.2862 & 12276 & Wakker     & LBI613 & 2011 06 12 &   5.1\\ 
SBS1537+577      & 15 38 10.0 & +57 36 13 & 0.0734 & 12276 & Wakker     & LBI606 & 2011 10 19 &   5.2\\ 
SBS1551+572      & 15 52 32.7 & +57 05 17 & 0.3660 & 12276 & Wakker     & LBI615 & 2011 03 13 &   5.1\\ 
SBS1624+575      & 16 25 26.5 & +57 27 27 & 0.0670 & 12276 & Wakker     & LBI616 & 2011 10 17 &   5.1\\ 
\enddata
\end{deluxetable*}

\begin{figure*} \figurenum{\Fqsomap}
\begin{center}$\begin{array}{c} \includegraphics[width=\rotwidth, angle=270]{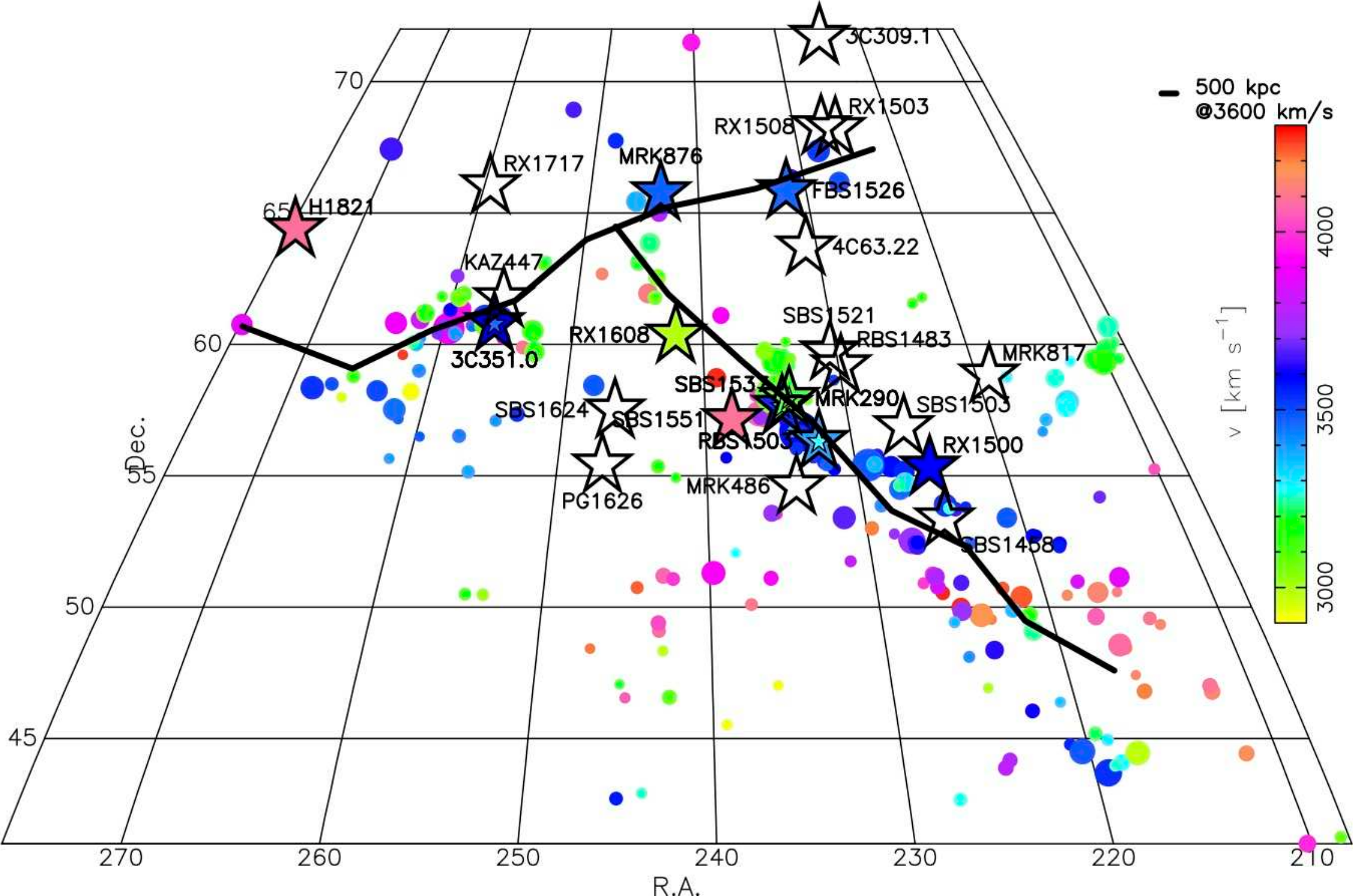} \end{array}$ \end{center} 
\caption{%
Distribution of galaxies (colored circles) and AGN (open and colored stars) in
the galaxy filament. The sizes of the galaxy circles are proportional to the
square of their diameters (i.e.\ area), while the color indicates their
velocities, following the scale bar on the right. AGN toward which \Lya\
absorption is seen are shown by filled colored stars, while non-detections are
indicated by open stars. The axes of the two filament segments are shown by the
wiggly black lines. See the text for the derivation of these axes.}
\end{figure*}


\subsection{AGN sampling the filament} 
\par Figure~\Fqsomap\ shows the directions to the 24 relevant AGN relative to
the galaxy filament found in Sect.~\Sfiladef. This figure also shows cases where
no \Lya\ absorption is found in the velocity range \vminfilmap\ to
\vmaxfilmap~\kms\ as open stars and the 10 sightlines with detected \Lya\ shown
as colored stars. These measurements will be discussed in Sect.~\Sresults.
\par Table~\Timpact\ presents the properties of the galaxies near each AGN
sightline. It includes all galaxies within the relevant velocity range
(\vminfiltab\ to \vmaxfiltab~\kms, see below) that lie within 1~Mpc of the AGN
sightline; if there are fewer than three such galaxies, it gives the three
galaxies with the smallest impact parameter.
\par For AGN toward which a \Lya\ line is seen (see Sect.~\Sresults), Col.~2
shows the velocity of the absorber. Column~4 then gives the difference between
that velocity and the filament velocity. For example, 3C\,351.0 has a \Lya\ line
at 3597~\kms, while the nearest axis segment (at impact parameter 569~kpc) is
centered at 3600~\kms. This implies the absorption is offset by $-$3~\kms\ from
the filament velocity. In the case of galaxies near the sightline, this column
also gives the difference between $v$(\Lya) and $v$(galaxy), but only when that
difference is less than $\pm$400~\kms\ and either the impact parameter is less
than 300~kpc or the ratio of impact parameter to virial radius (Col.~8) is less
than two. That is, a non-blank entry in this column for a galaxy row means that
in a conventional approach the \Lya\ absorption could be associated with the
galaxy.
\par We note that a galaxy's halo probably does not have a well-defined edge.
Gas beyond the virial radius may be falling in if it happens to be moving
radially. Other gas beyond the virial radius may be moving mostly transversely
and never reach the galaxy. In order to better separate possible halo gas from
possible filament gas, we use the arguments in the papers by Oort (1970), Maller
\& Bullock (2004) and Shull (2014). They find that for a Milky Way sized galaxy
($\sim$\Lstar) the boundary between gas falling in and being part of the IGM is
at about 230 kpc. In particular, Shull (2014) discusses the difference between
this ``accretion radius'', a ``gravitational radius'' and the virial radius. The
approximation formula given in Sect.~\Sgalsample\ gives 140~kpc as the virial
radius for an \Lstar\ galaxy. Thus a number on the order of 1.5~R$_{vir}$ is
justifiable as giving the border between gas in a galaxy halo and gas in the
filament.
\par Using this shows that there are no $L$$>$0.05\,\Lstar\ galaxies within
these limits that can be associated with the \Lya\ detections toward 4C\,63.22,
FBS\,1526+659, H\,1821+643, Mrk\,486, RBS\,1483, RBS\,1503, RX\,J1500.5+5517 and
RX\,J1717.6+6559, so conventionally these would be called ``void absorbers''.
\par There are galaxies within the 400~\kms, and 300~kpc or 2\,R$_{vir}$ limit
near the \Lya\ absorbers seen toward 3C\,351.0, Mrk\,290, Mrk\,876,
RX\,J1608.3+6814, SBS\,1537+577 and SBS\,1551+572. However, only toward
SBS\,1537+577 is the nearest (small) galaxy within 1.5 virial radii, while for
the other sightlines the ratio of impact parameter to virial radius lies between
1.7 and 2.3. I.e., these sightlines still pass rather far from the galaxies and
they do not really sample the circumgalactic medium of these galaxies.
\par Thus, using generous standard criteria ($\Delta$$v$$<$400~\kms,
$\rho$$<$300~kpc) at most four of the thirteen \Lya\ absorbers would be
associated with a galaxy. Using even more lenient criteria (including galaxies
up to 2 virial radii from the sightline) six of the detections could be
associated. Using more physically plausible criteria ($\Delta$$v$$<$400~\kms\
and $\rho$/R$_{\rm vir}$$<$1.5) only one \Lya\ absorber samples a galaxy halo.
Therefore, basically all \Lya\ absorbers in our sample are unlikely to be
associated with the halos of galaxies brighter than the SMC
($\sim$0.05\,\Lstar).
\par We can estimate how likely it is that our sample of randomly placed
sightlines passes inside the virial radius of one of the galaxies. Taking the
implied virial radii of all galaxies in our filament and adding up the total sky
area that is covered gives a value of 14~Mpc$^2$. Since the lower branch of the
filament covers about 18\deg$\times$7\deg\ ($\sim$15.5$\times$6~Mpc), while the
upper branch covers about 15\deg$\times$7\deg\ ($\sim$13$\times$6~Mpc), the
total filament area is about 170~Mpc$^2$. Thus, only about 8\% of the filament
area is covered by galaxy halos, and only 1 in 12 \Lya\ absorbers should be
associated with a galaxy. This is indeed what we find.
\par As we will now argue, our \Lya\ detections are instead likely to be
associated with the IGM in the galaxy filament. To quantify the relation between
detections and the filament, we define a new quantity: the ``filament impact
parameter'' ($\rho$(fil)) as the separation between an AGN sightline and the
nearest axis segment. This is the product of the angular separation scaled by
the distance corresponding to the segment's recession velocity, i.e.\ the
central velocity of each filament strip, found in the manner described above.
The resulting impact parameters are given in Col.~4 Table~\Timpact\ under the
entry ``Filament Axis'' after each AGN's name.

\begin{deluxetable*}{llrrrrrrrl} \tablenum{2} \tablecolumns{9} \tablewidth{0pt}
\tabletypesize{\scriptsize} \tablecaption{Galaxies between \vminfiltab\ and \vmaxfiltab~\kms\ within 1~Mpc of each sightline} \tablehead{%
\ch{Target} & \ch{Galaxy} & \ch{$cz$}   & \ch{$\Delta$$v$} & \ch{$\rho$} & \ch{Diameter} & \ch{L/\Lstar} & \ch{$R_{\rm vir}$} & \ch{$\rho$/$R_{\rm vir}$} & \ch{Type}  \\
            &             & \ch{[\kms]} & \ch{\kms}        & \ch{[kpc]}  & \ch{[kpc]}    & \ch{}         &                    &                           &            \\
\ch{(1)}    & \ch{(2)}    & \ch{(3)}    & \ch{(4)}         & \ch{(5)}    & \ch{(6)}      & \ch{(7)}      & \ch{(8)}           & \ch{(9)}                  & \ch{(10)}   }
\startdata
\tableline
3C309.1                   & Filament Axis                & 3600 &           & 3908 & \\
3C309.1                   & IC1110                       & 3373 &           & 3854 &  24.6 & 0.86 &  158 &  24.3 &  Sa \\
3C309.1                   & UGC09734                     & 3341 &           & 4756 &  19.9 & 0.55 &  137 &  34.7 &  Sm \\
3C309.1                   & CGCG338-038                  & 3827 &           & 4737 &  15.9 & 0.34 &  117 &  40.3 &   \\
\tableline
3C351.0                   & Ly$\alpha$ @  3597,3459      &      &           &      & \\
3C351.0                   & Filament Axis                & 3600 & -3,-141   &  569 & \\
3C351.0                   & Mrk0892                      & 3617 & -20,-158  &  179 &  12.9 & 0.22 &  101 &   1.8 &  Pair \\
3C351.0                   & NGC6307                      & 3057 &           &  284 &  22.7 & 0.72 &  149 &   1.9 &  PSBS0P. \\
3C351.0                   & NGC6306                      & 2973 &           &  271 &  19.3 & 0.51 &  134 &   2.0 &  .SBS2P* \\
3C351.0                   & NGC6310                      & 3419 &           &  406 &  34.5 & 1.76 &  200 &   2.0 &  .S..3* \\
3C351.0                   & NGC6292                      & 3411 &           &  375 &  22.4 & 0.70 &  148 &   2.5 &  .S..4.. \\
3C351.0                   & SDSSJ170327.95+610631.5      & 3313 &           &  333 &   9.2 & 0.11 &   81 &   4.1 &   \\
3C351.0                   & 2MASXJ17071270+6055144       & 3099 & 360       &  280 &   3.6 & 0.01 &   42 &   6.5 &  HII \\
3C351.0                   & SBS1700+603                  & 3736 &           &  616 &   9.6 & 0.12 &   83 &   7.4 &   \\
3C351.0                   & SDSSJ171138.94+604341.8      & 3855 &           &  828 &   8.3 & 0.08 &   75 &  11.0 &   \\
3C351.0                   & SDSSJ171140.34+604115.6      & 3350 &           &  727 &   5.5 & 0.04 &   57 &  12.7 &   \\
3C351.0                   & UGC10745                     & 3059 &           &  700 &   2.7 & 0.01 &   35 &  19.8 &  Sdm \\
\tableline
4C63.22                   & Ly$\alpha$ @       2420      &      &           &      & \\
4C63.22                   & Filament Axis                & 3600 & -1180     & 2333 & \\
4C63.22                   & KHG1-C07                     & 2548 &           & 2413 &   1.0 & 0.00 &    3 & 679.2 &   \\
4C63.22                   & SBS1543+593                  & 2698 &           & 3564 &   1.0 & 0.00 &    3 & 1003.3 &  dwarf \\
4C63.22                   & KHG1-C09                     & 2848 &           & 3593 &   1.0 & 0.00 &    3 & 1011.5 &  Comp \\
\tableline
FBS1526+659               & Ly$\alpha$ @       3476      &      &           &      & \\
FBS1526+659               & Filament Axis                & 3600 & -124      &  369 & \\
FBS1526+659               & KHG1-C07                     & 2548 &           & 3693 &   1.0 & 0.00 &    3 & 1039.8 &   \\
FBS1526+659               & KHG1-C09                     & 2848 &           & 4593 &   1.0 & 0.00 &    3 & 1293.1 &  Comp \\
FBS1526+659               & SBS1543+593                  & 2698 &           & 4821 &   1.0 & 0.00 &    3 & 1357.3 &  dwarf \\
\tableline
H1821+643                 & Ly$\alpha$ @  2825,4087      &      &           &      & \\
H1821+643                 & Filament Axis                & 3600 & -775,487  & 3249 & \\
H1821+643                 & NGC6636NED01                 & 4393 &           & 2491 &  40.2 &  2.4 &  222 &  11.2 &  .S?.... \\
H1821+643                 & NGC6687                      & 3374 &           & 4239 &  28.6 & 1.18 &  175 &  24.1 &  SAd \\
H1821+643                 & NGC6701                      & 3965 &           & 4369 &  29.6 & 1.27 &  180 &  24.3 &  PSBS1.. \\
\tableline
Kaz447                    & Filament Axis                & 3600 &           &  184 & \\
Kaz447                    & NGC6310                      & 3419 &           &  762 &  34.5 & 1.76 &  200 &   3.8 &  .S..3* \\
Kaz447                    & NGC6292                      & 3411 &           &  669 &  22.4 & 0.70 &  148 &   4.5 &  .S..4.. \\
Kaz447                    & NGC6307                      & 3057 &           &  826 &  22.7 & 0.72 &  149 &   5.5 &  PSBS0P. \\
Kaz447                    & SDSSJ170327.95+610631.5      & 3313 &           &  485 &   9.2 & 0.11 &   81 &   6.0 &   \\
Kaz447                    & NGC6306                      & 2973 &           &  817 &  19.3 & 0.51 &  134 &   6.1 &  .SBS2P* \\
Kaz447                    & Mrk0892                      & 3617 &           &  949 &  12.9 & 0.22 &  101 &   9.3 &  Pair \\
Kaz447                    & 2MASXJ17071270+6055144       & 3099 &           &  699 &   3.6 & 0.01 &   42 &  16.3 &  HII \\
Kaz447                    & UGC10745                     & 3059 &           &  673 &   2.7 & 0.01 &   35 &  19.0 &  Sdm \\
\tableline
Mrk290                    & Ly$\alpha$ @       3089      &      &           &      & \\
Mrk290                    & Filament Axis                & 3600 & -511      &   97 & \\
Mrk290                    & NGC5987                      & 3010 & 79        &  475 &  46.4 &  3.3 &  245 &   1.9 &  .S..3.. \\
Mrk290                    & 2MASXJ15351422+5730529       & 3092 &           &  318 &   8.5 & 0.09 &   76 &   4.1 &  HII \\
Mrk290                    & SDSSJ153802.75+573018.3      & 3525 &           &  446 &  10.4 & 0.14 &   87 &   5.1 &  Sd(f) \\
Mrk290                    & CGCG297-017                  & 3282 &           &  506 &  10.4 & 0.14 &   87 &   5.8 &   \\
Mrk290                    & SDSSJ153733.00+583447.8      & 2932 &           &  542 &  10.7 & 0.15 &   89 &   6.0 &  Sc(f) \\
Mrk290                    & SBS1533+574A                 & 3348 &           &  566 &   6.8 & 0.06 &   66 &   8.6 &  HII \\
Mrk290                    & SDSSJ153040.88+575301.0      & 2896 &           &  519 &   5.7 & 0.04 &   58 &   8.9 &   \\
Mrk290                    & SDSSJ153742.05+570506.4      & 3469 &           &  762 &   7.2 & 0.06 &   68 &  11.2 &   \\
Mrk290                    & SBS1540+576                  & 3717 &           &  760 &   6.7 & 0.05 &   65 &  11.7 &   \\
Mrk290                    & SDSSJ153706.68+585651.5      & 2989 &           &  819 &   6.6 & 0.05 &   64 &  12.6 &   \\
Mrk290                    & SBS1533+574B                 & 3429 &           &  579 &   3.9 & 0.02 &   44 &  12.9 &  HII \\
Mrk290                    & 2MASXJ15335796+5650509       & 3260 &           &  915 &   3.6 & 0.01 &   42 &  21.5 &   \\
Mrk290                    & SDSSJ152956.69+582635.9      & 2908 &           &  718 &   2.5 & 0.01 &   33 &  21.7 &   \\
\tableline
Mrk486                    & Ly$\alpha$ @       4386      &      &           &      & \\
Mrk486                    & Filament Axis                & 3600 & 786       & 1910 & \\
Mrk486                    & SBS1543+593                  & 2698 &           & 3342 &   1.0 & 0.00 &    3 & 940.8 &  dwarf \\
Mrk486                    & KHG1-C07                     & 2548 &           & 3841 &   1.0 & 0.00 &    3 & 1081.3 &   \\
Mrk486                    & KHG1-C09                     & 2848 &           & 4615 &   1.0 & 0.00 &    3 & 1299.3 &  Comp \\
\tableline
Mrk817                    & Filament Axis                & 3650 &           & 6208 & \\
Mrk817                    & [DYC2005]443                 & 2998 &           & 4679 &   1.0 & 0.00 &    3 & 1317.4 &  Blue \\
Mrk817                    & RCS06100303673               & 3298 &           & 4717 &   1.0 & 0.00 &    3 & 1327.9 &   \\
Mrk817                    & RCS06100304426               & 3298 &           & 4742 &   1.0 & 0.00 &    3 & 1335.1 &   \\
\enddata
\end{deluxetable*}

\begin{deluxetable*}{llrrrrrrrl} \tablenum{2} \tablecolumns{9} \tablewidth{0pt}
\tabletypesize{\scriptsize} \tablecaption{Galaxies between \vminfiltab\ and \vmaxfiltab~\kms\ within 1~Mpc of each sightline} \tablehead{%
\ch{Target} & \ch{Galaxy} & \ch{$cz$}   & \ch{$\Delta$$v$} & \ch{$\rho$} & \ch{Diameter} & \ch{L/\Lstar} & \ch{$R_{\rm vir}$} & \ch{$\rho$/$R_{\rm vir}$} & \ch{Type}  \\
            &             & \ch{[\kms]} & \ch{\kms}        & \ch{[kpc]}  & \ch{[kpc]}    & \ch{}         &                    &                           &            \\
\ch{(1)}    & \ch{(2)}    & \ch{(3)}    & \ch{(4)}         & \ch{(5)}    & \ch{(6)}      & \ch{(7)}      & \ch{(8)}           & \ch{(9)}                  & \ch{(10)}   }
\startdata
\tableline
Mrk876                    & Ly$\alpha$ @       3476      &      &           &      & \\
Mrk876                    & Filament Axis                & 3600 & -124      &  540 & \\
Mrk876                    & UGC10294                     & 3504 & -28       &  274 &  25.6 & 0.93 &  162 &   1.7 &  Sm? \\
Mrk876                    & UGC10376                     & 3246 &           &  797 &  19.3 & 0.51 &  134 &   5.9 &  Sm: \\
Mrk876                    & NGC6135                      & 3644 &           &  685 &  13.9 & 0.25 &  107 &   6.4 &   \\
\tableline
PG1626+554                & Filament Axis                & 3550 &           & 5684 & \\
PG1626+554                & NGC6258                      & 3064 &           & 4753 &  22.4 & 0.70 &  148 &  31.9 &  .E..... \\
PG1626+554                & KHG1-C09                     & 2848 &           & 4912 &   1.0 & 0.00 &    3 & 1383.0 &  Comp \\
PG1626+554                & SBS1543+593                  & 2698 &           & 4919 &   1.0 & 0.00 &    3 & 1384.9 &  dwarf \\
\tableline
RBS1483                   & Ly$\alpha$ @       2726      &      &           &      & \\
RBS1483                   & Filament Axis                & 3600 & -874      & 2260 & \\
RBS1483                   & NGC5894                      & 2466 &           &  839 &  33.6 & 1.66 &  196 &   4.3 &  .SB.8?. \\
RBS1483                   & SDSSJ151827.06+582658.6      & 2555 &           &  469 &   7.8 & 0.07 &   72 &   6.5 &   \\
RBS1483                   & Mrk0847                      & 2540 &           &  658 &  10.2 & 0.13 &   86 &   7.6 &  .S?.... \\
RBS1483                   & SDSSJ152209.95+583819.5      & 3349 &           &  535 &   5.8 & 0.04 &   59 &   9.0 &   \\
RBS1483                   & UGC09837                     & 2657 &           &  880 &   7.3 & 0.06 &   69 &  12.7 &  SAB(s)c \\
RBS1483                   & SDSSJ151917.62+603435.9      & 2535 &           &  953 &   4.8 & 0.03 &   51 &  18.4 &   \\
\tableline
RBS1503                   & Ly$\alpha$ @  3269,3306      &      &           &      & \\
RBS1503                   & Filament Axis                & 3600 & -331,-294 &  306 & \\
RBS1503                   & NGC5965                      & 3412 &           &  677 &  55.6 &  4.9 &  277 &   2.4 &  .S..3.. \\
RBS1503                   & Mrk0482                      & 3357 &           &  640 &  13.2 & 0.23 &  103 &   6.2 &  ....... \\
RBS1503                   & CGCG274-047                  & 3355 &           &  517 &   9.3 & 0.11 &   81 &   6.3 &   \\
RBS1503                   & 2MASXJ15332453+5636315       & 3553 &           &  624 &  10.2 & 0.13 &   86 &   7.2 &   \\
RBS1503                   & Mrk0481                      & 3298 &           &  726 &   8.9 & 0.10 &   78 &   9.2 &  E \\
RBS1503                   & MCG+09-26-001                & 3366 &           &  793 &   9.1 & 0.10 &   80 &   9.9 &   \\
RBS1503                   & SDSSJ153325.63+564156.5      & 3362 &           &  636 &   6.1 & 0.04 &   61 &  10.4 &   \\
RBS1503                   & SDSSJ153007.14+553432.8      & 3356 &           &  612 &   5.1 & 0.03 &   54 &  11.3 &   \\
RBS1503                   & SBS1524+554                  & 3409 &           &  978 &   8.5 & 0.09 &   76 &  12.8 &  BlueCG \\
RBS1503                   & SDSSJ153558.83+564108.2      & 3213 &           &  859 &   6.1 & 0.04 &   61 &  14.0 &  BLAGN \\
RBS1503                   & 2MASXJ15335796+5650509       & 3260 &           &  744 &   3.6 & 0.01 &   42 &  17.5 &   \\
\tableline
RX\,J1500.5+5517           & Ly$\alpha$ @       3592      &      &           &      & \\
RX\,J1500.5+5517           & Filament Axis                & 3650 & -58       & 2031 & \\
RX\,J1500.5+5517           & UGC09737                     & 3367 &           &  911 &  16.1 & 0.35 &  118 &   7.7 &  S \\
RX\,J1500.5+5517           & SDSSJ150621.06+550842.1      & 3374 &           &  738 &   7.7 & 0.07 &   71 &  10.3 &   \\
RX\,J1500.5+5517           & SDSSJ150804.21+551954.0      & 3400 &           &  948 &   7.4 & 0.07 &   69 &  13.6 &   \\
RX\,J1500.5+5517           & SDSSJ150654.40+553218.2      & 3295 &           &  805 &   4.8 & 0.03 &   51 &  15.5 &   \\
RX\,J1500.5+5517           & SDSSJ145718.28+543105.9      & 3216 &           &  750 &   2.8 & 0.01 &   36 &  20.7 &   \\
\tableline
RX\,J1503.2+6810           & Filament Axis                & 3600 &           & 1016 & \\
RX\,J1503.2+6810           & IC1110                       & 3373 &           & 1003 &  24.6 & 0.86 &  158 &   6.3 &  Sa \\
RX\,J1503.2+6810           & UGC09855                     & 3480 &           & 2538 &  22.2 & 0.69 &  147 &  17.2 &  Im: \\
RX\,J1503.2+6810           & CGCG297-009                  & 2500 &           & 4633 &   1.0 & 0.00 &    3 & 1304.4 &  Sd(f) \\
\tableline
RX\,J1508.8+6814           & Filament Axis                & 3600 &           & 1202 & \\
RX\,J1508.8+6814           & IC1110                       & 3373 &           &  799 &  24.6 & 0.86 &  158 &   5.0 &  Sa \\
RX\,J1508.8+6814           & UGC09855                     & 3480 &           & 2249 &  22.2 & 0.69 &  147 &  15.2 &  Im: \\
RX\,J1508.8+6814           & CGCG297-009                  & 2500 &           & 4624 &   1.0 & 0.00 &    3 & 1301.9 &  Sd(f) \\
\tableline
RX\,J1608.3+6018           & Ly$\alpha$ @  2983,2886      &      &           &      & \\
RX\,J1608.3+6018           & Filament Axis                & 3550 & -567,-664 &  880 & \\
RX\,J1608.3+6018           & UGC10247                     & 2995 & -12,-109  &  199 &  14.3 & 0.27 &  109 &   1.8 &  SBm: \\
RX\,J1608.3+6018           & UGC10279NED01                & 4421 &           &  563 &  14.1 & 0.26 &  108 &   5.2 &  Sb \\
RX\,J1608.3+6018           & UGC10279NED02                & 4429 &           &  566 &  13.1 & 0.22 &  102 &   5.5 &  SBc \\
RX\,J1608.3+6018           & KHG1-C10                     & 4317 &           &  814 &   1.0 & 0.00 &    3 & 229.2 &   \\
\tableline
RX\,J1717.5+6559           & Ly$\alpha$ @       4705      &      &           &      & \\
RX\,J1717.5+6559           & Filament Axis                & 3600 & 1105      & 4330 & \\
RX\,J1717.5+6559           & NGC6310                      & 3419 &           & 4436 &  34.5 & 1.76 &  200 &  22.2 &  .S..3* \\
RX\,J1717.5+6559           & HIJASSJ1720+71               & 2495 &           & 3526 &   1.0 & 0.00 &    3 & 992.8 &   \\
RX\,J1717.5+6559           & SSTXFLSJ171614.7+602439      & 3298 &           & 4669 &   1.0 & 0.00 &    3 & 1314.4 &   \\
\tableline
SBS1458+535               & Filament Axis                & 3650 &           &  563 & \\
SBS1458+535               & NGC5820                      & 3335 &           &  514 &  24.6 & 0.86 &  158 &   3.2 &  .L.... \\
SBS1458+535               & UGC09663                     & 2420 &           &  498 &  18.5 & 0.47 &  130 &   3.8 &  Im: \\
SBS1458+535               & NGC5821                      & 3376 &           &  540 &  19.1 & 0.50 &  133 &   4.0 &  S? \\
SBS1458+535               & SDSSJ145753.64+534602.5      & 3080 &           &  428 &   7.2 & 0.06 &   68 &   6.2 &   \\
SBS1458+535               & SDSSJ145620.69+534336.8      & 3458 &           &  591 &  10.1 & 0.13 &   86 &   6.9 &   \\
SBS1458+535               & UGC09632                     & 3206 &           &  454 &   4.9 & 0.03 &   53 &   8.6 &  .SA.7.. \\
SBS1458+535               & SBS1452+540                  & 3351 &           &  903 &   6.7 & 0.05 &   64 &  13.9 &  E \\
\tableline
SBS1503+570               & Filament Axis                & 3600 &           & 2395 & \\
SBS1503+570               & CGCG297-009                  & 2500 &           & 2993 &   1.0 & 0.00 &    3 & 842.6 &  Sd(f) \\
SBS1503+570               & KHG1-C07                     & 2548 &           & 3578 &   1.0 & 0.00 &    3 & 1007.2 &   \\
SBS1503+570               & SBS1543+593                  & 2698 &           & 4020 &   1.0 & 0.00 &    3 & 1131.7 &  dwarf \\
\tableline
SBS1521+598               & Filament Axis                & 3600 &           & 2390 & \\
SBS1521+598               & NGC5894                      & 2466 &           &  919 &  33.6 & 1.66 &  196 &   4.7 &  .SB.8?. \\
SBS1521+598               & SDSSJ151827.06+582658.6      & 2555 &           &  871 &   7.8 & 0.07 &   72 &  12.1 &   \\
SBS1521+598               & SDSSJ151917.62+603435.9      & 2535 &           &  637 &   4.8 & 0.03 &   51 &  12.3 &   \\
SBS1521+598               & SDSSJ152209.95+583819.5      & 3349 &           &  894 &   5.8 & 0.04 &   59 &  15.0 &   \\
\enddata
\end{deluxetable*}

\begin{deluxetable*}{llrrrrrrrl} \tablenum{2} \tablecolumns{9} \tablewidth{0pt}
\tabletypesize{\scriptsize} \tablecaption{Galaxies between \vminfiltab\ and \vmaxfiltab~\kms\ within 1~Mpc of each sightline} \tablehead{%
\ch{Target} & \ch{Galaxy} & \ch{$cz$}   & \ch{$\Delta$$v$} & \ch{$\rho$} & \ch{Diameter} & \ch{L/\Lstar} & \ch{$R_{\rm vir}$} & \ch{$\rho$/$R_{\rm vir}$} & \ch{Type}  \\
            &             & \ch{[\kms]} & \ch{\kms}        & \ch{[kpc]}  & \ch{[kpc]}    & \ch{}         &                    &                           &            \\
\ch{(1)}    & \ch{(2)}    & \ch{(3)}    & \ch{(4)}         & \ch{(5)}    & \ch{(6)}      & \ch{(7)}      & \ch{(8)}           & \ch{(9)}                  & \ch{(10)}   }
\startdata
\tableline
SBS1537+577               & Ly$\alpha$ @  3541,3257      &      &           &      & \\
SBS1537+577               & Filament Axis                & 3600 & -59,-343  &  293 & \\
SBS1537+577               & SDSSJ153802.75+573018.3      & 3525 & 16,-268   &   91 &  10.4 & 0.14 &   87 &   1.0 &  Sd(f) \\
SBS1537+577               & NGC5987                      & 3010 & 247       &  444 &  46.4 &  3.3 &  245 &   1.8 &  .S..3.. \\
SBS1537+577               & NGC5965                      & 3412 &           &  914 &  55.6 &  4.9 &  277 &   3.3 &  .S..3.. \\
SBS1537+577               & 2MASXJ15351422+5730529       & 3092 &           &  322 &   8.5 & 0.09 &   76 &   4.2 &  HII \\
SBS1537+577               & CGCG297-017                  & 3282 &           &  521 &  10.4 & 0.14 &   87 &   5.9 &   \\
SBS1537+577               & SBS1540+576                  & 3717 &           &  389 &   6.7 & 0.05 &   65 &   6.0 &   \\
SBS1537+577               & UGC10002                     & 4052 &           &  875 &  20.7 & 0.59 &  140 &   6.2 &  SB? \\
SBS1537+577               & SDSSJ153742.05+570506.4      & 3469 &           &  466 &   7.2 & 0.06 &   68 &   6.8 &   \\
SBS1537+577               & SBS1533+574A                 & 3348 &           &  534 &   6.8 & 0.06 &   66 &   8.1 &  HII \\
SBS1537+577               & SDSSJ153733.00+583447.8      & 2932 &           &  745 &  10.7 & 0.15 &   89 &   8.3 &  Sc(f) \\
SBS1537+577               & MCG+10-22-037                & 3969 &           &  992 &  15.7 & 0.33 &  116 &   8.5 &  Scd(f) \\
SBS1537+577               & SDSSJ154434.41+571243.8      & 3568 &           &  868 &  10.8 & 0.15 &   90 &   9.6 &   \\
SBS1537+577               & SDSSJ154331.89+571434.0      & 3975 &           &  821 &   7.1 & 0.06 &   67 &  12.1 &   \\
SBS1537+577               & SBS1533+574B                 & 3429 &           &  546 &   3.9 & 0.02 &   44 &  12.2 &  HII \\
SBS1537+577               & SDSSJ153558.83+564108.2      & 3213 &           &  801 &   6.1 & 0.04 &   61 &  13.1 &  BLAGN \\
SBS1537+577               & SDSSJ153040.88+575301.0      & 2896 &           &  780 &   5.7 & 0.04 &   58 &  13.4 &   \\
SBS1537+577               & SBS1542+573C                 & 4027 &           &  885 &   5.8 & 0.04 &   58 &  15.1 &   \\
SBS1537+577               & SDSSJ153325.63+564156.5      & 3362 &           &  962 &   6.1 & 0.04 &   61 &  15.8 &   \\
SBS1537+577               & SDSSJ154054.24+565139.2      & 3408 &           &  728 &   3.3 & 0.01 &   40 &  18.0 &  Irr(sa) \\
SBS1537+577               & 2MASXJ15335796+5650509       & 3260 &           &  797 &   3.6 & 0.01 &   42 &  18.7 &   \\
\tableline
SBS1551+572               & Ly$\alpha$ @       4097      &      &           &      & \\
SBS1551+572               & Filament Axis                & 3600 & 497       & 1778 & \\
SBS1551+572               & SDSSJ155235.47+565604.2      & 4104 & -7        &  161 &   7.3 & 0.06 &   69 &   2.3 &   \\
SBS1551+572               & SBS1553+573                  & 3610 &           &  290 &   9.6 & 0.12 &   83 &   3.5 &  E \\
SBS1551+572               & SDSSJ154434.41+571243.8      & 3568 &           &  997 &  10.8 & 0.15 &   90 &  11.0 &   \\
\tableline
SBS1624+575               & Filament Axis                & 3550 &           & 4103 & \\
SBS1624+575               & KHG1-C09                     & 2848 &           & 3635 &   1.0 & 0.00 &    3 & 1023.4 &  Comp \\
SBS1624+575               & SBS1543+593                  & 2698 &           & 3973 &   1.0 & 0.00 &    3 & 1118.6 &  dwarf \\
SBS1624+575               & KHG1-C10                     & 4317 &           & 4084 &   1.0 & 0.00 &    3 & 1149.7 &   \\
\enddata
\end{deluxetable*}


\section{COS data preparation} 

\begin{figure*}\figurenum{\Fshifts}
\begin{center}$\begin{array}{c} \includegraphics[width=\textwidth, angle=0]{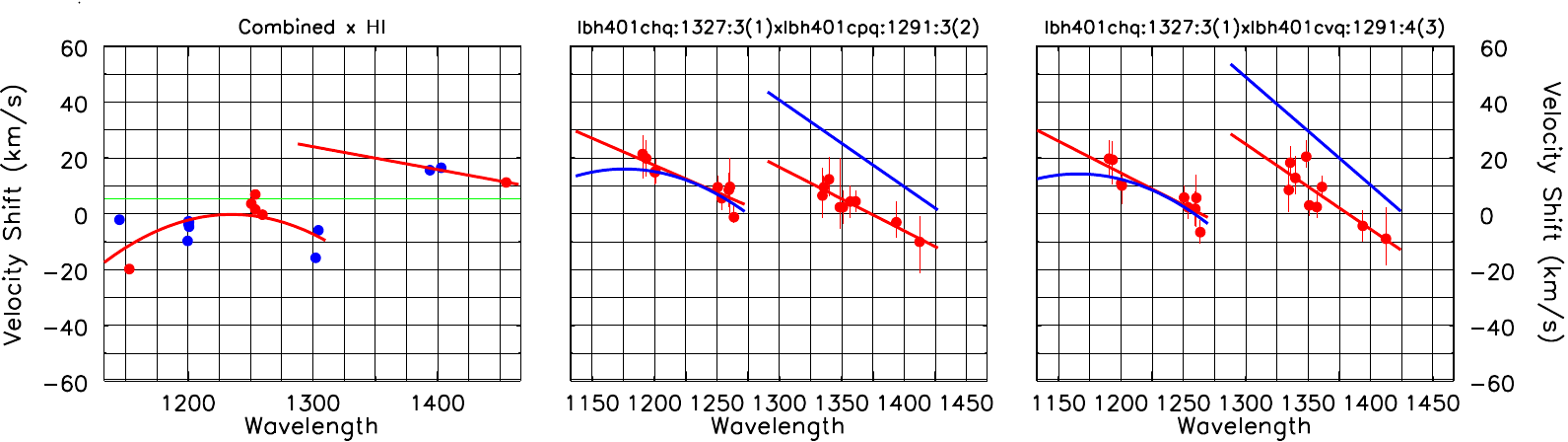} \end{array}$ \end{center} 
\caption{%
In each panel red points show the cross-correlation offsets for individual
absorption lines between a spectrum and the chosen reference spectrum (blue
points in panel 1 are for lines with high optical depth for which the centroid
determination is more uncertain). The error bars indicate the uncertainties. 
The left panel gives the differences between the centroids of the ISM lines
fitted in the 21-cm spectra vs.\ the combined UV spectra. The red lines are
least-squares fits through these points. The blue lines in the
spectrum-vs-spectrum panels show the final offsets implied for each spectrum,
found by combining the relative offsets of the UV spectra with the offset needed
to align the spectrum with the 21-cm data. The green line shows the offset
needed to bring the nominally calibrated heliocentric velocities in line with
LSR velocities. The label (e.g. lbi603o51:1327:3(1)) contains, respectively, the
observation ID, central wavelength, FP-POS setting and exposure ordinal number.
Clearly, spectrum with 1327:3 show shifts relative to spectra with 1291:3 that
vary by up to 60~\kms\ across one side.
}\end{figure*}

\begin{figure} \figurenum{\Ferror}
\begin{center}$\begin{array}{c} \includegraphics[width=\colwidth, angle=0]{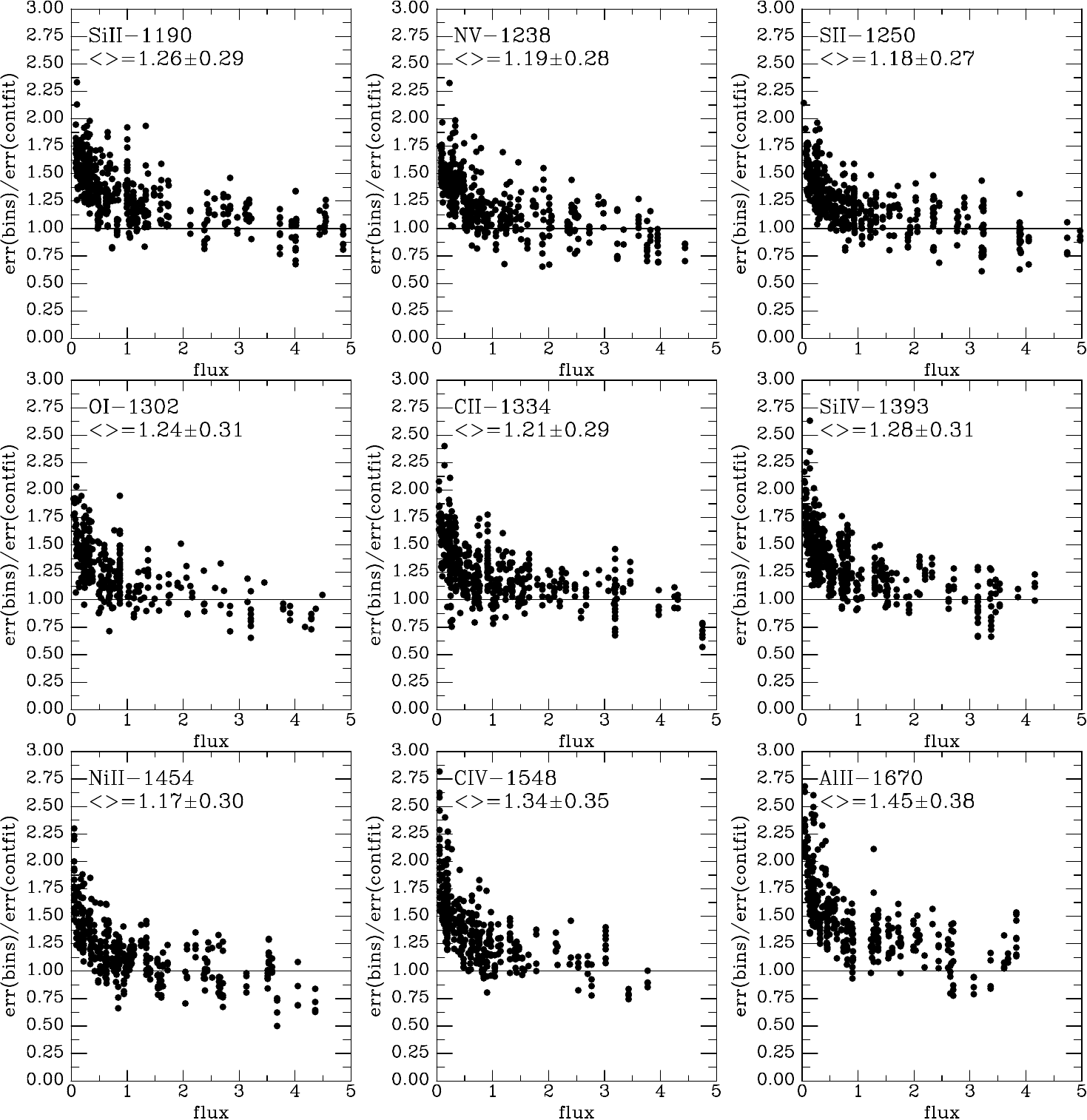} \end{array}$ \end{center} 
\caption{%
Ratio of fitted error to CALCOS error in individual 1-orbit COS exposures as
function of target flux, in units of
\dex{-14}\,erg\,\cmm2\,s$^{-1}$\,\AA$^{-1}$, near nine ISM absorption lines. The
fitted error is found as the rms around the polynomial fit to the continuum,
while the CALCOS error is found by averaging the error array in the original
datasets over a 300~\kms\ wide line-free region near each ISM line. We include
only spectra for which a 1st or 2nd order polynomial sufficies, and for which
the continuum is relatively flat (variation $<$20\% in a window several \AA\
wide).
}\end{figure}


\subsection{Correcting the \COS\ wavelength scale} 
\par As we were preparing COS spectra for a study of \OVI\ absorbers at high S/N
ratio (which turned into Savage et al.\ 2014) we discovered that in many cases
spectral lines in different spectra of the same target did not align, with
misalignments of up to $\pm$40~\kms, and with the misalignment varying as
function of wavelength. A summary of the \COS\ wavelength calibration procedure
is given by Oliveira et al.\ (2010). It is based on spectra of 6 bright targets
taken in program 11474 and 11487, tying absorption lines in their spectra to
\STIS-E140H data of the same targets. The workshop paper states that dispersion
relations were derived only using the FP-POS=3 setting and that a linear
dispersion relation was used for the G130M and G160M gratings. Thus, the offsets
may be due to the fact that there is no explicit calibration spectrum for every
combination of central wavelength and FP-POS settings, so that small distortions
in the image on the detector are not accounted for by CALCOS, which assumes the
same dispersion relation for different FP-POS settings at the same central
wavelength. Offsets can also occur if the target is not at precisely the same
spot in the aperture during different visits; this can not be accounted for in
the calibration code even in principle. Since the offsets can be as large as a
resolution element, this produces an artificial smearing of the lines in the
combined spectra. Moreover, absorption lines in a single absorption-line system
can appear misaligned.
\par Other authors have also noticed this issue. E.g., Savage et al.\ (2011),
Meiring et al.\ (2011) and Tumlinson et al.\ (2011). They fix the offsets by
cross correlating each spectral line separately, determining a centroid and
adjusting the wavelengths for each line in each exposure individually. Different
lines in the same absorption system are also explicitly forced to be aligned,
although that is of course an assumption. Danforth et al.\ (2014) released a
code in which they assume that there is a constant (wavelength-independent)
offset between exposures, which is determined by a blind cross-correlation of a
single 10\,\AA\ wide spectral region around a strong interstellar line,
separately for each detector segment (1255--1266\,\AA\ for G130M segment B,
1330--1340\,\AA\ for G130M side A, 1520--1533\,\AA\ for G160M segment B and
1664--1676\,\AA\ for G160M segment A). \par Both of these methods suffer from
problems, especially for spectra with S/N ratios below $\sim$10. In particular,
due to the random noise, the offsets found from cross correlations have
uncertainties of a few km/s, so shifting lines to match their centroids may not
yield the correct answer and gives rise to derived shifts that bounce around
over short wavelength ranges. When applying the Danforth et al.\ (2014) method,
in most cases the shifts are similar to the shifts we find using our method
(described below), but  we have found a few instances where the selected
wavelength ranges contain a complicated set of IGM lines and this leads to
wildly incorrect shifts -- in one case a shift of 100~\kms\ was derived.
Furthermore, even after aligning in this fashion, the ISM lines in the combined
spectra will not always be aligned with the Galactic 21-cm emission, as there
may be general offsets in the spectrum chosen to provide the reference
wavelengths. Thus, the absolute velocity scale remains suspect.
\par To solve these problems with wavelength offsets, we instead take the
following approach:
\par\noindent (a) First, display all exposures, separately for each grating and
each detector side. Then, from the set of ISM lines usually visible in the
spectrum (\FeII \ll1144, 1608 \PII \l1152, \SiII \ll1190, 1193, 1260, 1304,
1526, \NI \ll-1199/1199/1200, \SiIII \l1206, \SII \ll1250, 1253, 1259, OI
\l1302, \NiII \ll1317, 1370,1454, \CII \l1334, \CII* \l1335, SiIV \ll1393, 1402,
\CIV \ll1548,1550, \AlII \l1670) select the ones that are strong and not
contaminated by IGM lines. We note that the \NI, \OI\ and SiII-1304 lines
usually are not selected because of the presence of strong geocoronal \NI, \OI\
and \OI* emission. To this set we add all strong IGM lines.
\par\noindent (b) Cross-correlate each line in each exposure pair, using a
3~\AA\ wide region. One spectrum (the first in the list) is chosen as a
wavelength reference.
\par\noindent (c) For each exposure, plot the offsets relative to the reference
exposure as function of wavelength and fit a 1st (usually), 2nd (sometimes) or
3rd (if the S/N ratio permits it) order polynomial through these points,
separately for side A and B of each grating. This step is illustrated in
Fig.~\Fshifts.
\par\noindent (d) Apply the shift as function of wavelength to each spectrum and
then combine (see below for more details on this step).
\par\noindent (e) Fit the centroids of interstellar lines that are not
saturated, not blended, not too weak and not otherwise distorted. Well-separated
HVCs and IVCs are fitted separately.
\par\noindent (f) Fit the centroid of a 21-cm emission spectrum in the direction
of the target, with velocities referenced to the LSR. Comparing the centroids of
the 21-cm emission components with those of the ISM lines (see leftmost panel in
Fig.~\Fshifts) then gives the wavelength offsets needed to align the reference
spectrum. The 21-cm data are from the LAB survey (Kalberla et al.\ 2005) or the
GASS dataset (Kalberla et al.\ 2010). If higher angular resolution data is
available, these are used instead. The centroid determination is done separately
for weak and strong ISM lines and separately for low- and high-velocity (if any)
gas. For strong lines multiple components can blend. Usually the strong lines
are not used in the end, since weak emission components can give substantial
absorption components and thus the centroids of emission and absorption no
longer align.
\par This procedure yields a final spectrum with wavelengths such that
velocities will be on the LSR scale, which can be converted to a heliocentric
wavelength scale. This method works in all cases and does not require the
presence of STIS data. Figure~\Fshifts\ shows examples of the offsets as
function of wavelength for one of the targets used in this paper.


\subsection{Combining individual exposures} 
\par To combine the aligned spectra, we add the total counts in each pixel and
then convert back to flux, using the average flux/count ratio at each wavelength
that was used in the original retrieved datafiles. This is the same procedure
that was followed by Meiring et al.\ (2011) and Tumlinson et al.\ (2011), but
stands in contrast to the procedure of Danforth et al.\ (2014), who instead
calculate an inverse-variance weighted average flux. For cases with equal
exposure time for different exposures and non-varying target flux, the different
methods give the same answer. However, it is easily shown analytically that if
the target flux varies between exposures the latter method gives the wrong
answer. For instance, combining an exposure with one in which the source
brightened by a factor two, the inverse-variance-weights give a final flux in
the continuum that is 4/3 times the lower value. But at low count rates (e.g.\
in the darker part of spectral lines), the inverse-variance-weight gives a flux
more like 3/2 times the lower value. I.e., this weighting scheme changes the
shape of the spectral lines, which is not the case when combining counts.


\subsection{Error calculation} 
\par A study of the error array in the 1-dimensional spectra that CALCOS
provides revealed that at low fluxes the pipeline gives incorrect numbers. This
is found by comparing the CALCOS errors to measured errors. CALCOS errors were
calculated as the average of the error arrays in a 300~\kms\ wide line-free
region near nine ISM absorption lines (in a spectrum rebinned by seven pixels).
Measured errors are given by the rms found when fitting a polynomial through the
line-free regions near the same nine ISM absorption lines, also in spectra
binned to seven pixels. Figure~\Ferror\ shows the ratio of the error array in
the datafile to the fitted error. Clearly, for targets with fluxes below
$\sim$\dex{-14}\,\fu\ the measured rms typically is lower than the value implied
by CALCOS, by up to a factor 2 near a flux of \dex{-15}\,\fu. We have been
unable to track down the source of this problem, but it seems possible that at
low count rates the CALCOS pipeline overestimates one of the contributions to
the error calculation.
\par We find a better match between calculated and measured error if we just
estimate the error from the Poisson noise implied by the total count rate. In
that case the calculated and measured error remain similar even at the lowest
fluxes. Our final errors are thus calculated using this approach, and ignore the
expected, but apparently not actually seen, contributions from dark count,
background subtraction and other such items.


\section{Simulations} 

%

\begin{figure*} \figurenum{\Fsimulgals}
\begin{center}$\begin{array}{c} \includegraphics[width=\srotwidth, angle=270]{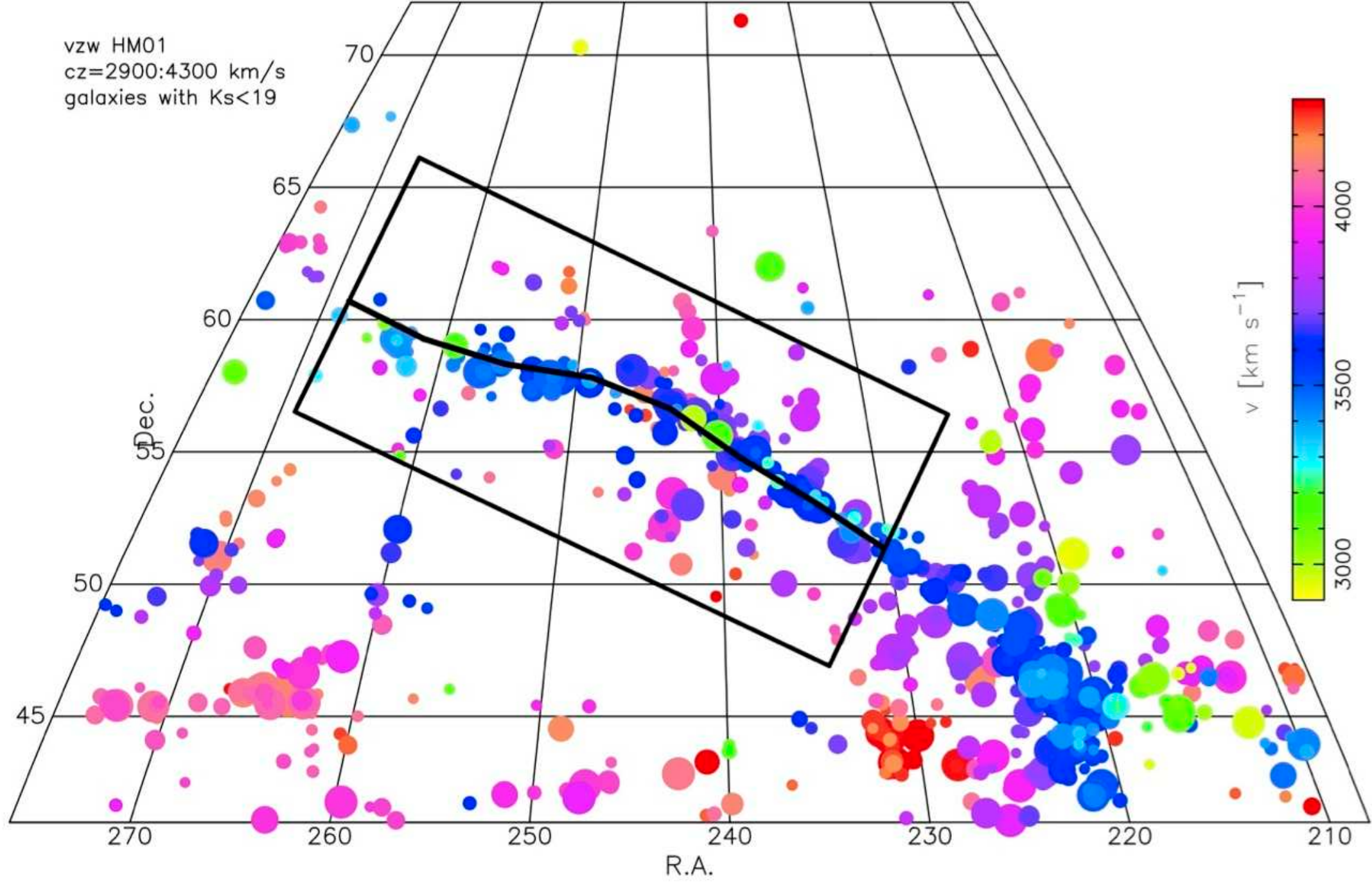} \end{array}$ \end{center} 
\caption{
View of the galaxies as seen from one viewpoints in the vzw simulation. Galaxies
are shown by colored circles, with the same scalings as were used for
Figs.~\Fboxmap\ and \Fqsomap. The filament axis is shown as a thick black line,
found by treating the simulated galaxies in the same manner as the observations.
Finally, an rectangular outline box shows the area of simulated sky used to
measure $N$(\HI) as function of filament impact parameter.
}\end{figure*}

\begin{figure} \figurenum{\Fgalprofile}
\begin{center}$\begin{array}{c} \includegraphics[width=\colwidth, angle=0]{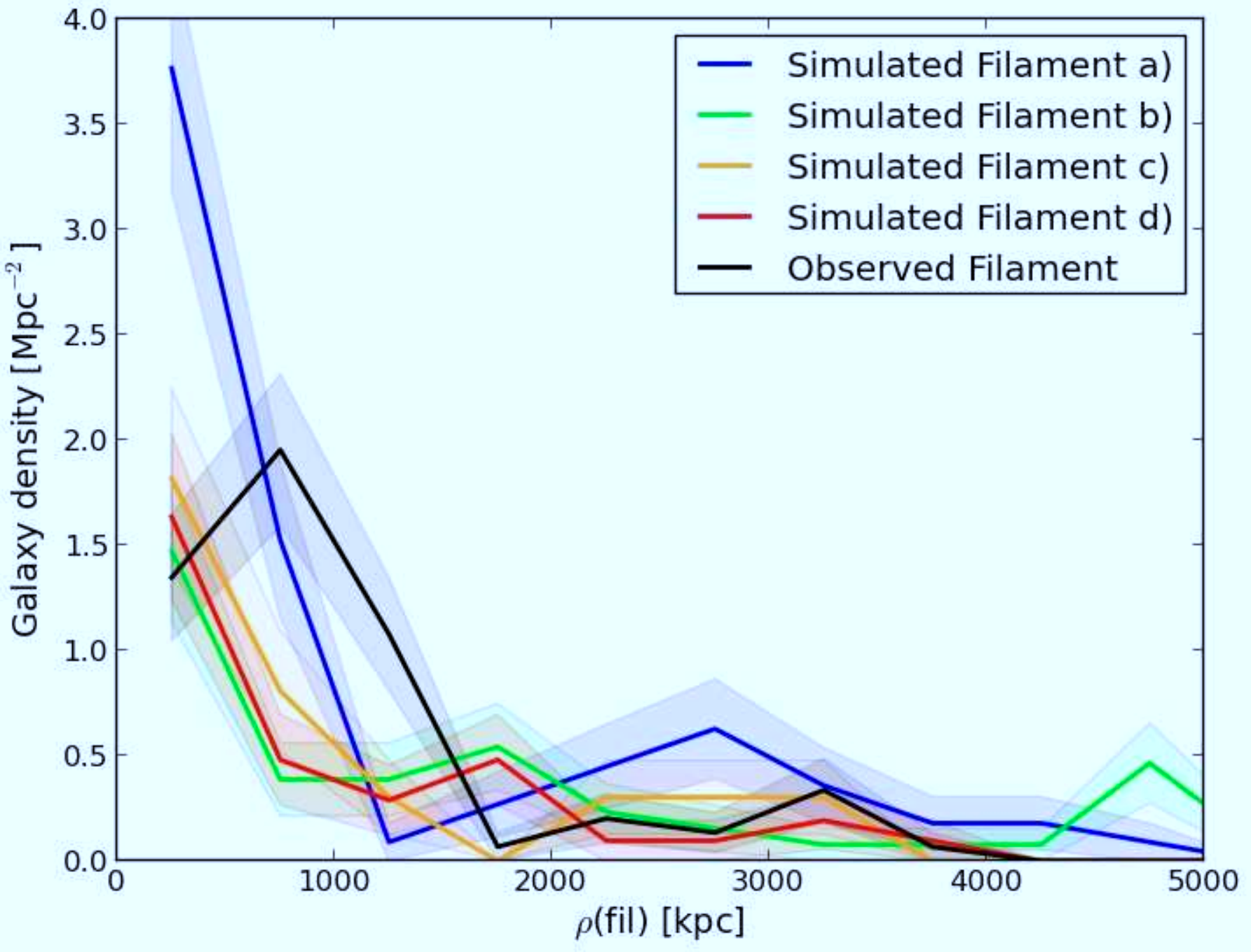} \end{array}$ \end{center} 
\caption{%
Density of galaxies brighter than absolute K$_s$=$-$16.3 as a function of
filament impact parameter for each of the four simulated viewpoints (colors) and
the observations (black). All galaxies out to 5~Mpc from the filament axis are
included and binned every 0.5~Mpc. The vertical scale gives the number of
galaxies per square Mpc in each bin.
}\end{figure}

\begin{figure*} \figurenum{\Fsimullocs a}
\begin{center}$\begin{array}{c} \includegraphics[width=\srotwidth, angle=270]{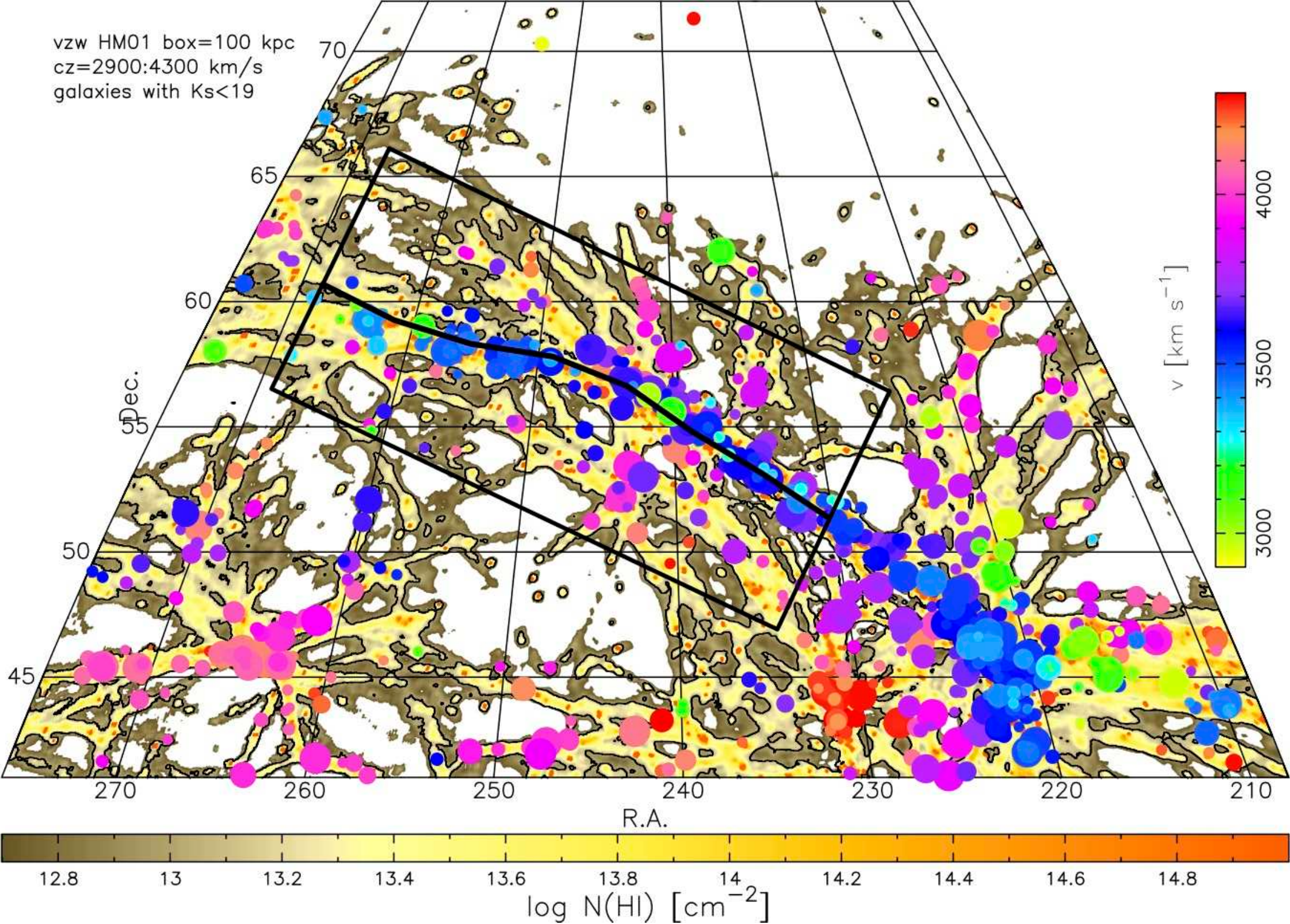} \end{array}$ \end{center} 
\caption{%
View of the vzw simulation (with 100~kpc voxels) from four different viewpoints
in the cube. \HI\ column densities calculated using the Haardt \& Madau (2001)
extragalactic background radiation field are shown by the grey-yellow-red
colors, with a contour at log\,$N$(\HI)=13, which is our detection limit.
Galaxies are shown by colored circles, with the same scalings as were used for
Figs.~\Fboxmap, \Fqsomap\ and \Fsimulgals. The filament axis is shown as a thick
black line, found by treating the simulated galaxies in the same manner as the
observations. A rectangular outline box shows the area of simulated sky used to
measure $N$(\HI) as function of filament impact parameter.
}\end{figure*}

\begin{figure*} \figurenum{\Fsimullocs b}
\begin{center}$\begin{array}{c} \includegraphics[width=\srotwidth, angle=270]{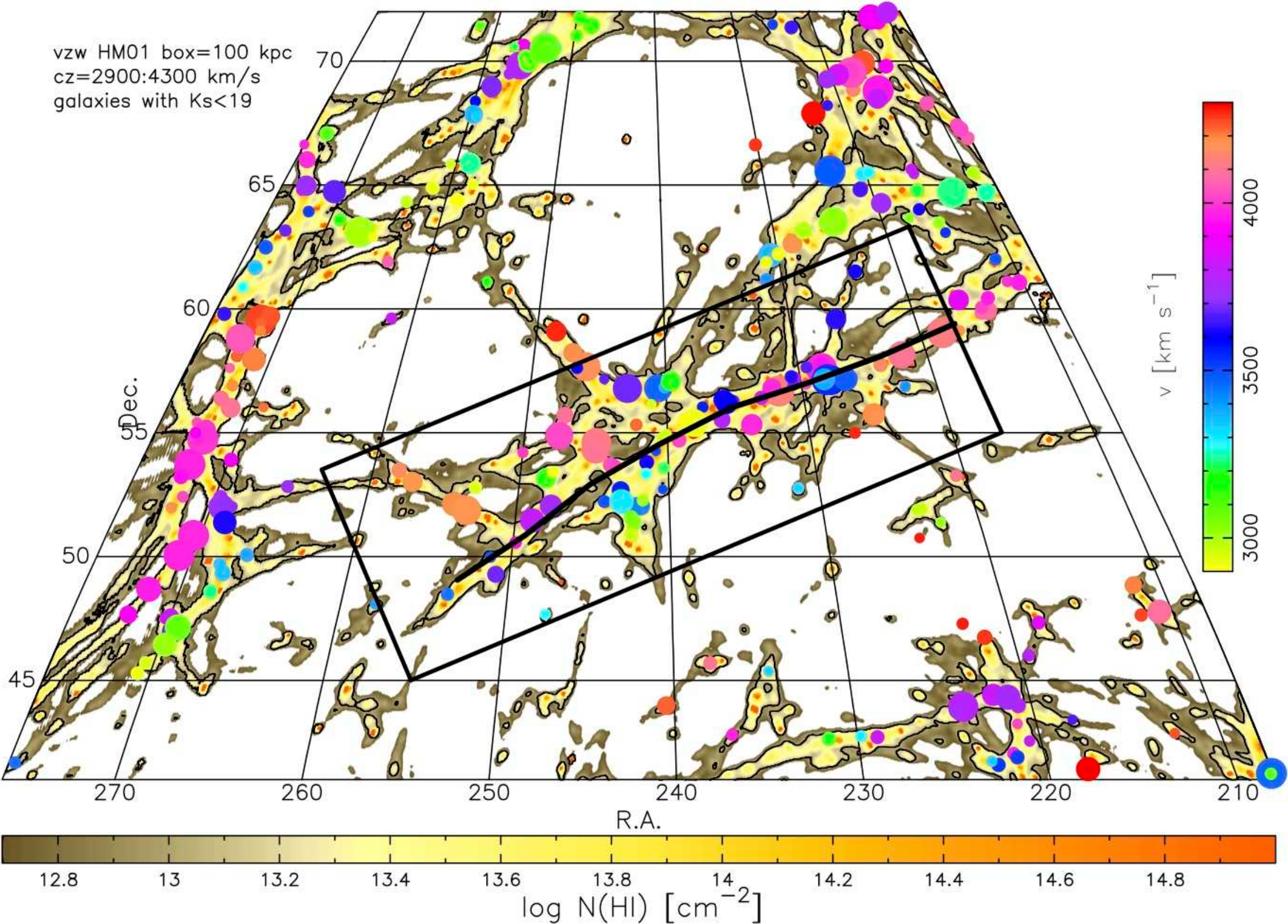} \end{array}$ \end{center} 
\caption{%
Continued.
}\end{figure*}

\begin{figure*} \figurenum{\Fsimullocs c}
\begin{center}$\begin{array}{c} \includegraphics[width=\srotwidth, angle=270]{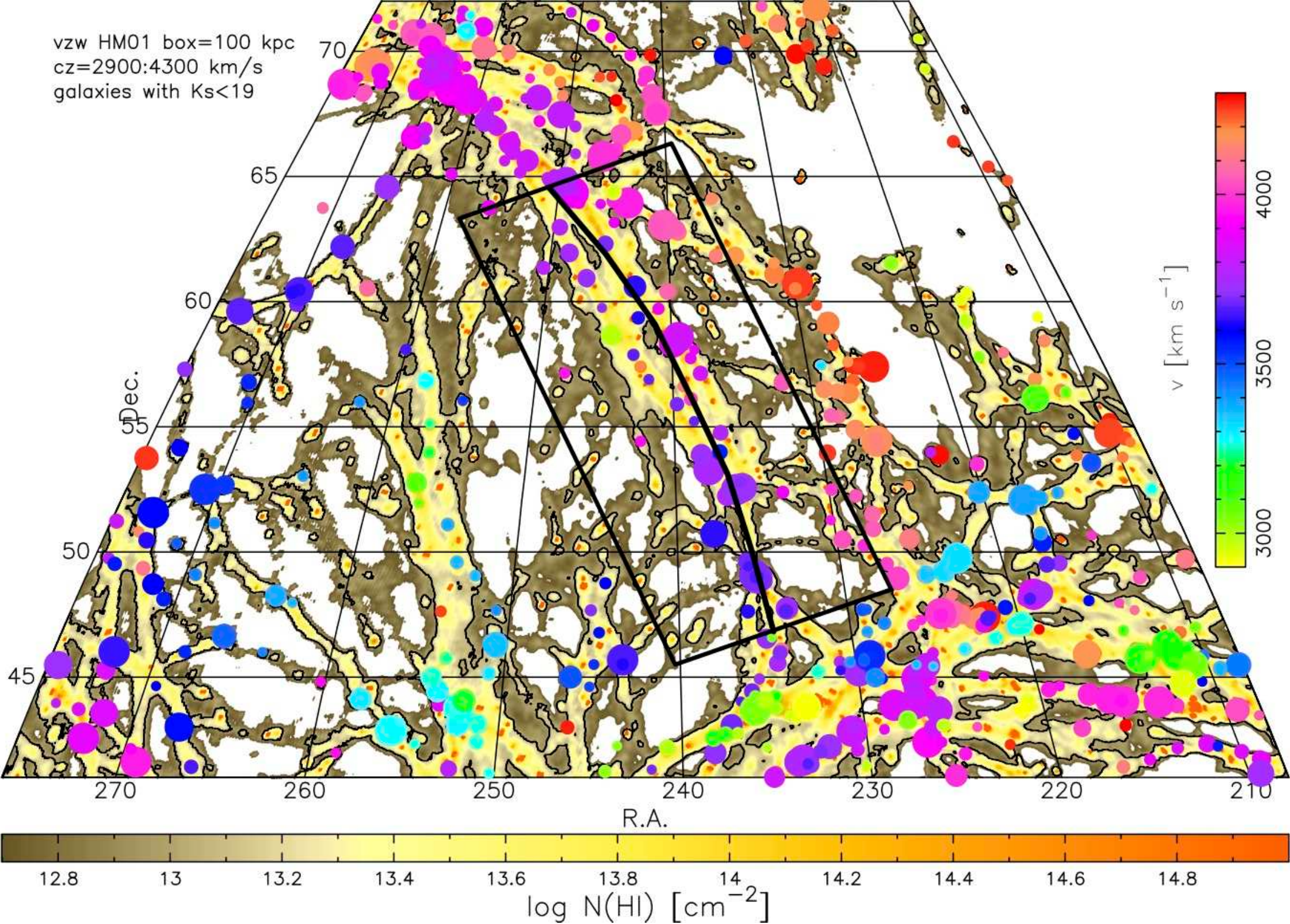} \end{array}$ \end{center} 
\caption{%
Continued.
}\end{figure*}

\begin{figure*} \figurenum{\Fsimullocs d}
\begin{center}$\begin{array}{c} \includegraphics[width=\srotwidth, angle=270]{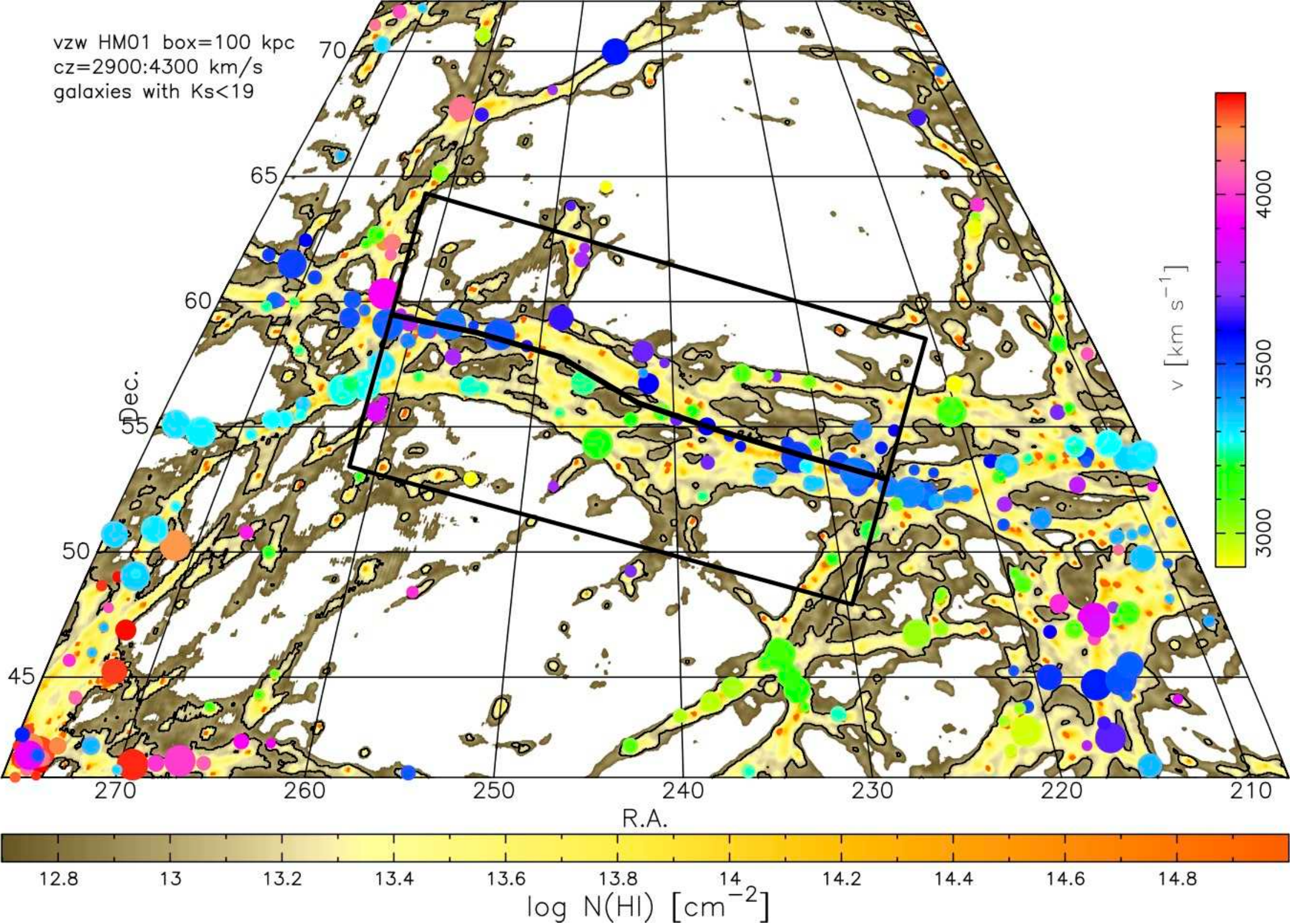} \end{array}$ \end{center} 
\caption{%
Continued.
}\end{figure*}

\begin{figure*} \figurenum{\Fsimulegb a}
\begin{center}$\begin{array}{c} \includegraphics[width=\srotwidth, angle=270]{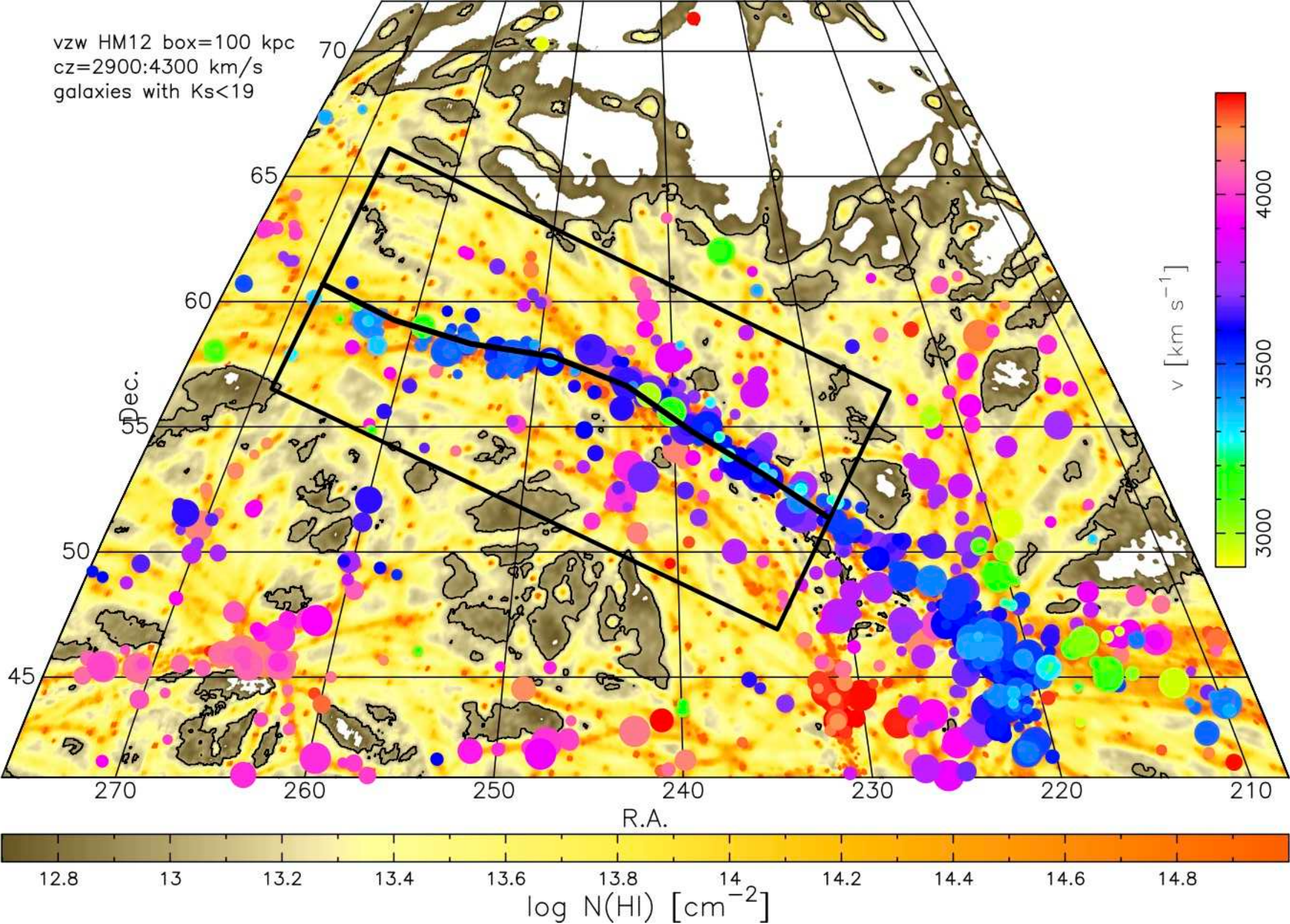} \end{array}$ \end{center} 
\caption{%
View of the vzw simulation from one viewpoint, with four different versions of
the extragalactic background radiation (EGB): Haardt \& Madau (2012) (panel a),
Haardt \& Madau (2012) times 2 (panel b), Haardt \& Madau (2001) (panel c) and
Haardt \& Madau (2001) times 2 (panel d). Otherwise the meaning of the colors,
contours and labels is the same as for Fig.~\Fsimullocs.
}\end{figure*}

\begin{figure*} \figurenum{\Fsimulegb b}
\begin{center}$\begin{array}{c} \includegraphics[width=\srotwidth, angle=270]{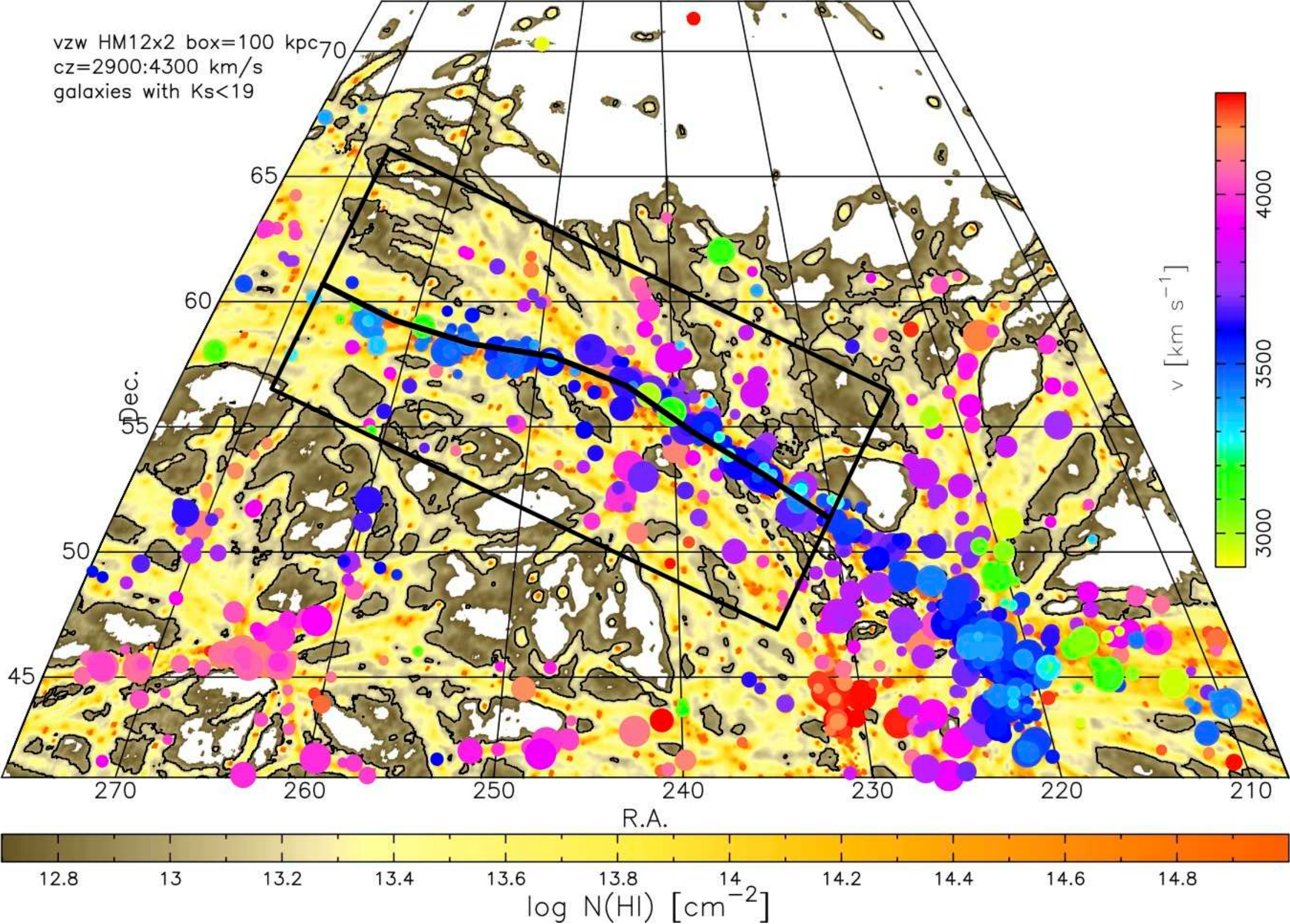} \end{array}$ \end{center} 
\caption{%
Continued.
}\end{figure*}

\begin{figure*} \figurenum{\Fsimulegb c}
\begin{center}$\begin{array}{c} \includegraphics[width=\srotwidth, angle=270]{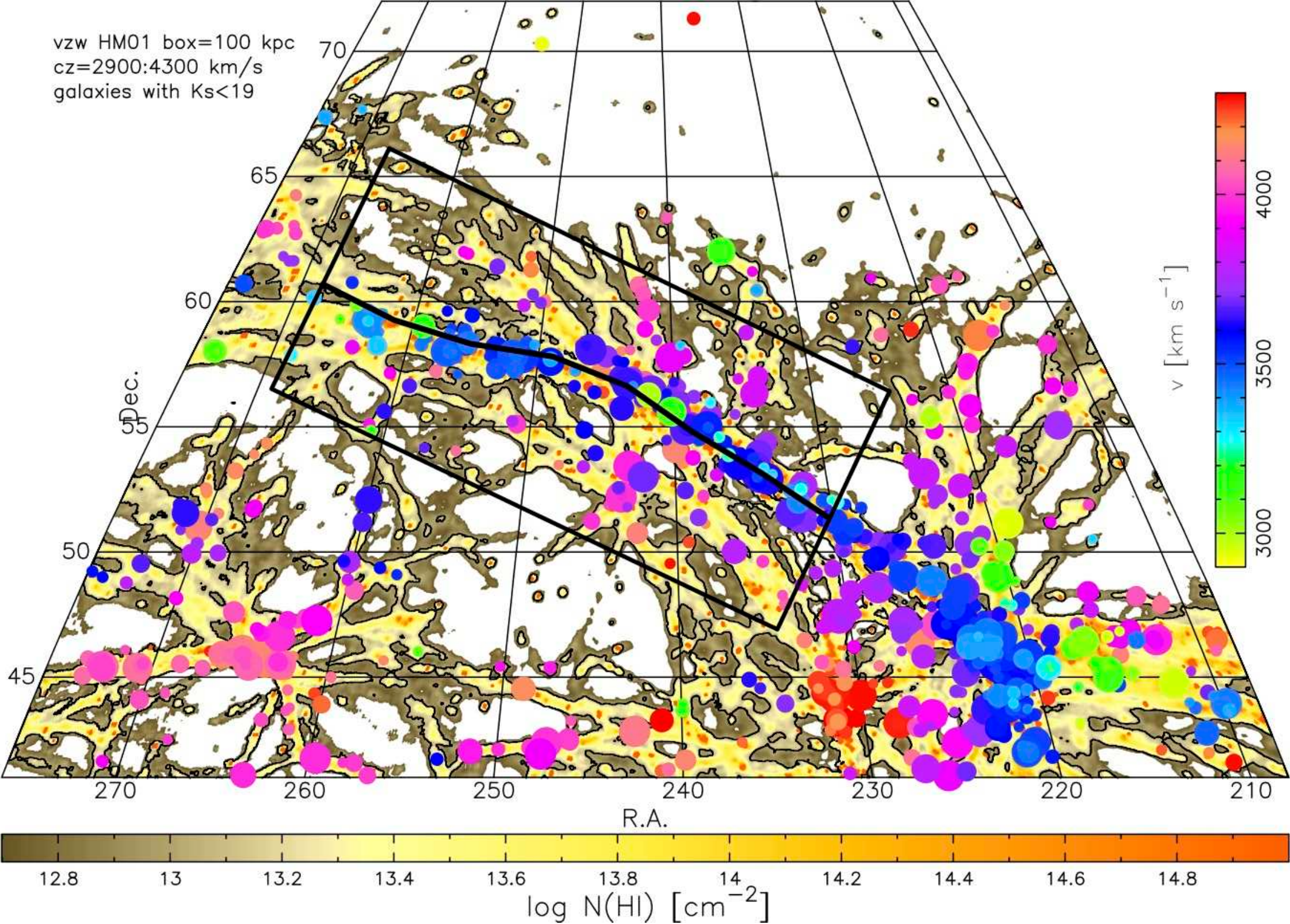} \end{array}$ \end{center} 
\caption{%
Continued.
}\end{figure*}

\begin{figure*} \figurenum{\Fsimulegb d}
\begin{center}$\begin{array}{c} \includegraphics[width=\srotwidth, angle=270]{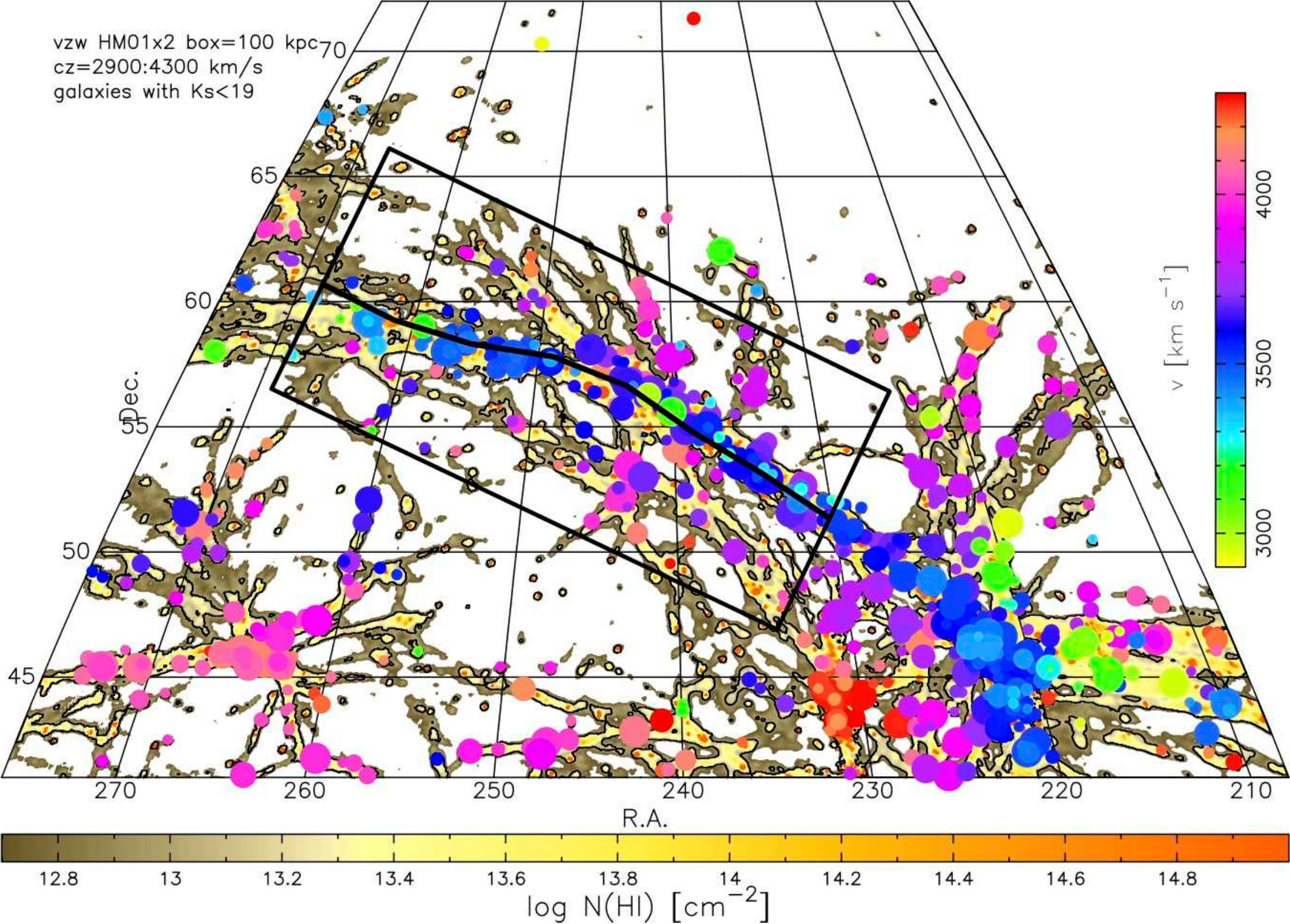} \end{array}$ \end{center} 
\caption{%
Continued.
}\end{figure*}


\subsection{Description} 
\par In order to interpret the observations, we analyzed the cosmological
hydrodynamical simulations of Oppenheimer et al.\ (2010). In this section we
first describe these simulations and the information that we extracted from
them, so that we can refer to this in the results section. The simulations are
{\it Gadget-2} smoothed particle hydrodynamic simulations run with 384$^3$ gas
and dark matter particles in a random periodic volume of 48\,$h^{-1}$~Mpc. The
cosmology used is $\Omega_{\rm M}$=0.28, $\Omega_{\Lambda}$=0.78,
$H_0$=70~\kms\,Mpc$^{-1}$, $\Omega_{\rm b}$=0.046, $n$=0.96, and
$\sigma_8$=0.82. The gravitational softening length is 2.5 $h^{-1}$\,kpc and the
gas mass resolution is 3.5\tdex7~M$_{\odot}$, both of which are sufficient to
resolve the structure of the \Lya\ forest. Dav\'e et al.\ (2010) used a series
of these simuluations with different feedback models to model the \Lya\ forest
statistics, and checked that the \Lya\ forest statistics are resolution
converged by using a 96\,$h^{-1}$~Mpc box with 3.375$\times$ lower resolution.
We use the ``vzw'' simulation, because this simulation's galaxy mass function
shown in Oppenheimer et al.\ (2010) is most similar to the observed galaxy mass
function of Bell et al.\ (2003), which used 2MASS K$_s$-band magnitudes along
with SDSS magnitudes. This simulation mimics momentum-driven winds from
starbursts as described in Oppenheimer \& Dav\'e (2008). However, the other
models described in that paper give very similar results.
\par We use SKID (Spline Kernel Interpolative Denmax)
(http://www-hpcc.astro.washington.edu/tools/ skid.html) to identify galaxies
(Kere\v{s} et al.\ 2005; Oppenheimer et al.\ 2010). We determine the broadband
photometric properties of each galaxy by summing up the single stellar
population models of Bruzual \& Charlot (2003) within each star particle, for
which we have an age and metallicity, assuming a Chabrier (2003) initial mass
function, and adding a dust correction based on the galaxy's metallicity as
explained in Finlator et al.\ (2006). 


\subsection{Finding filaments in the simulations} 
\par To compare these simulations to the observations we take a diffferent
approach than usual. Usually, simulations are displayed by taking a 3-D cube and
collapsing a slice onto a side of the cube. Here, however, we take the
simulations and project the galaxies and \HI\ volume density onto the sky,
taking into account the fact that a sky pixel represents a cone through the
simulation cube, not a rectangular bar. We identify simulated filaments that
look like the observed filament based on how the galaxies appear in the
simulated skies from various locations chosen within the simulation, which we
call ``viewpoints''.
\par To find viewpoints, we started with the SKID-identified galaxies,
projecting their locations onto the same region of the sky as our observed
filament ($\sim$30\deg x30\deg\ centered on R.A.=240\deg, Dec=55\deg). We then
calculated each galaxy's apparent velocity, combining its Hubble flow velocity
with its peculiar velocity and selected the galaxies in a 1400~\kms\ wide window
between \vminfilmap\ and \vmaxfilmap~\kms. Stepping through the simulation cube
in 2\,Mpc intervals in $x$, $y$, and $z$ (i.e., 32 steps each) and plotting the
resulting distribution of simulated galaxies on the sky, we found the viewpoints
where a galaxy filament stood out visually. Next, we refined the search around
each of these viewpoints to get the best viewpoint to within 1~Mpc. Using the
galaxy distances and simulated K$_s$ magnitudes, we find apparent magnitudes.
Applying a magnitude cut of K$_s$$<$19 and using just the galaxy locations, we
derive filament axes for each viewpoint, using the same method as was used for
the observations. 
\par Figure~\Fsimulgals\ presents the galaxy distribution for one viewpoint, in
order to show that the visual impression given by the simulation is similar to
that given by the data. To determine how well simulated galaxy filaments
represent our observed galaxy filament, we compare the galaxy density along the
filament and the K$_s$-band luminosity function (Fig.~\Fgalprofile), using the
boxes around the filaments seen in Fig.~\Fsimullocs. These have lengths of
18-22~Mpc. Within these boxes we find 336, 157, 156 and 127 galaxies with
K$_s$$<$19 for filaments a, b, c and d, respectively. The lower branch of our
observed filament has 137 galaxies and a length of 20.9~Mpc, i.e.\ similar to
the simulated filaments.
\par When comparing simulated and observed filament galaxy statistics, we apply
a luminosity cut of K$_s$=$-$16.3 which corresponds to $M_*$=\dex9\,M$_{\odot}$
and which is the quoted simulation resolution limit. Even though the vzw
simulation does not quench galaxies as required by observations, this does not
matter, because Oppenheimer et al.\ (2010) show that the number density of
galaxies above \dex{9.0}\,M$_{\odot}$ is in agreement with Bell et al.\ (2003)
within a factor of 30\% (see their Figure 6). While we did explore other wind
models including a no wind model and a constant wind model, there is
statistically no difference in filament statistics for a given ionization
background.
\par Figure~\Fgalprofile\ shows the galaxy filament density as a function of
impact parameter for the luminosity cut. Even though more galaxies were used to
identify the observed filament and make the initial identification of simulated
filaments, we plot the galaxy densities with absolute K$_s$$<$$-$16.3 binned
into 500~kpc bins perpendicular to the filament in areal density units. The
observed filament has a total density of 2.6 galaxies Mpc$^{-1}$ within 5~Mpc,
while simulated filaments a), b), c), and d) have 3.8, 2.0, 1.9, and 1.7, thus
bracketing the observed filament density. Based on binned areal density, no
filament exactly matches the observed filament, but filament a) is generally
denser, and filaments b), c), and d) are less dense within 1.5~Mpc and have
similar areal densities as the observations out to 4~Mpc. We also checked the
galaxy luminosity functions along the filaments and find that the simulated
luminosity functions are shaped like we expect from Oppenheimer et al.\ (2010).
They have a similar density of galaxies, but more bright galaxies than the
observations.

\subsection{Ionizing radiation in the simulations} 
\par To calculate an \HI\ column density map as seen from the selected
viewpoints, we used four different levels for the intensity of the extragalactic
background radiation. First, we took the model of Haardt \& Madau (2001; HM01),
in which the
contribution of galaxies and quasars to the ionizing flux is comparable at all
redshifts. Second, we used the purportedly improved model of Haardt \& Madau
(2012; HM12), which used more observational constraints. In this version the
escape fraction of ionizing radiation from galaxies evolves over time such that
the contribution of galaxies at $z$$\sim$0 is minimal. As Kollmeier et al.\
(2014) show, the predicted \HI\ column density distribution at $z$=0 then fails
to match the distribution observed by Danforth et al.\ (2014), giving five times
too many \HI\ clouds at any given column density. The HM01 version of the
radiation field matches the observations better. Kollmeier et al.\ (2014)
discuss the possible causes and remedies to improve the match between the
observed $z$$\sim$0 \Lya\ column density distribution and the distribution
implied by the HM12 model, but are unable to come to a conclusion. Shull et al.\
(2015) propose to solve the discrepancy by using a different simulation, which
has stronger heating and then a higher escape fraction (5\%) of ionizing photons
from starforming galaxies gives a factor two more ionizing photons. Khaire \&
Srianand (2015), however, redo the modeling of the EGB from scratch and suggest
that the QSO luminosity function alone implies a factor two more ionizing
photons than found by HM12, which combined with a 4\% escape fraction resolves
the discrepancy. In Sect.~\Scrosscuts\ and \Sdetfrac\ we describe our new way of
constraining the EGB models, now based on the spatial distribution of the \Lya\
forest in relation to galaxy filaments, rather than using the column density
distribution.
\par We now describe our method to calculate \HI\ column densities, which depend
on the assumed intensity of ionizing radiation. Our method produces a more
realistic description of the observations than the standard method of collapsing
a simulation cube onto one of the side planes. We start by calculating the
distance to each (3D) simulation voxel (i.e\ a 100$\times$100~kpc grid cell),
given a viewpoint. This is converted to its Hubble flow velocity, using a Hubble
constant of 71~\kms\,Mpc$^{-1}$. To this recession velocity we add the peculiar
velocity of the material in that simulation voxel to get its apparent velocity.
Selecting a range of velocities, we then draw a diverging bundle of sightlines
through the cube for each individual sky pixel, with the sightlines in a bundle
close enough that several span the most distant voxel used. Sampling each
sightline in steps of about 1/3 the size of a voxel, we sum the \HI\ volume
density that was calculated using value of the total gas density in that voxel
combined with the model for the EGB. This is multiplied by the pathlength to
find the column density in each sky pixel. Although we use SPH simulations, we
transform the particles onto a 3D grid with 100~kpc cells, which corresponds to
6\farcm5 at redshifts of 3500~\kms. We also tried using 50~kpc cells, and find
no statistical difference, which is to be expected since structures in the \Lya\
forest structures are larger than this (Dav\'e et al.\ 2010).
\par The resulting \HI\ column densities are shown in Fig.~\Fsimullocs. From
these figures it is clear that with a velocity range of $\sim$1400~\kms\ a
single filament as seen in the distribution of galaxies is obvious, but in the
\HI\ column density distribution other filaments ``spill over'' into the
selected velocity range. Figures~\Fsimulegb a to d show the effect of varying
the intensity of the ionizing background on the \HI\ column density maps.


\begin{figure*} \figurenum{\Fionization}
\begin{center}$\begin{array}{c} \includegraphics[width=\ccolwidth, angle=0]{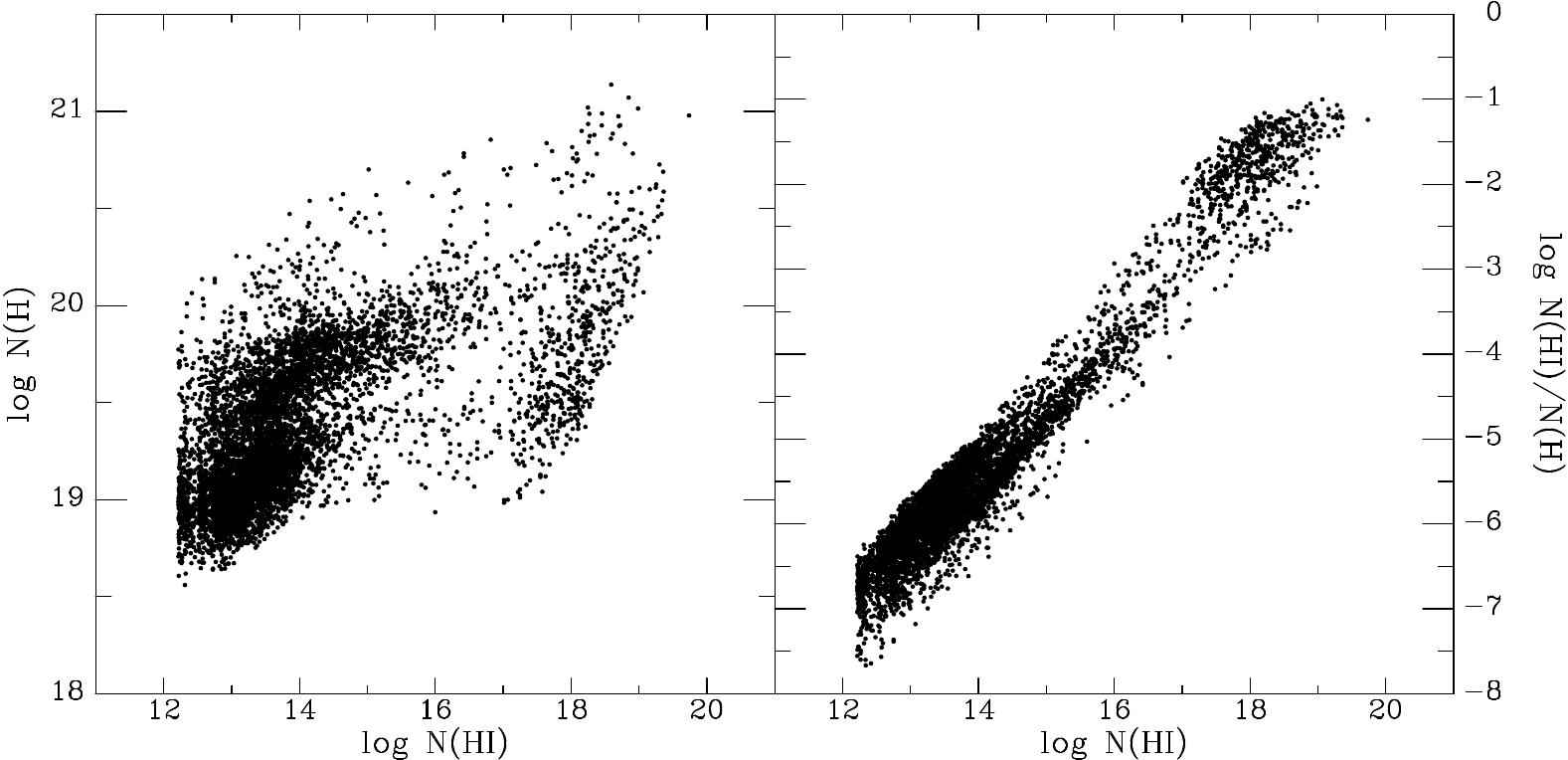} \end{array}$ \end{center} 
\caption{%
Scatter plot comparing the total baryon column density (log\,$N$(H)) against the
column density of neutral hydrogen (log\,$N$(\HI)) for the rectangular box
outlining the filament in Fig.~\Fsimullocs a. I.e., the column density when
integrating from $cz$=2900 to 4300~\kms\ (a 20~Mpc pathlength), which is
dominated by the voxels inside the filament. The concentration of points around
log\,$N$(\HI)$\sim$18 in the left panel corresponds to pixels with a galaxy
present. The panel on the right shows how the fraction of hydrogen that is
neutral varies with the observed \HI\ column density; i.e.\ it represents the
inverse of the ionization correction.
}\end{figure*}

\begin{figure*} \figurenum{\Fdensityhist}
\begin{center}$\begin{array}{c} \includegraphics[width=\ccolwidth, angle=0]{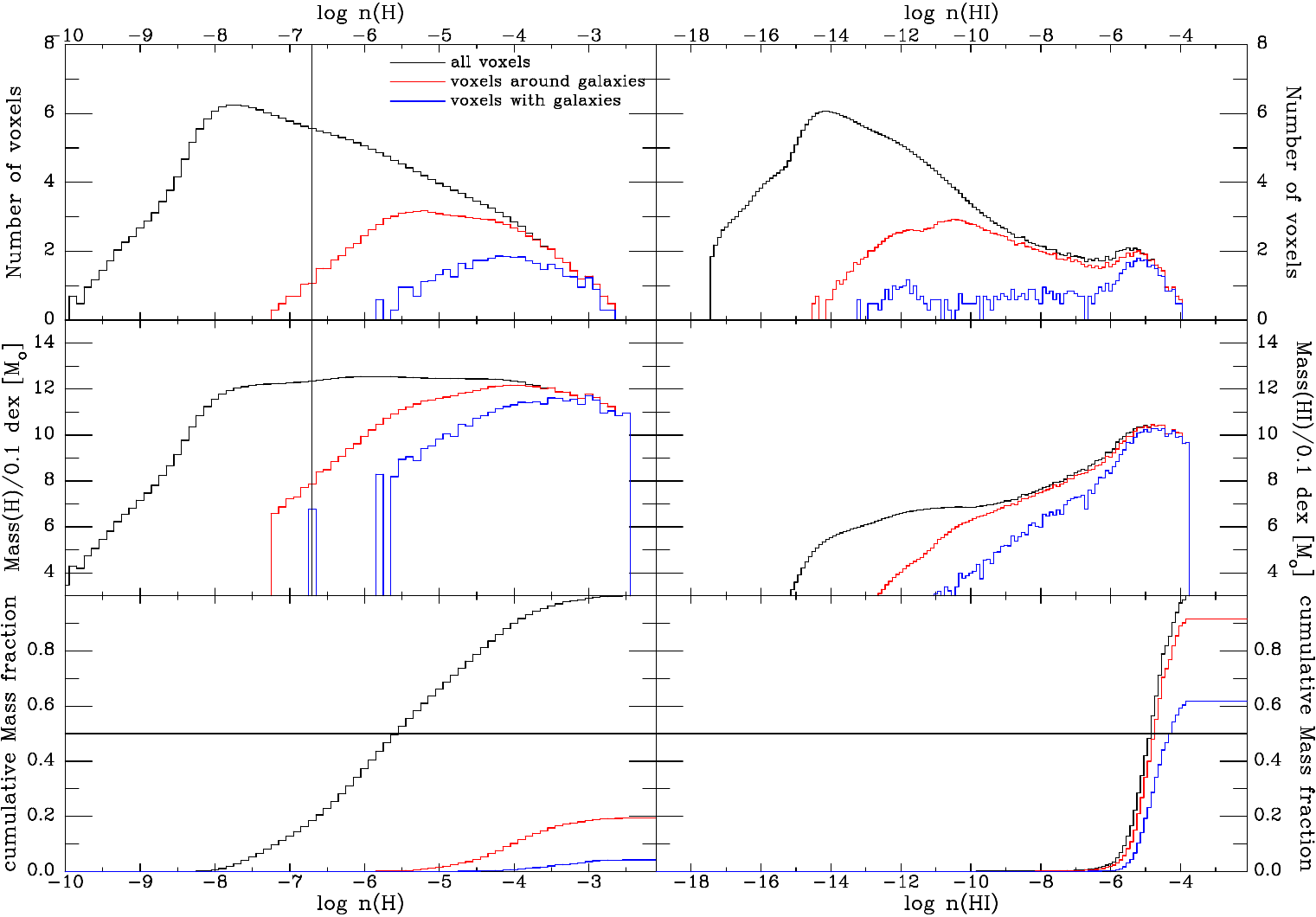} \end{array}$ \end{center} 
\caption{%
Top panels: number of 100~kpc$^3$ voxels as function of total or neutral
hydrogen volume density in the rectangular box outlining the filament in
FIg.~\Fsimullocs a. Middle panels: total mass of hydrogen in each 0.1 dex wide
bin in log\,$n$(H) (left) or log\,$n$(H) (right). Bottom panels: cumulative mass
fraction as function of volume density. The black curves are for the case that
includes all voxels in the box. The blue curves are for voxels in which SKID
identified a galaxy, while the red curves are for the 27 voxels around those
galaxy-containing voxels (i.e.\ a 300x300~kpc box). The vertical line in the
left two panels indicates the average density of hydrogen (log\,$n$(H)=$-$6.6,
while the vertical line in the right panels indicates the density that
corresponds to a \HI\ column density of \dex{13}\,\cmm2\ in the 20~Mpc deep box.
}\end{figure*}


\subsection{Properties of the gas in the simulation} 
\par To conclude this discussion of the simulations we present some of the
characteristics of the neutral hydrogen. In Fig.~\Fionization\ we show the
relation between the total and neutral amounts of hydrogen for the pixels inside
the selection box in the first filament (see Fig.~\Fsimullocs a), using the HM01
EGB model. The left panel shows that within the about 8~Mpc wide strip along the
filament the total hydrogen column density (integrated from $cz$=2900 to
4300~\kms) only varies from about log\,$N$(H)=18.5 to 20.0. The concentration of
dots on the right is caused by sightlines with high $N$(\HI), which occur in
voxels that contain a galaxy. Panel (b) reveals that the ratio of total to
neutral hydrogen (i.e.\ the ionization correction) ranges from about \dex{+6.5}
for absorbers with log\,$N$(\HI)$\sim$13 to about \dex{+1} when
log\,$N$(\HI)$\sim$19.5, with a spread of only about 1 dex at any given value of
$N$(\HI). This plot clearly shows that (a) the \HI\ column density can serve as
a proxy for the total hydrogen column density and (b) the \HI\ traces only a
very small fraction of the baryons, which was demonstrated by Dav\'e et al.\
(1999), although they did not explicitly show this in a figure.
\par The top and middle histograms in Figure~\Fdensityhist\ show the number of
pixels and the total or neutral hydrogen mass as function of the total and
neutral hydrogen volume density, for the pixels inside the outlined box for the
first viewpoint (see Fig.~\Fsimullocs a). The vertical line in the left panels
represents the average hydrogen density of the universe: $n$(H) =
0.75$\times$0.046\,(3\,H$_o^2$)/(8$\pi$G) = \dex{-6.7}\,\cmm3. The bottom
histograms give the cumulative mass fraction as function of volume density.
Three different curves are shown, with the black curve including all voxels
included in the outlined and the blue curve showing the voxels in which a
simulated galaxy is present. The red curve is for the 27 voxels containing each
galaxy (not double counting cases where these overlap when there are multiple
galaxies in the list that are close together), i.e.\ they represent the
circumgalactic medium out to about 150~kpc.
\par This figure shows that in a given random direction through the filament
most voxels have low volume density. The circumgalactic and galactic voxels have
high density, as can be expected. The middle left panel shows that the baryon
mass is fairly evenly distributed between all volume densities. In the 0.1 dex
bins near log\,$n$(H)=$-$8, $-7$, $-$6, $-$5, $-$4, and $-$3 there are 1, 3, 6,
5, 4 and 1\tdex{12} solar masses of baryons, respectively. Thus, the total mass
of baryons is dominated by the extended intergalactic medium. In contrast, the
distribution of neutral hydrogen mass is dominated by the voxels around galaxies
(see right middle panel).




\section{Results} 

\begin{deluxetable*}{lrrrcrrrrrrr} \tablenum{3} \tablecolumns{11} \tablewidth{0pt}
\tabletypesize{\scriptsize} \tablecaption{\Lya\ parameters and detection limits for absorption between \vminfiltab\ and \vmaxfiltab\ \kms} \tablehead{%
\ch{Target} & \ch{Flux}  & \ch{S/N} & \ch{$\rho$(fil)} & \ch{v(\Lya)}    & \ch{EW}     & \ch{log $N$(\HI)} & \ch{$b$}        & \ch{v(\Lya)} & \ch{log $N$(\HI)} & \ch{$b$}    & \ch{Note} \\
            &            &          &                  & \ch{($N_a(v)$)} &             & \ch{($N_a(v))$}   & \ch{($N_a(v)$)} & \ch{fit}     & \ch{(fit)}        & \ch{(fit)}  &           \\
            & \ch{[f.u.]}&          & \ch{kpc}         & \ch{[km/s]}     & \ch{[m\AA]} & \ch{[\cmm2]}      & \ch{[\kms]}     & \ch{\kms}    & \ch{[\cmm2]}      & \ch{[\kms]} &           \\
\ch{(1)}    & \ch{(2)}   & \ch{(3)} & \ch{(4)}         & \ch{(5)}        & \ch{(6)}    & \ch{(7)}          & \ch{(8)}        & \ch{(9)}     & \ch{(10)}         & \ch{(11)}   & \ch{(12)}  }
\startdata
3C309.1          &  0.12 &  13.3 & 3908  &         &   $<$39  &    $<$12.85 &           &          &             &           &      \\ 
3C351.0          &  1.28 &   8.4 &  569  & 3596\e2 & 185\e12  & 13.68\e0.03 & 33.1\e2.2 &  3597\e3 & 13.77\e0.05 & 28.9\e4.0 & b    \\ 
                 &       &       &       & 3456\e3 & 140\e15  & 13.49\e0.05 & 44.6\e2.5 &  3459\e5 & 13.53\e0.06 & 39.9\e8.0 &      \\ 
4C63.22          &  0.33 &   8.2 & 2333  & 2421\e2 & 228\e26  & 13.83\e0.07 & 35.7\e2.7 &  2420\e4 & 13.90\e0.08 & 32.6\e6.3 & a    \\ 
FBS1526+659      &  0.59 &  12.0 &  369  & 3462\e4 & 136\e19  & 13.48\e0.06 & 46.4\e2.3 &  3476\e7 & 13.48\e0.08 & 38.6\e9.7 &      \\ 
H1821+643        &  3.72 &  52.4 & 3249  & 2824\e4 &  44\e4   & 12.93\e0.04 & 65.0\e2.2 &  2825\e6 & 12.96\e0.04 & 62.5\e8.4 & a    \\ 
                 &       &       &       & 4087\e2 &  36\e3   & 12.86\e0.03 & 19.4\e2.2 &  4087\e1 & 12.91\e0.03 & 18.7\e2.6 &      \\ 
Kaz447           &  0.30 &  11.5 &  184  &         &   $<$48  &    $<$12.94 &           &          &             &           &      \\ 
Mrk290           &  2.58 &  30.4 &   97  & 3085\e2 & 505\e6   & 14.29\e0.01 & 56.9\e2.9 &  3089\e1 & 14.37\e0.01 & 52.9\e1.5 &      \\ 
                 &       &       &       & 3202\e2 & 318\e5   & 14.04\e0.01 & 37.8\e3.4 &  3204\e1 & 14.07\e0.02 & 32.1\e1.4 &      \\ 
Mrk486           &  0.18 &  10.0 & 1910  & 4387\e3 & 161\e17  & 13.63\e0.06 & 25.4\e1.4 &  4386\e3 & 13.74\e0.06 & 25.0\e4.7 & a    \\ 
Mrk817           & 10.83 &  47.1 & 6208  &         &   $<$9   &    $<$12.19 &           &          &             &           &      \\ 
Mrk876           &  4.00 &  48.8 &  540  & 3472\e2 & 280\e4   & 13.90\e0.01 & 53.2\e3.9 &  3476\e0 & 13.92\e0.02 & 24.7\e0.9 & c    \\ 
                 &       &       &       &         &          &             &           &  3470\e5 & 13.22\e0.07 & 80.9\e11.1 &      \\ 
PG1626+554       &  2.66 &  28.6 & 5684  &         &   $<$21  &    $<$12.52 &           &          &             &           &      \\ 
RBS1483          &  0.76 &  11.2 & 2260  & 2721\e2 & 259\e19  & 13.85\e0.04 & 44.6\e3.1 &  2726\e2 & 13.90\e0.03 & 35.0\e2.9 & a,d  \\ 
RBS1503          &  0.78 &  13.2 &  306  & 3272\e2 & 270\e16  & 13.88\e0.03 & 44.1\e2.2 &  3269\e3 & 13.81\e0.08 & 26.9\e5.3 & e    \\ 
                 &       &       &       & 3388\e2 &  56\e15  & 13.06\e0.10 & 39.9\e3.0 &  3306\e23 & 13.59\e0.12 & 116.8\e28.8 &      \\ 
RX\,J1500.5+5517  &  0.22 &  13.8 & 2031  & 3595\e2 & 120\e13  & 13.44\e0.05 & 28.0\e2.7 &  3592\e4 & 13.51\e0.06 & 26.8\e6.3 & f    \\ 
RX\,J1503.2+6810  &  0.75 &  12.7 & 1016  &         &   $<$51  &    $<$12.96 &           &          &             &           &      \\ 
RX\,J1508.8+6814  &  0.24 &   8.0 & 1202  &         &   $<$102 &    $<$13.27 &           &          &             &           &      \\ 
RX\,J1608.3+6018  &  0.81 &  20.2 &  880  & 2983\e4 & 388\e8   & 14.15\e0.02 & 43.0\e2.6 &  2983\e1 & 14.26\e0.03 & 37.0\e2.2 &      \\ 
                 &       &       &       & 2877\e2 & 117\e10  & 13.42\e0.04 & 37.7\e6.0 &  2886\e5 & 13.46\e0.06 & 37.4\e7.0 &      \\ 
RX\,J1717.5+6559  &  0.23 &   9.6 & 4330  & 4705\e2 & 234\e16  & 13.86\e0.04 & 29.6\e3.3 &  4705\e2 & 13.86\e0.03 & 32.3\e2.4 & a    \\ 
SBS1458+535      &  0.25 &  11.4 &  563  &         &   $<$57  &    $<$12.99 &           &          &             &           &      \\ 
SBS1503+570      &  0.31 &  13.5 & 2395  &         &   $<$51  &    $<$12.94 &           &          &             &           &      \\ 
SBS1521+598      &  0.18 &  10.0 & 2390  &         &   $<$66  &    $<$13.08 &           &          &             &           &      \\ 
SBS1537+577      &  0.14 &   8.8 &  293  & 3541\e2 & 436\e26  & 14.20\e0.05 & 49.4\e3.0 &  3541\e3 & 14.43\e0.18 & 37.6\e6.1 &      \\ 
                 &       &       &       & 3260\e2 & 360\e34  & 13.98\e0.05 & 70.9\e3.9 &  3257\e7 & 14.03\e0.05 & 73.2\e9.8 &      \\ 
SBS1551+572      &  0.11 &   7.3 & 1778  & 4097\e3 & 195\e21  & 13.74\e0.07 & 28.3\e1.6 &  4097\e5 & 13.93\e0.10 & 30.9\e7.6 &      \\ 
SBS1624+575      &  0.31 &  12.4 & 4103  &         &   $<$48  &    $<$12.94 &           &          &             &           &      \\ 
\enddata
\tablecomments{%
Col.\ 1: Target name;
Col.\ 2: Flux in units of \dex{-14} erg\,\cmm2\,s$^{-1}$\,m\AA$^{-1}$;
Col.\ 3: Signal to noise ratio at $cz$=3000~\kms;
Col.\ 4: $cz$ of \Lya\ detection as found from $N_a(v)$ integral;
Col.\ 5: Equivalent width of \Lya\ detection;
Col.\ 6; Column density of \Lya\ detection found as the integral of the $N_a(v)$ profile;
Col.\ 7: $b$-value of \Lya\ detection found from the second moment of the $N_a(v)$ profile;
Col.\ 8: $cz$ of \Lya\ detection as found from profile fitting;
Col.\ 9: Column density of \Lya\ detection as found from profile fitting;
Col.\ 10: $b$-value of \Lya\ detection as found from profile fitting;
Notes:
(a) Component not considered to be associated with the filament;
(b) The 3C351.0 spectrum was taken using the STIS-E140M echelle grating; 
(c) The very high S/N spectrum of Mrk\,876 is best fitted using both a narrow and a broad component, but with the AOD method only a single measurement can be made; 
(d) This line may have a second component at lower velocities, although a two component fit is not easily justified - yet the AOD 2nd moment is increased; 
(e) The properties of the second component are uncertain because it is weak and blended with the stronger component, see discussion in the text; 
(f) This line lies 0.6\,\AA\ above the Lyman limit of an absorber with log\,$N$(\HI)$\sim$16.5, i.e.\ at a wavelength where the individual Lyman lines are blended and weak. 
}
\end{deluxetable*}

\begin{figure*} \figurenum{\Fspec}
\begin{center}$\begin{array}{c} \includegraphics[width=6in, angle=0]{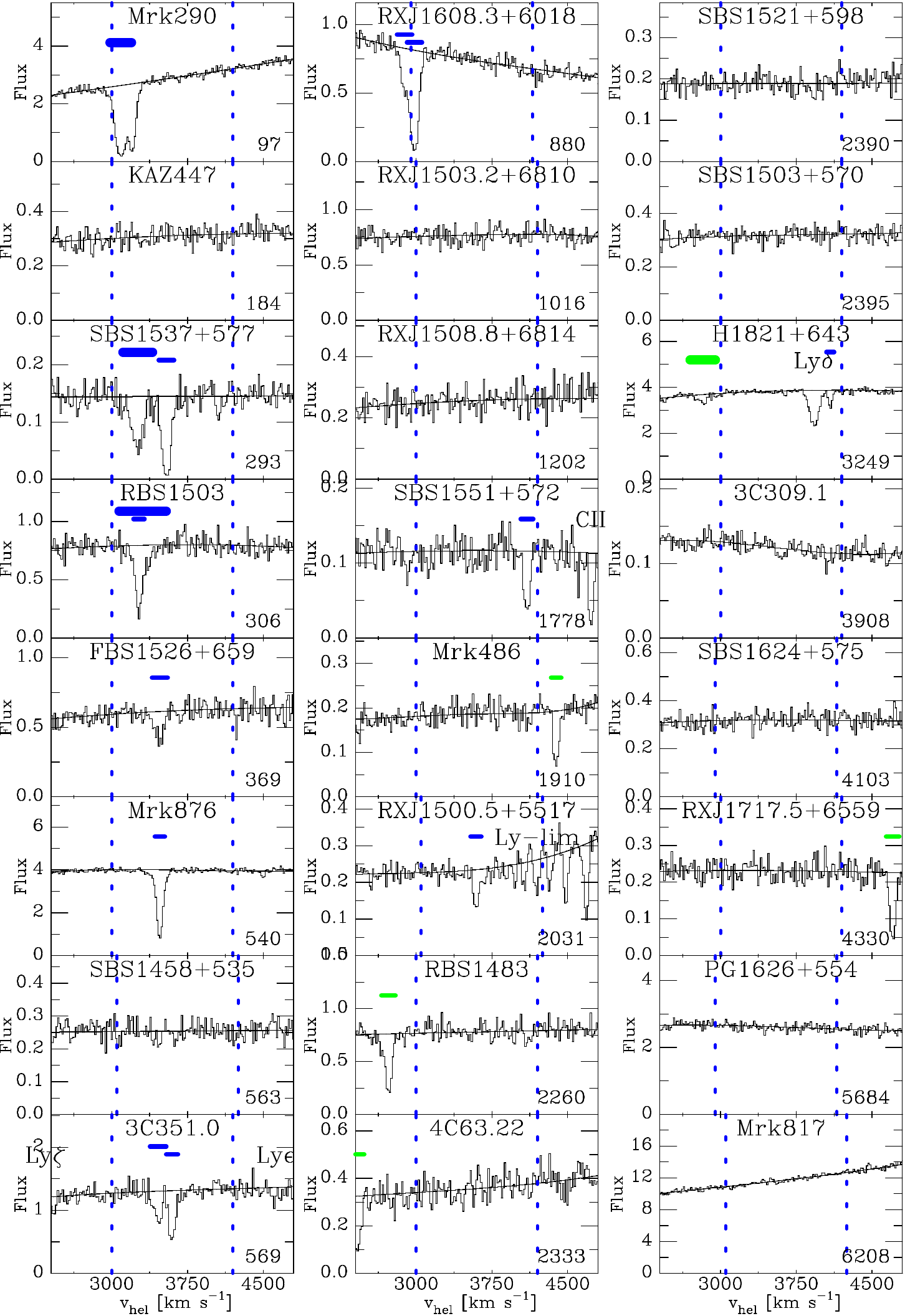} \end{array}$ \end{center} 
\caption{%
Relevant sections of the COS spectrum for each of the 24 targets analyzed. The
targets are sorted by filament impact parameter, given in units of kpc in the
bottom right corner. All absorption features are \Lya, except for the five
labeled \Lyd, \Lye, \Lyz, Ly-lim and \CIII. The blue dashed vertical lines shows
a 1200~\kms\ wide window around the velocity of the nearest filament axis
segment (see Sect.~\Sfilament). Only lines falling {\it inside} this window are
considered to be associated with the filament. These are indicated with blue
horizontal bars above the \Lya\ absorption. \Lya\ lines that are not considered
associated with the filament are indicated by green horizontal bars. The four
BLAs are indicated by the thicker blue bars.
}\end{figure*}

\begin{figure*} \figurenum{\Fbcomp}
\begin{center}$\begin{array}{c} \includegraphics[height=\textwidth, angle=270]{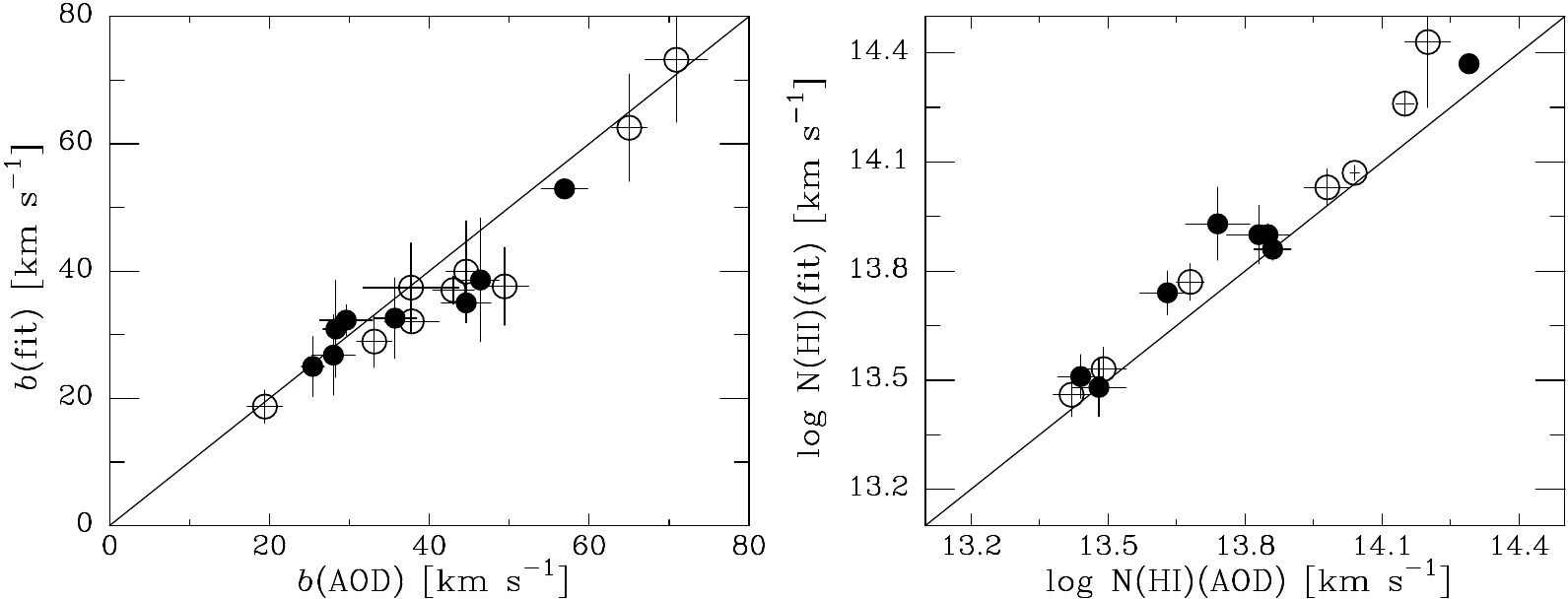} \end{array}$ \end{center} 
\caption{%
Comparison of linewidth (left) and column density (right) derived using the
apparent optical depth (AOD) method (horizontal axes) and Voigt profile fitting
(vertical axes). Closed symbols are for single-component \Lya\ absorption lines,
while open circles are for individual components in multi-component lines.
}\end{figure*}


\subsection{Spectral analysis} 
\par We now describe all the results we derive from the data and the comparison
with simulations.
\par We applied our wavelength-correction and error-calculation code to all the
spectra listed in Table~\Tobslist\ and then identified the absorption lines in
each, making sure to identify each feature between 1216 and 1240\,\AA, i.e., 0
to 6000~\kms\ relative to \Lya. Figure~\Fspec\ shows the relevant section of
each spectrum, including a continuum that was determined by fitted a low-order
(1st or 2nd) polynomial to line-free regions near the window shown. The targets
are sorted by filament impact parameter (given in the bottom right corner of
each panel). It is obvious that absorption is present close to the filament
(left column) and absent far from the filament (right column).
\par Since most of our targets have relatively low redshift ($z$$<$0.4), there
are few hard-to-identify lines and most features seen are clearly \Lya\ or metal
lines associated with a higher redshift system. In fact, there are only a
handful of non-\Lya\ lines that fall in the window of interest for this paper:
Ly$\epsilon$ at $z$=0.316 toward 3C\,351.0, Ly$\delta$ at $z$=0.297 toward
H\,1821+643, Ly$\mu$-Ly$\rho$ in a Lyman limit system at $z$=0.347 toward
RX\,J1500.5+5517, and CII $\lambda\lambda$903.6, 903.9 in a Lyman limit system
at $z$=0.367 toward SBS\,1551+572. These lines are labeled in Fig.~\Fspec. All
other visible features are identified as \Lya. 
\par We used two different methods to measure the column densities and
linewidths in the \Lya\ lines, using the same continuum fits. First, we
calculated column density using the apparent optical depth method (see Savage \&
Sembach 1991): $$
N = \int N_a(v) dv = \int {m_e c\over\pi e^2} f \lambda \ln {C(v)\over F(v)}, $$
with $F(v)$ the observed profile, $C(v)$ the fitted continuum, $f$ the
oscillator strength (0.4164 for \Lya) and $\lambda$ the rest wavelength
(1215.67\,\AA). We also calculated the second moment of the apparent optical
depth profile to derive a linewidth. We then deconvolved this linewidth for
instrumental broadening (assuming an instrumental resolution of 20~\kms\ FWHM).
Second, we used the VPFIT package (see Carswell et al.\ 2002, Kim et al.\ 2007)
to make a Voigt profile fit to the \Lya\ lines. VPFIT version 10.2 (http://
www.ast.cam.ac.uk/\~rfc/vpfit.html) was used. The theoretically calculated line
spread function for the COS G130M grating at Lifetime Position 1 (Kriss 2011)
was used to convolve with the model fit profile. This line-spread function
accounts for both scattering in the far wings due to microroughness in the
primary mirror and zonal polishing errors in the primary and secondary mirrors,
which results in a \COS\ LSF with reduced core intensity and non-Gaussian strong
wings.
\par Equivalent width detection limits for non-detections were found as three
times the error in the equivalent width for a line that is 50~\kms\ wide. This
was then converted to a column density detection limit by calculating the
$N_a(v)$ integral of a line with the equivalent width equal to the detection
limit and an FWHM of 50~\kms. On average this limit is about \dex{13}\,\cmm2.
\par Table~\Tmeasure\ presents the measured equivalent widths, column densities
and linewidths determined using both the apparent optical depth and profile
fitting methods. For the sake of completeness, the table includes \Lya\
detections for the extended velocity range shown in Fig.~\Fspec\
(\vminfiltab~\kms\ to \vmaxfiltab~\kms). In this figure the \Lya\ components
that we associate with the galaxy filament are shown by blue horizontal lines
extending from $v$(\Lya)-2$b$(\Lya) to $v$(\Lya)+2$b$(\Lya). Four components can
be considered Broad \Lya\ Absorbers (BLAs; $b$$>$40~\kms; see Richter et al.\
2004; Lehner et al.\ 2007). The four components in Table~\Tmeasure\ that are not
in the velocity range where filament galaxies are found are shown by green
horizontal bars.
\par Toward Mrk\,290 (filament impact parameter 97~kpc) two components are
clearly visible. Associated \OVI\ is also seen in the FUSE spectrum of this
target. These components were analyzed in detail by Narayanan et al.\ (2010),
who concluded that they originate from a cloud with $N$(H)=4\tdex{19}\,\cmm2\ at
$T$=1.4\tdex5~K which is also photoionized. Where Narayanan et al.\ (2010)
focused on a possible association with NGC\,5987 (impact parameter 475~kpc
(1.9~R$_{\rm vir}$ see Table~\Timpact), it is quite likely that in fact this
absorber provides evidence for the presence of WHIM gas in the filament.
\par The spectrum of Kaz\,447 (filament impact parameter 164~kpc) appears to
show a small dip near 4300~\kms. This feature is not significant, however,
measuring as 2$\sigma$.
\par For SBS\,1537+577 our S/N ratio is insufficient to be sure that the
apparent broad component can only be fitted with a single gaussian or whether it
is a combination of two narrower gaussians. We proceed assuming it is a single
broad component.
\par Toward RBS\,1503 two components are clearly present. The profile fitting
method gives $b$-values of 26.9$\pm$5.3~\kms\ and 116.8$\pm$28.8~\kms, but the
two components are heavily blended and the S/N ratio of the spectrum is not
sufficient to make a reliable fit. Therefore, in Table~\Tmeasure\ we list a very
different velocity for the second component in Col.~4 vs Col.~8. This is due to
the fact that when using the $N_a(v)$ method we can only measure the line wing,
which has a centroid velocity of 3388~\kms.
\par The BLA component toward Mrk\,876 overlaps with a narrower component at the
same velocity, but because the S/N ratio of the spectrum is very high
($\sim$50), it is clearly significant and not due to the wings on the \COS\ line
spread function.
\par It is clear from Fig.~\Fspec\ that the strongest and widest absorbers occur
toward sightlines passing closest to the filament axis (within 660~kpc), as is
the case for multi-component absorbers. The exceptions are a weak line toward
FBS\,1526+659 and non-detections toward Kaz\,447 and SBS\,1458+535. The former
maybe due to the fact that FBS\,1526+659 is at the far end of the upper filament
and the galaxy density near it is rather low. So even though it is nominally
close to the filament axis, it is not in a region with a lot of gas. In the
other two cases we must be looking through a hole, with the Kaz\,447
non-detection being especially interesting and revealing small-scale structure
because it is close to the sightline to 3C\,351.0 where two components are seen.
\par A map of the galaxies between 1500 and 2900~\kms\ shows that the
non-filament absorbers toward 4C\,63.22, RBS\,1483, and H\,1821+643 can be
associated with another structure, which is clearly visible, but not quite as
well defined as the filament shown in Fig.~\Fqsomap. Similarly, the component at
4705~\kms\ toward RX\,J1717.5+6559 is at the end of yet another filament that
extends from (R.A, Dec.)$\sim$(260\deg, 65\deg) to $\sim$(240\deg, 45\deg) with
$cz$ between 5000 and 6200~\kms. Only the absorber at 4387~\kms\ toward Mrk\,486
is an orphan that is not clearly associated with any filaments, although it is
not too far off from our main filament (off by 1.9~Mpc and 786~\kms) and a case
could be made that it just represents an outlier and should be included.
\par We note that all our detections have \HI\ column densities below about
log\,$N$(\HI)=14.5, whereas absorbers that are associated with galaxy halos
typically have higher column densities.
\par In Fig.~\Fbcomp\ we compare the column densities and linewidths derived
using both the apparent optical depth and profile fitting methods. This reveals
that the deconvolved apparent-optical-depth linewidths are typically wider than
those derived from profile fitting -- the ratio $b$(AOD)/$b$(fit) is
1.08$\pm$0.11. On the other hand, column densities measured using the apparent
optical depth method are consistently lower than those found from profile
fitting, with $\Delta$log$N$=0.07$\pm$0.06 dex. Both of these discrepancies can
be understood as a consequence of the non-gaussian shape of the COS linespread
function (LSF). Since the LSF has strong line wings, the apparent line width
becomes larger than what it would be if the LSF was a single gaussian, even
after deconvolution. Similarly, the LSF smearing increases the apparent flux in
the center of the line, lowering the apparent optical depth. As our comparison
shows, this ends up as an about 10\% increase in the apparent linewidth and an
about 0.1 dex decrease in the derived column density.


\begin{figure*} \figurenum{\FimpEWb}
\begin{center}$\begin{array}{c} \includegraphics[width=\textwidth, angle=0]{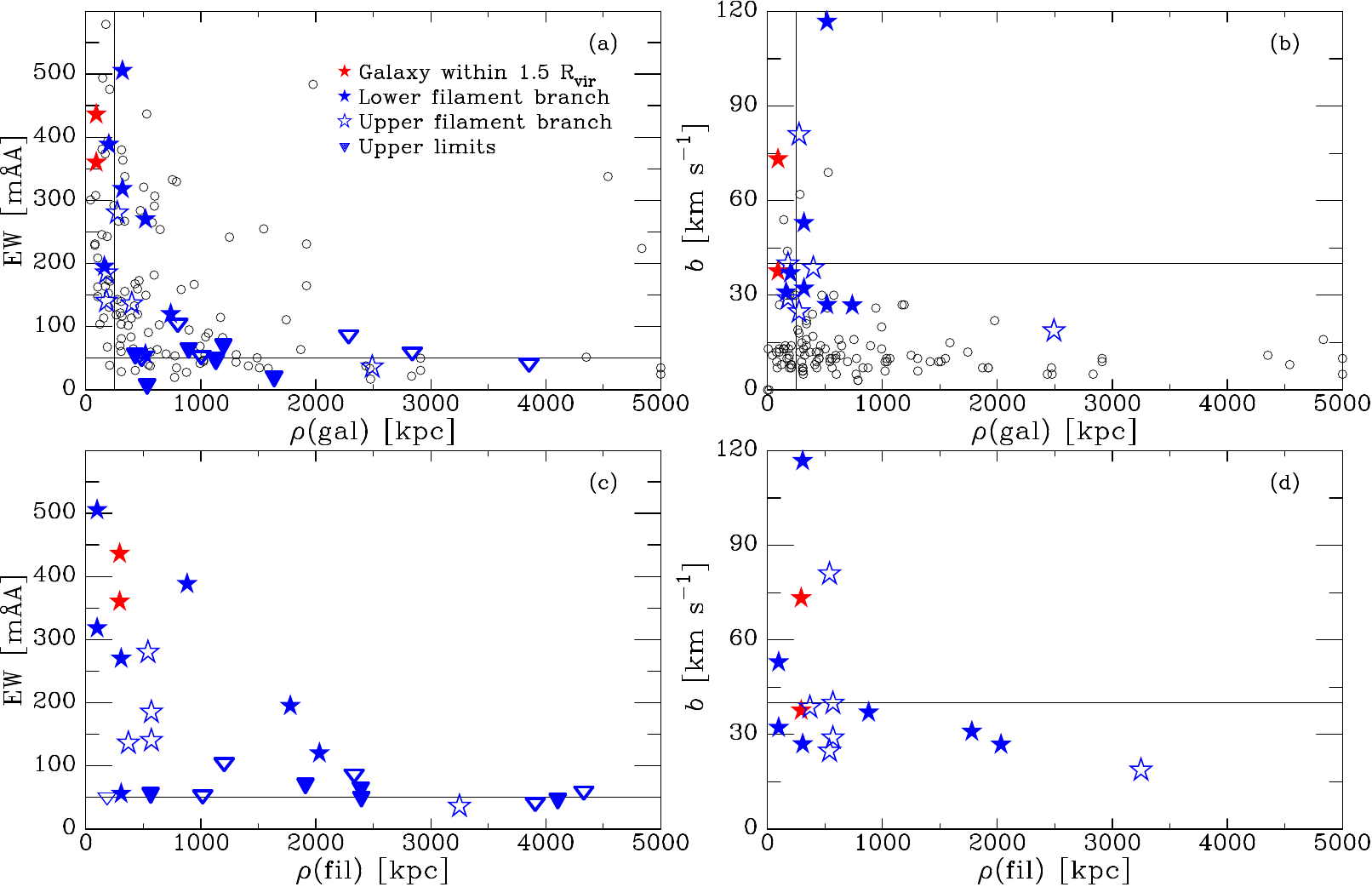} \end{array}$ \end{center} 
\caption{%
\Lya\ equivalent width (left) linewidth (right) vs filament impact parameter to
galaxies (top) and to the filament (bottom). Five sightlines have two
components, which are shown separately. For the top panels, the impact parameter
is that to the nearest galaxy of any size with velocity within $\pm$500~\kms\ of
a \Lya\ detection. Open black circles show the results from Wakker \& Savage
(2009). The red star is for the one sightline that passes within one virial
radius from a galaxy. The closed blue stars are for sightlines in the ``lower
branch'', running from (RA,Dec)=(220,47) to (247,64), while open blue stars are
for the ``upper branch'', running from (270,60) to (230,68). Blue downward
pointing triangles are upper limits for sightlines with no detected \Lya\ in the
velocity range \vminfilmap\ to \vmaxfilmap~\kms. The vertical lines in panels
(a) and (b) are at an impact parameter of 250~kpc. The horizontal line in the
left panels indicates the typical equivalent width detection limit of 50\,m\AA,
while the one in the right panels show the canonical $b$=40~\kms\ separation
between narrow and broad lines. Two non-detections from Table~3 are missing in
the bottom plot since they have filament impact parameter $>$5~Mpc. Note that
the detection at $\rho$(fil)=3300~kpc does not count for the statistic shown in
Fig.~\Fdetfrac\ as it has log\,$N$(\HI)$<$13. The trend of increasing equivalent
and line width with decreasing filament impact parameter is clear. The
exceptions (the points at 50~m\AA\ and 28~\kms\ at 400~kpc) come from the
multi-component feature toward RBS\,1503 that is difficult to fit. Note that for
all but four of the blue points the nearest galaxy is at least 500~kpc distant,
or more than 3\,$R_{\rm vir}$. For three the nearest galaxy is at 200-350~kpc
(1.5-2.5\,$R_{\rm vir}$) and the red point is at 91~kpc (1.0\,$R_{\rm vir}$).
}\end{figure*}

\subsection{Equivalent width and line width vs filament impact parameter} 
\par Using the galaxy and filament impact parameters given in Table~\Timpact, we
can now plot the distribution of \Lya\ lines relative to galaxies and relative
to the filament axis. This is done in Fig.~\FimpEWb. We use the equivalent width
instead of the column density of the \Lya\ absorption to facilitate a comparison
of the results for the 24 sightlines in this paper with the much larger sample
of 125 \Lya\ detections given in Wakker \& Savage (2009).
\par The two top panels (a,b) of Fig.~\FimpEWb\ correspond to the conventional
way of analyzing the relation between \Lya\ and galaxies. For non-detections we
find the galaxy impact parameter as that to the nearest galaxy with velocity
between \vminfilmap\ and \vmaxfilmap~\kms, i.e.\ galaxies shown in
Fig.~\Fqsomap. For \Lya\ absorbers the relevant impact parameter is that to the
nearest galaxy whose velocity is within $\pm$400~\kms\ of that of the absorber.
As Table~\Timpact\ shows, there are only a few cases where the nearest galaxy is
very small ($L$$<$0.1~\Lstar; $D$$<$7.2~kpc) (Mrk\,486, SBS\,1503+570,
SBS\,1521+598, SBS\,1624+575), but in all of those cases the impact parameter to
the nearest galaxy with $L$$>$0.1\,\Lstar\ is not very different, so we avoid
the possibly confounding problem of whether to choose the small impact parameter
to a dwarf galaxy or the much larger value to a substantial galaxy. 
\par In this figure, we use a red symbol if the sightline passes within 1.5
virial radii of a galaxy (which happens in just one case, for SBS\,1537+577),
and a blue symbol otherwise. There are four sightlines (3C\,351.0, Mrk\,876,
RX\,J1608.3+6018 and SBS\,1551+572) where the nearest galaxy is between 150 and
250~kpc, but the impact parameter also is between 1.7 and 2.0 virial radii, so
it is not likely that the absorption is associated with the galaxy halo.
\par Panels (a) and (b) of Fig.~\FimpEWb\ show that the sample of sightlines in
this paper has similar properties as the larger sample in Wakker \& Savage
(2009). I.e., (1) the average equivalent increases for smaller impact
parameters; (2) every sightline with a galaxy impact parameter below 250~kpc
shows a \Lya\ line; (3) for some high equivalent width absorbers the nearest
galaxy is separated by more than 250~kpc; (4) non-detections become more common
at larger impact parameters; (5) the typical linewidth is larger at smaller
impact parameters. Detections at large impact parameters are typically noted,
but their properties are mostly ignored because they can't easily be interpreted
in terms of galaxy halos. Such absorbers only play a role when interpreting the
column density distribution of \Lya\ lines in order to estimate the baryon
content of the \Lya\ forest.
\par Panels (c) and (d) give a different way of looking at these \Lya\ absorbers
and non-detections. Here the red symbols indicate the components toward
SBS\,1537+577 which originate at 1.0 times the virial radius from a galaxy and
thus conceivable could be associated with that galaxy. The closed blue stars
show the components in the lower filament branch seen in Fig.~\Fqsomap, while
the open blue stars are for the upper branch. It is clear that

\begin{figure*} \figurenum{\Fcrossviewp}
\begin{center}$\begin{array}{c} \includegraphics[width=\textwidth, angle=0]{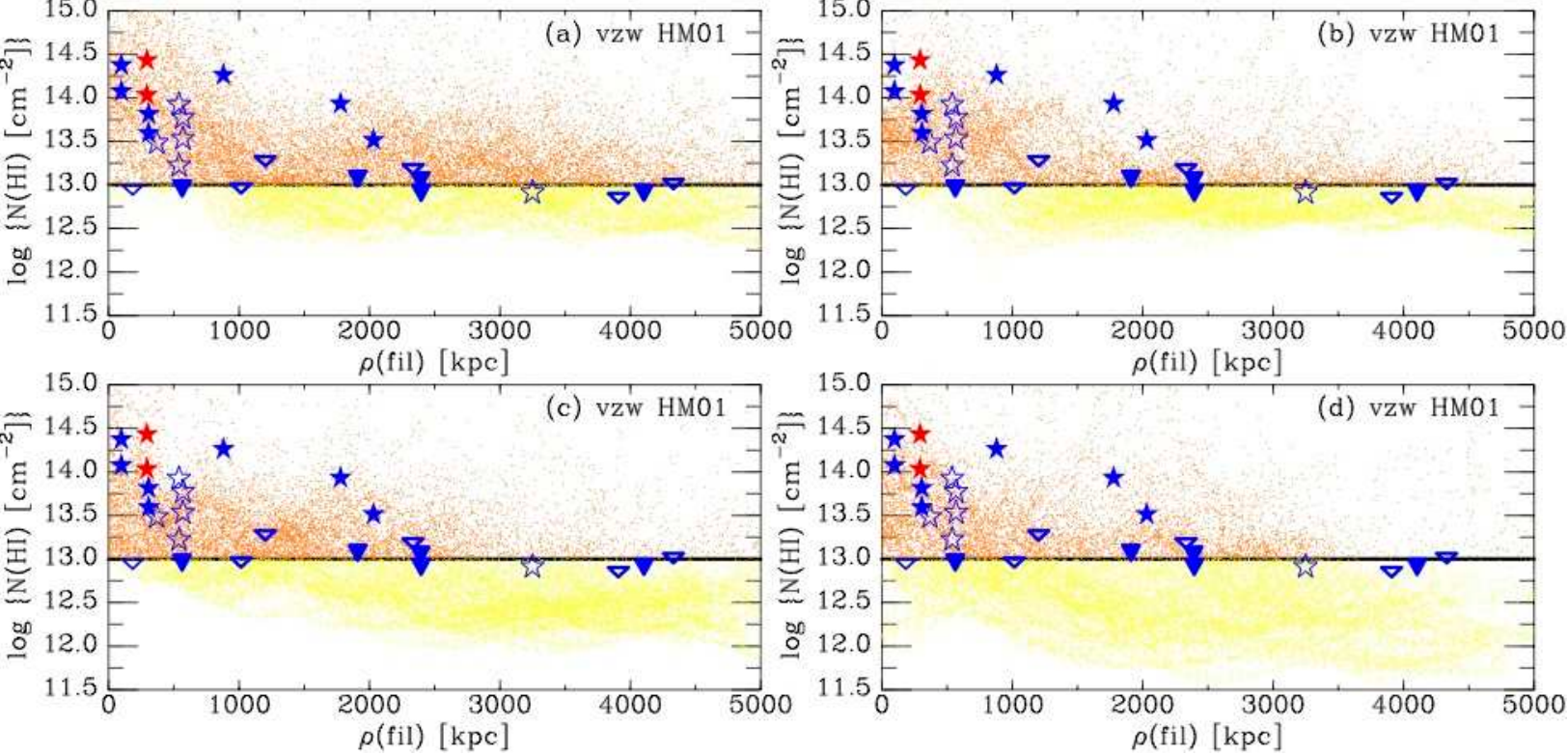} \end{array}$ \end{center} 
\caption{%
Scatter diagram of \HI\ column density vs filament impact parameter for the four
different viewpoints in the vzw model shown in Fig.~\Fsimullocs. The nominal
Haardt \& Madau (2001) EGB prescription was used. Only points inside the
rectangular outline box in that figure are shown. Column densities above
\dex{13}\,\cmm2\ are shown by orange points, lower column densities by yellow
points. The downward pointing triangles give observed upper limits as function
of filament impact parameter, while stars show the detections, with the red star
for the sightline within 150~kpc of a galaxy. The black lines give the 10th,
25th, 50th, 75th and 90th percentile of column densities in intervals of
200~kpc.
}\end{figure*}

\begin{figure*} \figurenum{\Fcrossegb}
\begin{center}$\begin{array}{c} \includegraphics[width=\textwidth, angle=0]{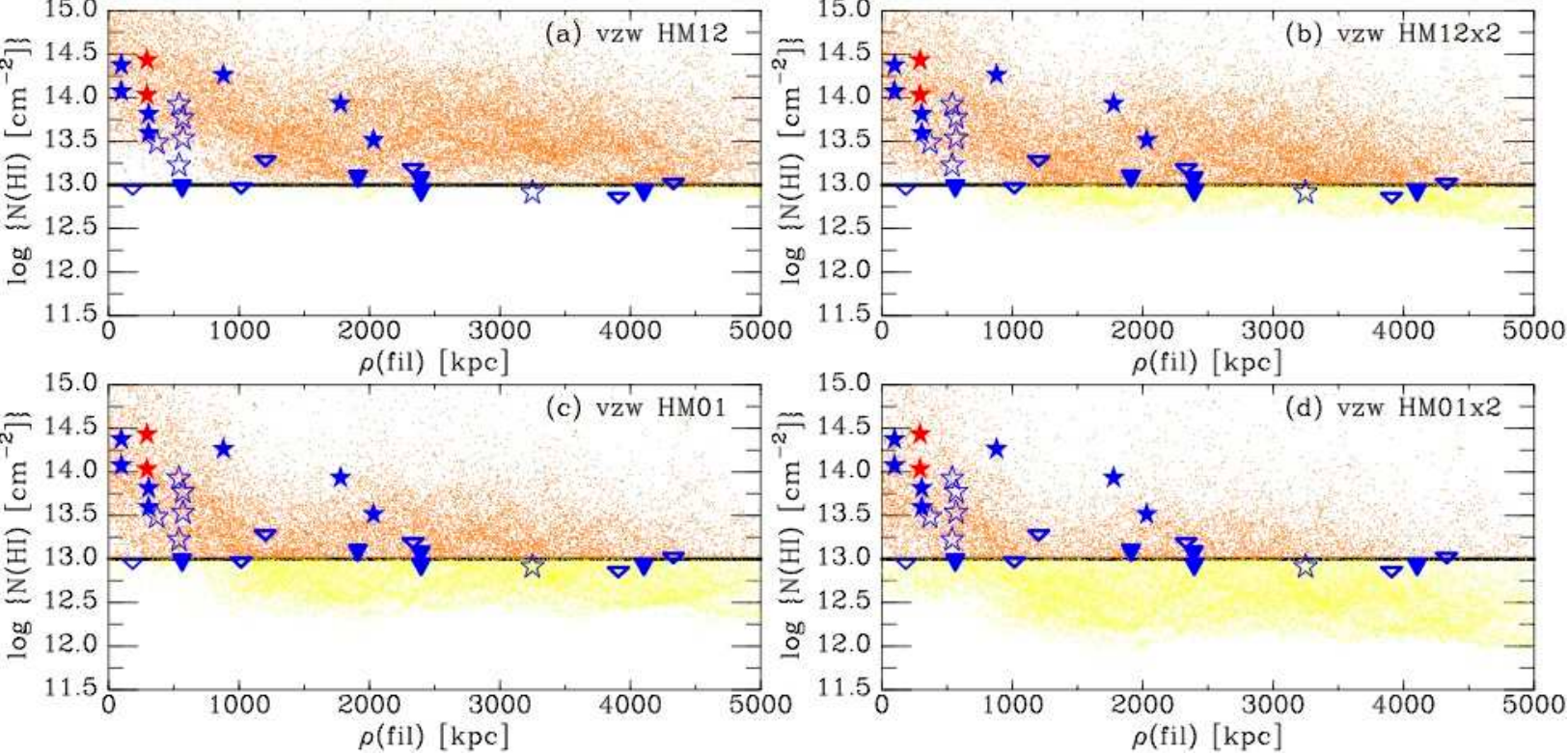} \end{array}$ \end{center} 
\caption{%
Same as Fig.~\Fcrossviewp, but comparing the vzw model from one viewpoint but
with five different versions of the intensity of the EGB, as labeled.
}\end{figure*}

\par\noindent (1) The high equivalent width absorbers occur close to the
filament axis, {\it whether or not there is a galaxy within one virial radius of
the sightline}.
\par\noindent (2) There are {\it no} detections with equivalent width
$>$50~m\AA\ further than 2~Mpc from the filament axis, unlike what is the case
in panel (a) where for some absorbers with equivalent width $>$100~m\AA\ there
is no galaxy within 2~Mpc.
\par\noindent (3) There is a strong correlation between equivalent width and
filament impact parameter. This is especially obvious for the lower branch
(closed blue stars). Only the broad component toward RBS\,1503
($\rho$(fil)=373~kpc, EW=56\,m\AA) does not follow this pattern, but its
equivalent width cannot be measured properly and the linewidth is very
uncertain.
\par (4) The increase in linewidth with decreasing filament impact parameter is
more systematic than the increase in linewidth with decreasing galaxy impact
parameter.
\par (5) The four \Lya\ absorbers that fit the definition of a BLA
($b$$>$40~\kms) all occur in sightlines with two components and they only occur
at the smallest filament impact parameters. This indicates that there the
structure of the gas gets more complicated and/or that the gas gets hotter. With
just four cases out of twelve sightlines we will need more sightlines and more
filaments to make this a statistically sound conclusion, however.


\begin{figure*} \figurenum{\Fdetfrac}
\begin{center}$\begin{array}{c} \includegraphics[width=\textwidth, angle=0]{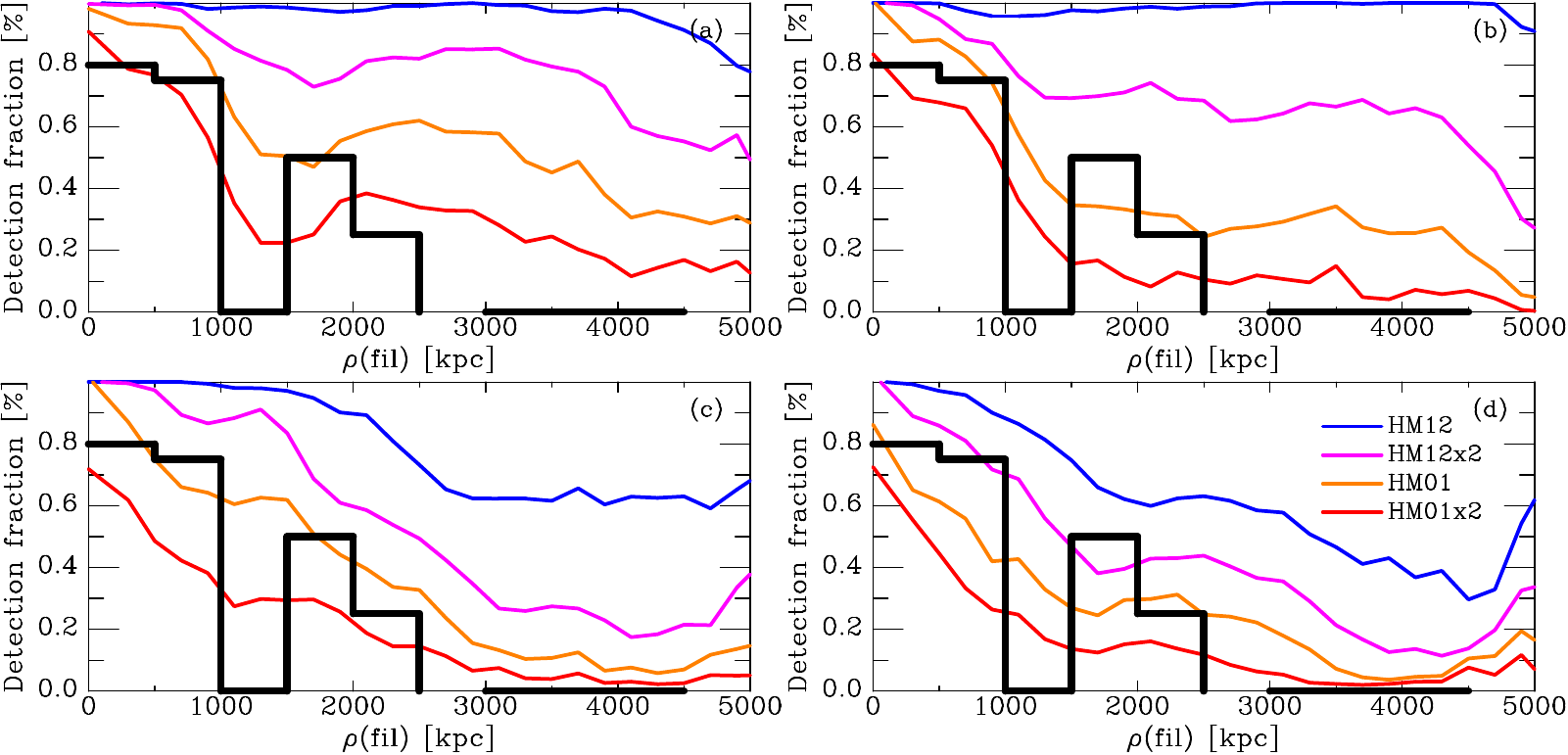} \end{array}$ \end{center} 
\caption{%
Detection fraction as function of filament impact parameter for different four
viewpoints inside the simulation cube. Differently colored lines indicate four
different versions of the EGB, as shown by the label in the bottom left panel.
The detection fraction is the fraction of sightlines in a 200~kpc (for
simulations) or 500~kpc (for observations) interval where
$N$(\HI)$>$\dex{13}\,\cmm2. The observed detection fractions (black line) are
80\% (4/5) for $\rho$=0--0.5~Mpc, 75\% (3/4) for $\rho$=0.5--1~Mpc, 0\% (0/2)
for $\rho$=1--1.5~Mpc, 50\% (1/2) for $\rho$=1.5--2.0~Mpc, and 25\% (1/4) for
$\rho$=2--2.5~Mpc. No detections are found for seven sightlines with
$\rho$=2.1--5.0~Mpc, as indicated by the thick line at zero.
}\end{figure*}


\subsection{$N$(\HI) vs filament axis: observations vs simulations} 
\par To compare observed \HI\ column densities against the simulations, we first
show a scatter plot of $N$(\HI) against filament impact parameter, for four
viewpoints for the vzw model (Fig.~\Fcrossviewp) using the HM01 version of the
EGB. The four viewpoints are sorted by the density of galaxies along the
filament axis (3.8, 2.0, 1.9 and 1.7 galaxies per Mpc for panels a, b, c and d,
respectively). Compare to 2.6 galaxies per Mpc for the observed filament. In
Fig.~\Fcrossegb\ we show the effect of varying the EGB for one viewpoint for the
vzw model. Observed values are included as colored stars, using the same coding
as in Fig.~\FimpEWb.
\par The orange points in Fig.~\Fcrossviewp\ represent \HI\ column densities
that lie above the average detection limit of the observations (\dex{13}\,\cmm2,
see Sect.~\Sspectra), for directions that lie inside the rectangular box outline
in Fig.~\Fsimullocs. The black lines give the 10th, 25th, 50th, 75th and 90th
percentile of column density. The comparison between the different viewpoints
shows that there are differences in detail between the four simulated filaments,
but in general the column density is always above the detection limit at low
filament impact parameters ($<$$\sim$500~kpc, while it drops at higher impact
parameters. For three of the viewpoints almost 100\% of the sightlines have
$N$(\HI)$>$\dex{13}\,\cmm2\ at zero impact parameter. Across the filament the
10th to 90th percentile of column densities have a spread of about a factor of
100. For three viewpoints column densities $>$\dex{13}\,\cmm2\ correspond to the
50\% or 75\% line at several Mpc from the axis, while for one viewpoint the 75\%
percentile is crossed at 2~Mpc. The observed sightlines appear to provide a fair
sample of the expected column densities.
\par We note here that we adopted the momemtum-driven wind (`vzw'') version of
the simulation of Oppenheimer \& Dav\'e (2008) as a reference, rather than the
no-wind, constant-wind or energy-driven wind version. However, although in these
simulations the galactic winds heat and disturb the IGM close to galaxies, at
the Mpc scales of the galaxy filament they have little influence. We did make
the equivalent version of Fig.~\Fcrossviewp\ using different wind models, but in
all cases the distribution of $N$(\HI) vs filament impact parameter is
essentially the same, varying only by at most 10\% in any individual sightline.
\par In Fig.~\Fcrossegb\ we analyze the impact of the assumed intensity of the
extragalactic background radiation showing the scatter plot of \HI\ column
density vs filament impact parameter from the viewpoint shown in
Fig.~\Fsimullocs a and the top left in Fig.~\Fcrossviewp\ (i.e.\ the filament
with the best match between observed and simulated galaxy density). We use four
different versions of the Haardt \& Madau model: HM01 and HM12, and both scaled
up by a factor two. As expected, the assumed ionizing radiation field has a huge
impact on the predicted column densities, with the HM12 version predicting
almost no column densities below \dex{13}\,\cmm2\ and the HM01x2 version
predicting thin filaments at the \dex{13}\,\cmm2\ level.



\subsection{Filament gas vs galaxy mass } 
\par Combining our measurements of \HI\ column density with the ionization
corrections implied by the simulation, we can estimate the total mass of baryons
in the filament. Fitting the relation between ionization correction and \HI\
column density that is shown in the right panel of Fig.~\Fionization\ yields
log\,$N$(H) = 0.36\,log\,$N$(\HI) + 14.37, with a typical variation on the order
of $\pm$0.3~dex. Note that this relation is implied when using the HM01 version
of the EGB. The coefficient will change in proportion to the strength of the
ionization radiation field. Using this conversion, the observed \HI\ column
densities listed in Table~\Tmeasure\ imply total hydrogen column densities
ranging from log\,$N$(H)=19.2 to 19.6, with an average of 19.35. We note that
for the absorber toward Mrk\,290 Narayanan et al.\ (2010) used the combination
of \OVI\ and \Lya\ and a hybrid ionization model to derive log\,$N$(H)=19.14.
For this cloud the total hydrogen column density implied by the simulation is
log\,$N$(H)=19.44, which is within the $\pm$0.3 dex range implied by the
log\,$n$(H) vs log\,$n$(\HI) correlation. If the ionizing background were twice
as strong, the two values would match.
\par In Sect.~\SfilAGN\ we estimated the total area of the filament as
170~Mpc$^2$. With the average log\,$N$(H)=19.35 this then implies a total
hydrogen mass in the gaseous filament of 3\tdex{13}\,\Msun, or
5.2\tdex{13}\,\Msun\ in baryons.
\par We can compare this to the mass of the galaxies in the filament. This can
be derived by using the simulated galaxies, from which we find that log\,(mass
in \Msun) = log\,(luminosity in \Lstar) + 11.1. Combining this with the galaxy
luminosities derived in Sect.~\Sgalsample\ and adding the implied masses for all
galaxies inside the stripe boxes shown in Fig.~\Fqsomap\ gives a total baryonic
galaxy mass of 1.4\tdex{13}\,\Msun.
\par Thus, although we use the simulations to derive the conversions between
luminosity and mass and between neutral and total hydrogen, scaling the {\it
observed} galaxy luminosities and \HI\ column densities implies that the 21
absorbers in our sample represent four times more baryons than all the baryons
inside all the galaxies that were used to define the filament. We previously
found that half of the \Lya\ absorbers are within 400~kpc of a galaxy (see
Table~9 in Wakker \& Savage 2009), implying that the baryonic mass of the
circumgalactic medium is comparable to that of the intergalactic gas, and that
both are several times more massive than the condensed baryons inside galaxies.


\subsection{Detection fraction} 
\par With the limited number of sightlines we have available at the moment, we
cannot make the full distribution of column densities in different impact
parameter intervals. But we {\bf can} compare the observed and predicted
detection fraction as function of filament impact parameter. This is shown in
Fig.~\Fdetfrac. To make this figure we calculated the fraction of sightlines for
which the vzw simulation predicts an \HI\ column density above \dex{13}\,\cmm2\
in intervals of 200~kpc, using the Haardt \& Madau (2001, 2012) EGB model. We
also calculated the fraction of sightlines for which we detected \HI, in
intervals of 500~kpc. The \dex{13}\,\cmm2\ boundary value is the average
detection limit for our data (see Sect.~\Sspectra). In all but two sightlines we
are sensitive to \Lya\ with log\,$N$(\HI) above this limit. There are only six
sightlines where the S/N ratio is $>$20 and the detection limit is lower than
log\,$N$(\HI)=12.9, (H\,1821+643, Mrk\,290, Mrk\,817, Mrk\,876, PG\,1626+554 and
RX\,J1608.3+6018). In this set there is just one \HI\ line with log\,$N$(\HI)
below the limit (12.9 toward H\,1821+643, see Table~\Tmeasure). For the
statistics discussed here we discount this line as it has log\,$N$(\HI) below
the 13.0 limit.
\par The observed detection rate (black histogram in Fig.~\Fdetfrac) is based on
relatively few sightlines, and therefore the detection fraction as function of
filament impact parameter remains fairly uncertain. See the figure caption for
the percentages and numbers of sightlines. The three robust conclusions that can
be drawn are that:
\par\noindent (1) At filament impact parameters below 1~Mpc the detection rate
is high: seven out of nine sightlines show \Lya.
\par\noindent (2) The detection rate regularly decreases with filament impact
parameter.
\par\noindent (3) No \Lya\ lines are found in the seven sightlines with impact
parameter $>$2.1~Mpc.
\par Comparing the observed detection fraction to the predicted fraction shows
that the HM12 model (blue line) grossly overpredicts the number of \Lya\
absorbers, i.e.\ the IGM is underionized in this model. Even scaling it up by a
factor two (magenta line), as proposed by Shull et al.\ (2015) does not help.
The orange line (corresponding to HM01, which has a factor 3.7 more ionizing
photons) corresponds much more closely to the observed detection rate at small
filament impact parameters ($<$2~Mpc). However, even then the detection fraction
along the axis is predicted to higher than the observed 80\%. Further, some
\Lya\ lines are expected at filament impact parameters $>$2~Mpc. For the best
matched filaments (a) and (b), the expected probability given the orange and
red lines in Fig.~\Fdetfrac\ ranges from $\sim$20\% to $\sim$40\%. Given this
expectation, the probability to find zero detections in a sample of seven
ranges from (1$-$0.4)$^7$ to (1$-$0.2)$^7$ or $\sim$3\%--20\%. Thus, the fact
that we do not find any \Lya\ lines at large impact parameter has a low
probability.
\par In summary: the HM12 version of the EGB has too few ionizing photons, while
the shape of the run of detection fraction with filament impact parameter may
not match the observations.


\section{Discussion} 
\par From our results it is clear that understanding the majority of \Lya\
absorbers is best done in terms of large-scale structure, rather than in terms
of galaxy halos. Although strong ($N$(\HI)$>$\dex{15}\,\cmm2) absorbers {\it
are} associated with the circumgalactic medium of galaxies, the weaker \Lya\
lines are more likely to be associated with filaments. Studies of the
galaxy-galaxy vs galaxy-absorber correlation function (e.g.\ Tejos et al.\ 2014
and references therein) also indicate that the two differ. As can be seen from
Table~\Timpact, only one of our sightlines passes within one virial radius of a
galaxy (SBS\,1537+577 with impact parameter 91~kpc to SDSS\,J153802.75+573018.3,
which has $R_{\rm vir}$=87~kpc according to our approximation to the relation of
Stocke et al.\ (2013). In general, we previously found that 50\% of \Lya\
absorbers occur more than 300~kpc away from the nearest galaxy. As
Fig.~\FimpEWb\ shows, however, the properties of \Lya\ absorbers far from
galaxies are not random, but they correlate well with the impact parameter to
the galaxy filament that we studied. Not only are the largest equivalent widths
seen closest to the filament axis, there is also a hint that the linewidth
decreases with filament impact parameter. However, our sample is not large
enough to come to a definitive conclusion on this.
\par With our relatively small sample, the cleanest way to compare observational
results to simulations is in terms of the detection fraction. As Fig.~\Fdetfrac\
shows, the fraction of sightlines that show \Lya\ absorption is maximal near the
filament axis and decreases rapidly away from the axis, while no detections are
found more than 2.1~Mpc away from the filament axis. Further, the expected
decrease in detection fraction depends strongly on the intensity of the ionizing
flux in the extragalactic background radiation field (EGB). Using the most
sophisticated model that fits the high redshift \Lya\ forest (Haardt \& Madau
2012) yields an ionizing flux that is far too low at $z$=0, by a factor four to
five. This conclusion was previously reached by Kollmeier et al.\ (2014), who
compared the observed column density distribution (from Danforth et al.\ 2014)
with the predictions. Where Shull et al.\ (2015) find a factor two more intense
ionizing flux by using a different hydrodynamical simulation and suggesting a
higher escape fraction from galaxies for Lyman continuum photons, Khaire \&
Srianand (2015) suggest that the contribution from QSOs is twice as high as
Haardt \& Madau (2012) had found and suggest a 4\% Lyman continuum escape
fraction to imply an ionizing flux at $z$=0 that is five times higher than HM12.
\par Our results show that the problem is even more severe, in that the models
predict the wrong shape for the run of detection fraction with filament axis.
Matching the detection fraction at low impact parameters means that there is too
much \HI\ at filament impact parameters $>$2~Mpc for simulated filaments (a) and
(b), which have galaxy density similar to that in the observed filament.
Matching to the non-detections at large impact parameters means that too few
absorbers are predicted at low impact parameters (see e.g.\ filament (d) in
Fig.~\Fdetfrac. This could indicate that ionizing radiation field is even
stronger than expected far from filament axis, or that the total volume density
of hydrogen is lower than the simulations predict.
\par To follow up on this work, we hope to collect other simulations, using
different prescriptions for galaxy formation and possibly different predictions
for the distribution of the gas around filaments. We are also working on
increasing our observational sample by analyzing the nine other nearby galaxy
filaments that we have located. None of those filaments is sampled by more than
10 sightlines, but cumulatively we will be able to more than double our sample,
using COS archival data. It would also be useful to observe of one these
filaments with a much denser pattern of sightlines.


\section{Summary} 
\par We present a study of a local filament in the \Lya\ forest in three
dimensions. We obtained spectra of 24 AGN obtained with the Hubble Space
Telescope, and we describe a new method to properly align multiple exposures
taken using the COS instrument. We measure the properties of 21 \Lya\ absorbers
with $cz$ between \vminfiltab\ and \vmaxfiltab~\kms, seen in 17 of the
sightlines. We associate 15 absorbers with a filament of galaxies that is
apparent in the distribution on the sky for galaxies with $cz$=\vminfilmap\ to
\vmaxfilmap~\kms. We present a method to objectively derive an axis for the
filament and analyze the \Lya\ absorbers with reference to this axis. We also
search for similar galaxy filaments in an SPH simulation and determine filament
axes in the same manner as was done for the real sky. We then compare the
observed and predicted \HI\ column density as function of the sightline and
transverse dimensions, for several different prescriptions for the strength of
the intergalactic ionizing radiation field. We derive the following conclusions.

\subsection{Data Handling} 
\par (1) We describe a method to correct the COS wavelength scale that aligns
individual exposures without assuming a constant offset, by crosscorrelating all
ISM and IGM lines in each exposure. Further, in contrast to other methods that
have been described, we properly determine an absolute wavelength scale for the
combined spectrum by aligning the ISM lines with a 21-cm spectrum. This prevents
smearing of absorption lines in misaligned spectra and allows us to properly
assess the alignment of lines in intergalactic absorption-line systems.
\par (2) An analysis of the error array produced by CALCOS shows that those
errors tend to be higher than the errors measured from the rms around a fit to
the continuum in the spectra. For target fluxes below $\sim$\dex{14}\,\fu\ this
results in an overestimate of the errors -- on average by 50\% for the faintest
targets. Thus, using CALCOS errors can lead to large discrepancies when
determining the significance of weak lines and of detection limits.
\par (3) Using the apparent optical depth method to measure column densities and
linewidths in COS spectra leads to values that are 10\% lower and 10\% higher,
respectively, when compared to measuring these quantities using profile fitting;
the latter more properly takes into account the complex COS line-spread
function.

\subsection{Properties of \Lya\ lines} 
\par (4) Our set of sightlines does not generally sample galaxy halos. All but
one of the \Lya\ lines originate far from galaxies -- 13 out of 15 absorbers (in
10 sightlines) in our sample do not pass within 1.5 virial radii of any galaxy
with luminosity greater than our completeness limit of 0.05\,\Lstar\ (about the
luminosity of the SMC). We use 1.5 virial radii as the criterion since this is
about the radius at which theoretical considerations (Oort 1970, Maller \&
Bullock 2004) place the boundary between infalling and intergalactic gas. A fair
number of fainter galaxies are also included in our sample, but none is close to
a sightline. Using a simple impact parameter limit, three more sightlines with a
\Lya\ absorber pass between 200 and 300~kpc from a galaxy, which is between 1.7
and 2 virial radii. In a very lenient approach, at best these might be
classified as sampling the outermost halo of those galaxies.
\par (5) We use the distribution of galaxies to define a filament at
$cz$$\sim$3500~\kms. The filament has two branches with sizes
18\deg$\times$7\deg\ ($\sim$15.5$\times$6~Mpc) and 15\deg$\times$7\deg\
($\sim$13$\times$6~Mpc), covering an area of 170~Mpc$^2$.
\par (6) All \Lya\ lines with $N$(\HI)$>$\dex{13}\,\cmm2\ originate within
2.1~Mpc of the filament axis. This includes detections that would be classified
as ``void'' detections in an approach based solely on comparing their location
to that of galaxies.
\par (7) There is a strong correlation between the equivalent width of the \Lya\
lines and filament impact parameter.
\par (8) Using the simulations as a guide, we apply an ionization correction to
each detected \Lya\ line and find an average total hydrogen column density of
log\,$N$(H)=19.35. Combining with the area of the filament this implies a
total baryonic mass of 5.2\tdex{13}\,\Msun. We also derive the combined mass of
the galaxies inside the filament, which is about 1.4\tdex{13}\,\Msun.
\par (9) There is a strong suggestion that the \Lya\ linewidth correlates with
filament impact parameter.
\par (10) The four broad \Lya\ components occur only in sightlines that pass
close to (within 400~kpc) the axis of the filament. Further, multi-component
absorbers preferentially occur within about 1~Mpc of the axis. Although the
number of sightlines in our sample is small, this strongly suggest an increase
in temperature and/or turbulence near the filament axis.
\par (11) We note that Narayanan et al.\ (2010) previously found that the BLA
toward Mrk\,290 containing O\,VI provides evidence for warm highly ionized gas
in the filament with $N$(H)/$N$(\HI)$\sim$1.8\tdex{5} and a very large total
column density of 1.4\tdex{19}\,\cmm2. This is consistent with the column
density expected from the simulations.

\subsection{Comparison with simulations} 
\par (12) Using the locations of galaxies in simulations we can find and define
galaxy filaments that resemble the ones seen on the real sky. These filaments
are then also seen in the \HI\ column density distribution.
\par (13) The predicted distribution of \HI\ column density as function of
filament impact parameter is strongly dependent on the assumed intensity of the
extragalactic background radiation field (EGB). Of the four versions of the
Haardt \& Madau (2001, 2012) EGB that we tested, we find that the best fit to
the data occurs when using their standard 2001 version, in which the
contribution to the ionizing flux at $z$=0 is about the same for quasars and
galaxies. This implies that the HM12 model underestimates the ionizing flux by a
factor four to five, in agreement with Kollmeier et al.\ (2014). Which is about
the factor proposed by Khaire \& Srianand (2015), but larger than the factor two
proposed by Shull et al.\ (2015) to match the column density distribution of
\Lya\ absorbers.
\par (14) Using the best matching EGB, the fraction of sightlines toward which
we see \Lya\ with $N$(\HI)$>$\dex{13}\,\cmm2\ matches the simulations only at
small filament impact parameter. For large filament impact parameters
($>$2.1~Mpc) the simulations predict a detection rate of about 20--40\% to find
\Lya\ absorbers with $N$(\HI)$>$\dex{13}\,\cmm2. The fact that we see 0
absorbers in 7 sightlines may be significant. Given the expected detection rate,
probability of finding 0 detections in 7 sightlines is 3--20\%. This suggests
that there may be a problem with the width of the filaments predicted by the
simulations and/or the ionization background is stronger than HM01, though more
data is needed.
\par In summary, we have shown that the properties of the majority of \Lya\
absorbers can be understood more easily with reference to the large scale
structure of the Cosmic Web rather than to individual galaxy halos. By analyzing
the three-dimensional distribution of \Lya\ forest absorbers it is possible to
constrain simulations, as well as the extragalactic radiation field. More
filaments need to be mapped to make our conclusions more statistically
robust.


\bigskip Acknowledgements
\par Support for Wakker and Hernandez was provided by NASA through grants
HST-GO-12276 and HST-GO-13444 from the Space Telescope Science Institute, which
is operated by the Association of Universities for Research in Astronomy,
Incorporated, under NASA contract NAS5-26555. Wakker and French were supported
by grant AST-1108913 from the National Science Foundation. Kim acknowledges
funding support from the European Research Council Starting Grant "Cosmology
with the IGM" through grant GA-257670. Oppenheimer was support by Space
Telescope Science Institute grant HST-AR-13262.

\bigskip
\def\ref{\par\noindent}
\noindent {\bf References}
\ref Bell E.F., McIntosh D.H., Katz N., Weinberg M.D., 2003, ApJS, 149, 289
\ref Bruzual G., Charlot S., 2003, MNRAS, 344, 1000
\ref Carswell, R., Schaye, J., \& Kim, T-S., 2002, ApJ, 578, 43 
\ref Cautun M., van de Weygaert R., Jones B.J.T., 2013, MNRAS, 429, 1286
\ref Cen R., Ostriker J., 1999, ApJ, 514, 1
\ref Cen R., 2013, ApJ, 770, 139
\ref Chabrier G., 2003, PASP, 115, 763
\ref Chen Y.-C., Ho S., Freeman P.E., Genovese C.R., Wasserman L., 2015, arXiv 1501.05303
\ref Croft R., Weinberg D., Katz N., Hernquist L., 1998, ApJ, 495, 44%
\ref Danforth C.W., Shull J., 2008, ApJ, 679, 643
\ref Danforth C.W., Tilton E.M., Shull J.M., et al.\ 2014, arXiv 1402.2655
\ref Danforth C.W. TiltonE.M., Shull J.M., Keeney B.A., Stevans M., et al., 2014, arXiv 1402.2655  IGM survey
\ref Dav\'e R., Hernquist L., Katz N., Weinberg D.H., 1999, ApJ, 511, 521
\ref Dav\'e R., Cen R., Ostriker J.P., Bryan G.L., Hernquist L., Katz N.,  Weinberg D.H., Norman M.L. O'Shea B., 2001, ApJ, 552, 473
\ref Dav\'e, R., Oppenheimer B.D., Katz N., Kollmeier J.A., Weinberg D.H., 2010, MNRAS, 408, 2051
\ref Finlator K., Dav\'e R., Papovich C., Hernquist L., 2006, ApJ, 639, 672
\ref Fukugita M., Peebles P.J.E., 2006, ApJ, 639, 590
\ref Green J.C., Froning C.S., Osterman S., Ebbets, D., Heap S.H. Leitherer C., et al., 2012, ApJ, 744, 60
\ref Haardt F., Madau P., 2001, in Clusters of Galaxies and the High Redshift Universe Observed in X-rays, Recent Results of XMM-Newton and Chandra, XXXVI Rencontres de Moriond, XXI Moriond Astrophysics Meeting, ed. D.M. Neumann \& J.T.T. Van (Saclay, France: CEA), 64
\ref Haardt F., Madau P., 2012, ApJ, 746, 125
\ref Jarrett T.H., Chester T., Cutri R., Schneider S.E., Huchra J.P., 2003, AJ, 125, 525
\ref Kalberla P.M.W., Burton W.B., Hartmann D., Arnal E.M., Bajaja E., Morras R., P\"oppel W.G.L., 2005, A\&A, 440, 775
\ref Kalberla P.M.W., McClure Griffiths N.M., Pisano D.J., Calabretta M.R., Ford H.A., Lockman F.J., Staveley-Smith L., Kerp J., Winkel B., Murphy T., Newton-McGee K., 2010, A\&A, 512, 17 
\ref Kere\v{s} D., Katz N., Weinberg D.H., Dav\'e R., 2005, MNRAS, 363, 2
\ref Khaire V., Srianand R., 2015, arXiv:1503.07168
\ref Kim T.-S., Bolton J., Viel M., et al., 2007, MNRAS, 382, 1657%
\ref Kollmeier J.A., Weinberg D.H., Oppenheimer B.D., Haardt F., Katz N., Dav\'e R., Fardal M., Madau P, Danforth C., Ford A.B., Peeple M.S., McEwen J., 2014, ApJL, 789, L32
\ref Kriss G.A., 2011, COS IRS 2011-01, http://www.stsci.edu/hst/cos/performance/spectra\_resolution/
\ref Lanzetta K.M., Bowen D.V., Tytler D., Webb J.K., 1995, ApJ, 442, 538 
\ref Lee K.-G., Hennawi J.F., Spergel D.N., Weinberg D.H., Hogg D.W., Viel M., Bolton J.S., Baile S., Pieri M.M., Carithers W., Schlegel D.J., Lundgren B., Palanque-Delabrouille N., Suzuki N., Schneider D.P., Y\'eche C., 2015, ApJ, 799, 196
\ref Lehner N., Savage B.D., Richter P., Sembach K.R., Tripp T.M., Wakker B.P., 2007, ApJ, 658, 680
\ref Luki\'c Z., Stark C.W., Nugent P., White M., Meiksin A.A., Almgren A., 2015, MNRAS, 446, 3697
\ref Maller A.H., Bullock J.S., 2004, MNRAS, 355, 695
\ref Marzke R.O., Huchra J.P., Geller M.J., 1994, ApJ, 428, 43
\ref Meiring J.D., Tripp T.M., Prochaska J.X., Tumlinson J., Werk J., Jenkins E.B., Thom C., O'Meara J.M., Sembach K.R., 2011, ApJ, 732, 35
\ref Morris S.L., Weymann R.J. Dressler A., McCarthy P.J., Smith  B.A., Terrile R.J., Giovanelli R., Irwin M., 1993, ApJ, 419, 524
\ref Morris S.L., Januzzi B.T. 2006, MNRAS, 367, 1261
\ref Narayanan A., Wakker B.P., Savage B.D., Keeney B.A., Shull J.M., Stocke J.T., Sembach K.R., 2010, ApJ, 721, 960 
\ref Oliveira C., Beland S., Keyes C.T., Aloisi A., Niemi S., Osterman S., Proffitt C., 2010, STScI Calibration Workshop, Space Telescope Science Institute, Susana Deustua and Cristina Oliveira, eds., p408
\ref Oort J.H., 1970, A\&A 7, 381
\ref Oppenheimer B.D. Dav\'e R.A., 2008, MNRAS, 387, 587
\ref Oppenheimer B.D., Dav\'e R., Kere\v{s} D., Fardal M., Katz N., Kollmeier J.A., Weinberg D.H., 2010, MNRAS, 406, 2325
\ref Penton S., Stocke J., Shull J., 2002, ApJ, 565, 720
\ref Paschos P., Jena Trividesh, Tytler D., Kirkman D., Norman M.L., 2009, MNRAS 399, 1934
\ref Prochaska J., Weiner B., Chen H.-W, Mulchaey J., 2006, ApJ 643, 680%
\ref Rahmati A., Pawlik A.H., Rai{\v c}evi{\'c} M., Schaye J., 2013, MNRAS, 430, 2427 %
\ref Richter P. Savage B.D., Tripp T.M., Sembach K.R., 2004, ApJS, 153, 165
\ref Ryan-Weber E.V., 2006, MNRAS, 367, 1251%
\ref Rudie G., Steidel C.C., Trainor R.F., Rakic O., Bogosavljevic M., Pettini M., Reddy N., Shapley A.E., Erb D.K., Law D.R., 2012, ApJ, 750, 67%
\ref Rudie G., Steidel C.C., Shapley A.E., Pettini M., 2013, ApJ, 769, 146
\ref Savage B.D., Sembach K.R., 1991, ApJ, 379, 245 
\ref Savage B.D., Narayanan A., Lehner N., Wakker B.P., 2011, ApJ, 731, 14 
\ref Savage B.D., Kim, T.-S, Wakker B.P., Keeney B., Shull J.M., Stocke J.T., Green J.C., 2014, ApJ, 212, 8
\ref Schaye J., 2015, ApJ 559, 507
\ref Shull J.M., Smith B.D., Danforth C., 2012, ApJ, 759, 23
\ref Shull J.M., 2014, ApJ, 784, 142
\ref Shull J.M., Moloney J., Danforth C.W., Tilton E.M., 2015, arxiv 1502.00637
\ref Smith B.D., Hallman E.J., Shull J.M., O'Shea B.W., ApJ, 2011 731, 6
\ref Sousbie T., 2011, MNRAS, 414, 350
\ref Stone A.M., Morris S.L., Crighton N., Wilman R.J., 2010, MNRAS, 402, 2520%
\ref Stocke J.T., Penton S.V., Danforth C.W., Shull J.M., Tumlinson J., McLin 2006, ApJ 641, 217 
\ref Stocke J., Keeney B., Danforth C., Shull J., Froning C., Green J., Penton S., Savage B., 2013, ApJ, 763, 148
\ref Tejos N., Morris S.L, Finn W., Crighton N.H.M., Bechtold J., et al., 2014, MNRAS, 437, 2017
\ref Tempel E., Stoica R.S., Martinez V.J., Liivam\"agi L.J., Castellan G., Saar E., 2014, MNRAS, 438, 3465
\ref Tripp T.M., Lu L., Savage B.D., 1998, ApJ, 508, 200
\ref Tumlinson J., Thom C., Werk J.K., Prochaska J.X., Tripp T.M., et al, 2011, Science, 334, 948; 
\ref Tumlinson J., Thom C., Werk J.K., Prochaska J.X., Tripp T.M., Katz N., et al., 2013, ApJ, 777, 59
\ref Wakker B.P., Savage B.D.,, 2009, ApJ, 182, 378
\ref Weymann R., Januzzi B., Lu L., Bahcall J.N., Bergeron J., Boksenberg A., et al., 1998, ApJ, 506, 1
\end{document}